\documentclass[3p,times,11pt,sort&compress]{elsarticle}

\usepackage[utf8]{inputenc}
\usepackage{graphicx}
\usepackage{enumerate}
\usepackage{wrapfig}
\usepackage{amsmath,amsthm,amssymb,mathrsfs}
\usepackage{caption,subcaption}

\usepackage{soul}
\usepackage{cancel}
\usepackage{ulem}
\usepackage{algorithm}
\usepackage{amsfonts}
\usepackage{mathtools}
\usepackage[version=4]{mhchem}
\usepackage{dsfont}
\usepackage{url}
\usepackage{algpseudocode}
\usepackage{hyperref}
\usepackage{xcolor}
\usepackage{tcolorbox}
\usepackage{colortbl}
\usepackage{booktabs}
\usepackage{multirow}
\usepackage{paralist}

\usepackage[numbers]{natbib}   
\usepackage{bibunits}
\defaultbibliography{biblio/global,biblio/hypersonics,biblio/ml_pde,biblio/rom}
\defaultbibliographystyle{elsarticle-num-names-nourl}

\allowdisplaybreaks
\makeatletter
\def\ps@pprintTitle{%
 \let\@oddhead\@empty
 \let\@evenhead\@empty
 \let\@oddfoot\@empty
 \let\@evenfoot\@empty
}
\makeatother



\newcommand{\vb}{\mathbf{b}}
\newcommand{\ve}{\mathbf{e}}
\newcommand{\vf}{\mathbf{f}}

\newcommand{\vq}{\mathbf{q}}

\newcommand{\vy}{\mathbf{y}}

\newcommand{\vz}{\mathbf{z}}
\newcommand{\vw}{\mathbf{w}}

\newcommand{\vmu}{\boldsymbol{\mu}}
\newcommand{\vchi}{\boldsymbol{\chi}}

\newcommand{\mA}{\mathbf{A}}

\newcommand{\mU}{\mathbf{U}}

\newcommand{\mS}{\mathbf{S}}

\newcommand{\mF}{\mathbf{F}}
\newcommand{\mD}{\mathbf{D}}
\newcommand{\mT}{\mathbf{T}}

\newcommand{\hmA}{\hat{\mathbf{A}}}
\newcommand{\hvb}{\hat{\mathbf{b}}}

\newcommand{\vn}{\mathbf{n}}

\newcommand{\mGamma}{\boldsymbol{\Gamma}}
\newcommand{\vGamma}{\boldsymbol{\gamma}}
\newcommand{\vTheta}{\boldsymbol{\theta}}

\newcommand{\phib}{\boldsymbol{\Phi}}

\newcommand{\atom}{\text{\tiny O}}
\newcommand{\mol}{\text{\tiny O$_2$}}

\newcommand{\bd}{{\scalebox{.6}{$BD$}}}
\newcommand{\abc}{{\scalebox{.6}{$ABC$}}}

\newcommand{\eqspace}{\,}

\usepackage{xcolor}

\newcommand\fulltitle{MENO: Hybrid Matrix Exponential–based Neural Operator for Stiff ODEs \\ Application to Thermochemical Kinetics}

\begin{document}


\begin{bibunit}  

\begin{abstract}
    We introduce MENO (``Matrix Exponential–based Neural Operator''), a hybrid surrogate framework for efficiently solving stiff systems of ordinary differential equations (ODEs) that exhibit a sparse nonlinear structure.
In such systems, only a few variables contribute nonlinearly to the dynamics, while the majority affect the equations linearly.
MENO leverages this property by decomposing the system into two components: the few nonlinear variables are modeled using conventional neural operators, while the linear time-varying subsystem, describing the dynamics of the remaining variables, is integrated using a novel, neural matrix exponential formulation.
This method combines the exact solution of linear time-invariant systems with learnable, time-dependent graph-based corrections applied to the linear operators.
Unlike black-box or soft-constrained physics-informed (PI) networks, MENO embeds the governing equations directly into its architecture, ensuring physical consistency (e.g., steady-state solutions), improved robustness, and more efficient training.
We validate MENO on three increasingly complex real-world thermochemical reactive systems: the POLLU atmospheric chemistry model, an oxygen mixture in thermochemical nonequilibrium, and a collisional-radiative argon plasma deployed in one- and two-dimensional shock-tube configurations.
MENO achieves relative errors below 2\% in zero-dimensional settings and maintains good accuracy even in multidimensional extrapolatory regimes.
In addition to its predictive fidelity, MENO delivers significant computational speedups, achieving up to 4\,800$\times$ on GPU and 185$\times$ on CPU compared to standard implicit ODE solvers.
Although MENO is an intrusive method requiring access to the governing equations, its physics-based design offers greater reliability and generalization than purely data-driven or soft-constrained PI alternatives.
This tight coupling of physics and machine learning enables scalable, real-time simulation of stiff reactive systems.

\end{abstract}


\begin{frontmatter}

    \title{\fulltitle}

    \cortext[cor1]{Corresponding Author}

    \author{Ivan Zanardi}
    \ead{zanardi3@illinois.edu}
    \address{Department of Aerospace Engineering,
	University of Illinois Urbana-Champaign,
	Urbana, IL 61801, USA}
    
    \author{Simone Venturi}
    \ead{simone@atomicmachines.com}
    \address{Atomic Machines, Emeryville, CA 94608, USA}
    
    \author{Marco Panesi\corref{cor1}}
    \ead{mpanesi@uci.edu}
    \address{Department of Mechanical and Aerospace Engineering,
    University of California, Irvine, CA 92697, USA}

\end{frontmatter}

\section{Introduction}
The accurate simulation of thermochemically reacting systems remains a fundamental challenge in high-speed aerothermodynamics, plasma dynamics, and combustion modeling, where the capabilities of computational fluid dynamics (CFD) are ultimately constrained by available computational resources.
Within the scientific community, it is widely recognized that the current limitations stem not from an absence of physical models, but from the prohibitive cost of resolving them at full fidelity.
A significant gap persists between the detailed, high-fidelity models present in the literature and the simplified approximations that can be feasibly employed in practical simulations.
This gap arises from the inherently stiff and multiscale nature of thermochemical systems, which involve a large number of reactive scalar variables, each requiring the solution of a separate transport equation, along with a broad spectrum of time scales spanning from picoseconds (associated with fast chemical kinetics) to micro- or milliseconds (associated with fluid transport).
Accurately resolving this wide range of temporal scales demands extremely fine time-stepping, dramatically increasing computational cost.
Although modern CFD solvers can, in principle, capture such dynamics, the associated wall-clock time and hardware requirements render high-resolution simulations impractical for anything beyond reduced-order or idealized test cases.
Therefore, the stiffness introduced by detailed kinetic mechanisms poses a formidable bottleneck that cannot be easily overcome by conventional parallelization strategies, thereby limiting the scope and fidelity of large-scale, realistic simulations.

Historically, several strategies have been developed to mitigate the computational burden associated with stiff chemical kinetics in CFD.
These systems are often integrated using implicit schemes such as backward differentiation formulas (BDFs)~\cite{Curtiss1952IntegrationEquations}, typically embedded within operator-splitting frameworks~\cite{Strang_SIAM_1968} to decouple chemistry from fluid dynamics.
While BDF schemes offer stability for stiff problems, they require solving a linear system at each time step, making them computationally expensive.
A common alternative is chemical mechanism reduction, where skeletal models~\cite{Nouri2022SkeletalModes} or quasi-steady state (QSS)~\cite{Mott2000AKinetics,Goussis2012QuasiValidity,Park_AIAA_1969,Panesi_JCP_2013} assumptions eliminate fast-reacting intermediates to simplify the system.
These approaches reduce stiffness but often require case-specific tuning and may compromise accuracy. Tabulation methods, such as in situ adaptive tabulation (ISAT)~\cite{Pope1997ComputationallyTabulation} and PRISM~\cite{Tonse2003ComputationalPRISM}, accelerate simulations by reusing precomputed chemistry data; however, their performance degrades in high-dimensional or highly transient regimes.
Adaptive chemistry techniques~\cite{Liang2009TheFuels,Contino2011CouplingSimulations,Sun2017ATransport,DAlessio2020ImpactSimulations} dynamically adjust the reaction mechanism during runtime, improving efficiency at the cost of increased algorithmic complexity.
Manifold-based methods like intrinsic low-dimensional manifolds (ILDM)~\cite{Maas1992SimplifyingSpace} and computational singular perturbation (CSP)~\cite{Lam1994TheKinetics,Lu2001ComplexAnalysis} project the system onto its essential dynamics but rely on strong assumptions about timescale separation and involve substantial mathematical overhead.
While all these methods aim to reduce the cost of integrating finite-rate chemistry, each introduces trade-offs in terms of generality, robustness, or physical fidelity.

Recent advances in scientific machine learning (SciML) have opened promising avenues for modeling stiff thermochemical systems, offering compelling alternatives to conventional approaches.
Machine learning (ML)-based surrogates are increasingly used to address stiffness in finite-rate chemistry by providing fast, stable evaluations that eliminate the need for costly implicit solvers.
Moreover, these models can learn and leverage the low-dimensional structure underlying thermochemical kinetics, despite the system’s typically large state space.
As a result, SciML methods yield memory-efficient, computationally scalable surrogates that maintain physical fidelity while significantly accelerating simulations.
Building on this potential, recent efforts have introduced a variety of ML-driven strategies for chemistry acceleration.
These can be broadly grouped into three categories: (i) direct surrogation of reaction source terms, using standard deep neural networks (DNNs) or neural ordinary differential equations (NODEs)~\cite{Chen2018NeuralEquations,Poli2019GraphEquations,Kim2021StiffEquations};
(ii) reduced-order models (ROMs), employing either linear or nonlinear architectures (e.g., autoencoders) to reduce the dimensionality of the thermochemical state space~\cite{Sutherland2009CombustionAnalysis,Parente2013PrincipalSensitivity,Zanardi2025Petrov-GalerkinMixtures,Zanardi_SciTech_2025_v2,Zanardi2025Petrov-GalerkinPlasma};
and (iii) operator learning approaches~\cite{Kissas_arXiv_2022,Lu_CMAME_2022,Lu_NMI_2021,Li_arXiv_2020,Kovachki2023NeuralPDEs,You_JComP_2022}, which aim to directly approximate the solution operator of the governing equations.
In practice, many recent efforts adopt hybrid strategies that blend elements of these categories to balance accuracy, computational efficiency, and generalization.
Within the first category, numerous studies have shown that neural surrogates can effectively accelerate the integration of stiff systems, particularly in the context of chemical kinetics (see~\cite{Ihme2022CombustionProspects} for a comprehensive review).
For example, Kim et al.~\cite{Kim2021StiffEquations} introduced a derivative-aware training strategy to enhance the stability of NODEs in stiff systems, while Owoyele and Pal~\cite{Owoyele2022ChemNODE:Solvers} developed ChemNODE, a chemistry-specific NODE architecture leveraging the law of mass action.
Similarly, Caldana and Hesthaven~\cite{Caldana2024NeuralSystems} proposed a time-reparameterized NODE to dynamically adjust integration steps and mitigate stiffness.
In the second category, ROM approaches aim to reduce system complexity either through linear or nonlinear projections. Linear methods typically rely on orthogonal projections~\cite{Sutherland2009CombustionAnalysis,Parente2013PrincipalSensitivity} or oblique Petrov-Galerkin formulations~\cite{Zanardi2025Petrov-GalerkinMixtures,Zanardi_SciTech_2025_v2,Zanardi2025Petrov-GalerkinPlasma} to reduce the dimensionality of the governing equations.
Nonlinear approaches often employ autoencoders (AEs) to identify a latent, low-dimensional manifold.
For example, Zhang and Sankaran~\cite{Zhang2022AUTOENCODERSYSTEMS} leveraged AEs to compress the state space of reacting systems into a latent manifold, where unclosed terms are modeled by auxiliary DNNs, allowing for more efficient integration.
Hybrid strategies that blend linear and nonlinear techniques have also been explored.
Notably, Zdybał et al.~\cite{Zdyba2023ImprovingOutputs} combined a linear encoder with a nonlinear decoder to enhance the topological qualities of the reduced representation.
The third category encompasses (physics-informed) operator learning methods tailored for stiff systems.
Ji et al.~\cite{Ji2021Stiff-PINN:Kinetics} employed physics-informed neural networks (PINNs) that incorporate QSS assumptions to enhance training stability for stiff chemical kinetics.
Galaris et al.~\cite{Galaris2021NumericalRPNNs} introduced random projection neural networks specifically designed for stiff initial value problems.
Venturi and Casey~\cite{Venturi_CMAME_2023} extended the standard DeepONet framework proposed by Lu et al.~\cite{Lu_NMI_2021}, addressing its limitations in capturing rigid transformations and scalings in the data, which are commonly associated with singular value decomposition-based formulations.
Their enhanced architecture demonstrated improved accuracy and computational efficiency in modeling combustion chemistry and was later applied to fuel-specific systems by Kumar and Echekki~\cite{Kumar2024CombustionDeepONets}.
Hybrid approaches that combine elements from the aforementioned categories are also gaining traction.
Goswami et al.~\cite{Goswami2024LearningOperators} combined autoencoders with DeepONet to simultaneously reduce dimensionality and learn dynamics in combustion systems.
Similarly, Kumar and Echekki~\cite{Kumar2025Physics-constrainedKinetics} combined linear dimensionality reduction with React-DeepONet, a DeepONet variant for combustion, to model representative species dynamics and reconstruct non-representative ones while enforcing mass and elemental conservation.
Liu et al.~\cite{Liu2025NeuralODE} proposed ChemNNE, a hybrid emulator combining Fourier Neural Operators (FNOs)~\cite{Li_arXiv_2020} with attention mechanisms to improve generalization and runtime efficiency.
Finally, Vijayarangan et al.~\cite{Vijayarangan2024ATraining} introduced a hybrid NODE-autoencoder architecture, leveraging Jacobian-based insights to construct latent spaces that filter out fast, non-essential modes and alleviate stiffness.

While the aforementioned efforts have significantly advanced the use of ML for stiff chemical kinetics, many rely on black-box architectures with limited physical grounding.
Typically, physical consistency is imposed only through soft constraints, such as physics-informed loss terms, resulting in models that often lack robustness, generalizability, and strict adherence to physical symmetries and laws.
To address these limitations, Zanardi et al.~\cite{Zanardi_SciRep_2023} introduced a physics-inspired deep learning framework for reduced-order nonequilibrium thermochemical kinetics.
Their approach leverages a hierarchical structure to reflect the multiscale nature of the problem and incorporates physically meaningful transformations, such as Boltzmann equilibrium distributions, to enhance both robustness and predictive accuracy.
Nevertheless, concerns remain about the reliability of ML-based physics surrogates, as recently highlighted by McGreivy and Hakim~\cite{McGreivy2024WeakEquations}, who cautioned against the proliferation of overly optimistic results based on weak or unrepresentative baselines.
We share the view that ML is a powerful modeling tool, but we argue it must be tightly coupled with physical and mathematical insight.
Rather than using ML as an end in itself, we advocate for its integration within frameworks that explicitly encode the governing equations and structural properties of the physical system.
With this philosophy, we introduce MENO (``Matrix Exponential–based Neural Operator''), a novel surrogate modeling framework tailored for stiff thermochemical kinetics.
MENO falls broadly within the operator learning category, yet it is readily compatible with reduced-order modeling strategies.
A key insight motivating our approach is that many thermochemical systems, though formally nonlinear, exhibit a sparse nonlinear structure, where nonlinearity stems from only a limited subset of variables.
This observation enables a decomposition strategy inspired by exponential time differencing (ETD) integrators~\cite{Cox2002ExponentialSystems}, which treat the dominant stiff linear terms exactly via the matrix exponential, while handling nonlinearities separately.
Following this idea, MENO decomposes the state vector into two components: a low-dimensional nonlinear part, modeled with standard neural operators, and a dominant linear time-varying (LTV) subsystem.
This design restricts the use of black-box deep learning to only a few dynamical variables, aiming to enhance the model’s reliability.
The LTV system is derived by removing the equations corresponding to the nonlinear terms, which then appear only as inputs modulating the newly derived linear operators.
The resulting system is integrated using a novel, learnable neural matrix exponential approach.
This formulation builds on the exact solution of linear time-invariant (LTI) systems, providing a physically grounded \textit{a priori} baseline that improves physical consistency and simplifies training.
To account for time-dependent variability, the formulation is augmented with learnable graph-based, adaptive corrections.
We validate the framework on three challenging real-world problems: the POLLU atmospheric chemistry model, an oxygen mixture in thermochemical nonequilibrium, and a collisional-radiative argon plasma.
In the latter case, we further demonstrate MENO's generalizability and accuracy in both one- and two-dimensional shock-tube simulations, underscoring its potential as a scalable and physically consistent surrogate modeling framework.
Although developed with thermochemical applications in mind, MENO is general and applies to a broad class of differential systems characterized by sparse nonlinear structure.

The remainder of this paper is organized as follows. Section \ref{sec:method} introduces the problem formulation, reviews exponential integration methods, and presents the design, training, and deployment of the MENO framework.
Section \ref{sec:num_exp} presents results from the three numerical experiments introduced earlier, and section \ref{sec:conclusions} concludes with final remarks and directions for future work.

\section{Methodology}\label{sec:method}

\subsection{Problem statement}\label{sec:method:problem}
We consider stiff systems of ordinary differential equations (ODEs) of the general form
\begin{equation}\label{eq:nlin_sys}
    \frac{d}{dt}\vq(t) = \vf(\vq(t))\eqspace,
    \quad \vq(0) = \vq_0\left(\vmu\right)\eqspace,
\end{equation}
where $\vq \in \mathbb{R}^n$ is the time-dependent state vector, and the initial condition $\vq(0)$ is defined by the function $\vq_0: \mathbb{R}^{n_\mu} \rightarrow \mathbb{R}^n$, parameterized by $\vmu \in \mathbb{R}^{n_\mu}$.
Accordingly, the solution can be written as $\vq(t;\vmu)$, although we omit the explicit dependence on $\vmu$ for brevity. In the simplest case, $\vq_0$ may be the identity map, yielding $\vq(0) = \vmu$ and $n_\mu=n$.
A defining feature of the systems under consideration is their sparsely distributed nonlinearities: only a small subset of state variables contributes nonlinearly to the dynamics, while the remaining variables influence the system linearly.
Inspired by exponential time differencing integrators~\cite{Cox2002ExponentialSystems}, this property enables us to decompose the state vector as $\vq = [\vq_l, \vq_{nl}]$, where $\vq_l \in \mathbb{R}^{n_l}$ denotes the ``linear'' states, and $\vq_{nl} \in \mathbb{R}^{n_{nl}}$ the ``nonlinear'' ones, with $n = n_l + n_{nl}$.
To maximize the potential of our approach, it is crucial that the dimensionality of the linear component is significantly greater than that of the nonlinear component, meaning $n_l \gg n_{nl}$.

Our goal is to learn the integral solution of \eqref{eq:nlin_sys} by leveraging the dense linear core of the system. Ultimately, we aim to replace conventional implicit time integration schemes with a more efficient surrogate model, enabling faster predictions.
While the methodology is broadly applicable to any system with this structural property, we focus here on thermochemical zero-dimensional reactors as a primary application. As in recent works \cite{Zanardi_SciTech_2023,Zanardi_SciTech_2024,Zanardi_SciRep_2023,Goswami2024LearningOperators}, the learned surrogate can be integrated into computational fluid dynamics (CFD) simulations of reactive flows, without limiting the integration of the hydrodynamic equations by the stiffness of the chemical operator.
We propose a reusable, flexible surrogate model that is compatible with diverse CFD solvers, including those employing adaptive time-stepping and adaptive mesh refinement.
These features allow our approach to dynamically accommodate local Courant-Friedrichs-Lewy (CFL) constraints and varying integration time steps while maintaining physical and numerical consistency.

\subsection{Exponential integrators}\label{sec:method:exp_int}
This section reviews key concepts of exponential integrators, which constitute a core component of our methodology. Although matrix exponential-based methods are often underutilized due to their computational cost, they become advantageous in scenarios involving highly stiff systems~\cite{Baake2011TheSeries,Cox2002ExponentialSystems}, such as those addressed in this study.
\par
Consider a linear time-varying (LTV) system governed by the following differential equation:
\begin{equation}\label{eq:lin_sys}
    \frac{d}{dt}\vq(t) = \mA(t) \, \vq(t) + \vb(t)\eqspace ,
\end{equation}
where $\vq(t) \in \mathbb{R}^n$ is the state vector, $\mA(t) \in \mathbb{R}^{n \times n}$ is a continuous, time-dependent matrix (the system operator), and $\vb(t) \in \mathbb{R}^n$ is a time-dependent forcing term. According to linear system theory~\cite{Rugh1996LinearTheory}, the unique solution to this system can be expressed using the \textit{variation of constants formula}:
\begin{equation}\label{eq:lin_sys.sol}
    \vq(t)=\phib\left(t, t_0\right) \vq(t_0)+\int_{t_0}^t \phib(t, \tau) \vb(\tau) d \tau\eqspace, \quad t \geq t_0\eqspace ,
\end{equation}
where $\phib\left(t, t_0\right) \in \mathbb{R}^{n \times n}$ is the \textit{state transition matrix}, which evolves the homogeneous system (i.e., with $\vb(t) = \mathbf{0}$) from time $t_0$ to $t$.
\par
Several techniques exist to compute $\phib\left(t, t_0\right)$ associated with LTV systems~\cite{Rugh1996LinearTheory,Wu1980SolutionSystems,VanderKloet2000DiagonalizationSystems,Farkas1994PeriodicMotions,Jain2012ComputationSystems}. For example, the Lyapunov-Floquet transformation is applicable to periodic systems~\cite{Sinha1996Liapunov-FloquetSystems}, while the more general \textit{Peano-Baker series}~\cite{Baake2011TheSeries} provides a formal infinite-series solution, which is generally infeasible to compute except for relatively simple cases.
For example, when the matrix \(\mA(t) = \mA\) is constant, an exact formulation is provided, and the series simplifies to:
\begin{equation}\label{eq:lin_sys.transmat}
    \phib\left(t, t_0\right) = e^{\mA(t-t_0)} = \sum_{k=0}^{\infty}\frac{1}{k!}\mA^k(t-t_0)^k \eqspace,
\end{equation}
that converges uniformly and absolutely on any time interval $[-T,T]$, where $T>0$ is arbitrary~\cite{Rugh1996LinearTheory}.
While the \textit{Peano-Baker series} is rarely used for numerical computation, it provides theoretical insight, suggesting that the transition matrix $\phib\left(t, t_0\right)$ approximates, but does not exactly follow, an exponential form~\cite{Rugh1996LinearTheory}. 
A notable examples occurs when $\mA(t)$ commutes with its integral $\int_{t_0}^t \mA(\tau) d \tau$, in which case $\phib\left(t, t_0\right)$ simplifies to $\exp\int_{t_0}^t \mA(\tau) d \tau$.
While this simplification is elegant, the commutativity condition is generally difficult to verify and holds only under specific circumstances, which are detailed in section~\ref{suppl:method:matexp_commute} of the Supplementary Material.

The second term in equation \eqref{eq:lin_sys.sol} accounts for the inhomogeneous contribution due to the external forcing. If both $\mA$ and $\vb$ are constant and $t_0 = 0$, this integral simplifies to:
\begin{equation}\label{eq:lin_sys.action.const}
    \int_0^t \phib(t, \tau) \vb(\tau) d \tau
        = \int_0^t e^{\mA (t-\tau)} \vb d \tau
        = \left(e^{\mA t}-\mathbf{I}\right) \mA^{-1} \vb\eqspace,
\end{equation}
which leads to the closed-form solution of \eqref{eq:lin_sys.sol} with constant operators and $t_0 = 0$:
\begin{equation}\label{eq:lin_sys.sol.const}
    \vq(t)=e^{\mA t} \vq(0)
    + \left(e^{\mA t}-\mathbf{I}\right) \mA^{-1} \vb
    \eqspace, \quad t \geq 0 \eqspace.
\end{equation}
This expression is particularly attractive due to its simplicity and ease of computation within standard numerical libraries.
The intermediate solution $\vq_b = \mA^{-1} \vb$ can be readily obtained by solving the linear system $\mA \vq_b = \vb$, and the matrix exponential $e^{\mA t}$ can be computed using Padé approximants~\cite{Higham2006TheRevisited,Al-Mohy2009AExponential} or more efficient methods for evaluating its action on vectors~\cite{Al-Mohy2011ComputingIntegrators}.

\subsection{Matrix Exponential–based Neural Operator (MENO)} \label{sec:method:meno}
To leverage the structural properties of the equations described in section \ref{sec:method:problem}, we introduce a hybrid surrogate called MENO (``Matrix Exponential–based Neural Operator'').
This model separately treats the nonlinear and linear components of the system, leveraging the relatively low dimensionality of the nonlinear variables and the predominantly linear nature of the system.
MENO yields a computationally efficient and accurate surrogate that preserves the essential physical properties of the original system.

\paragraph{Learning the nonlinear dynamics}
In the first stage, MENO targets the low-dimensional set of nonlinear variables, denoted by $\vq_{nl} \in \mathbb{R}^{n_{nl}}$. We employ conventional neural operators to predict their trajectories over time from parametric initial conditions $\vmu \in \mathbb{R}^{n_\mu}$.
Specifically, we adopt the \textit{flexDeepONet} model introduced by Venturi and Casey~\cite{Venturi_CMAME_2023}, an extension of the original DeepONet~\cite{Lu_NMI_2021}.
This architecture includes an additional component called the ``pre-net,'' which applies a shift and scaling transformation to the input coordinate, in our case time.
Given the similarity between DeepONet and other basis-decomposition methods like Proper Orthogonal Decomposition (POD) or, more broadly, Singular Value Decomposition (SVD), the purpose of the ``pre-net'' is to overcome the limitations of these techniques in capturing dynamics with intrinsic rigid transformations and scalings~\cite{Venturi_CMAME_2023}.

\paragraph{Neural exponential integrators for linear dynamics}
Once the few nonlinear variables \(\vq_{nl}(t)\) are learned, we use them to parameterize a LTV system governing the remaining linear states \(\vq_l(t) \in \mathbb{R}^{n_l}\). This results in the following differential equation:
\begin{equation}\label{eq:meno:ltv}
    \frac{d}{dt}\vq_l(t) = \mA(t)\, \vq_l(t) + \vb(t)\eqspace,
\end{equation}
where $\mA(t) \in \mathbb{R}^{n_l \times n_l}$ and $\vb(t) \in \mathbb{R}^{n_l}$ are time-dependent operators whose coefficients are functions of $\vq_{nl}(t)$. That is, $\mA(t) = \mA(\vq_{nl}(t))$ and $\vb(t) = \vb(\vq_{nl}(t))$.

In light of the considerations expressed in section \ref{sec:method:exp_int}, we propose a novel deep learning-based formulation of equation \eqref{eq:lin_sys.sol.const} to solve the LTV system \eqref{eq:meno:ltv}, while preserving the exponential nature of the transition matrix. Specifically, the solution is expressed as:
\begin{equation}\label{eq:lin_sys.sol.ml}
    \vq_l(t)=e^{\hmA(t) t} \vq_l(0)
    + \left(e^{\hmA(t) t}-\mathbf{I}\right) \hmA(t)^{-1} \hvb(t)
    \eqspace, \quad t \geq 0
\end{equation}
where
\begin{align}
    \hmA(t) & = \mGamma(t,\vmu,\mA(t)) \eqspace , \label{eq:lin_ops.corr.A} \\
    \hvb(t) & = \vGamma(t,\vmu,\vb(t)) \eqspace . \label{eq:lin_ops.corr.b}
\end{align}
Here, \(\mGamma: \mathbb{R}\times\mathbb{R}^{n_\mu}\times\mathbb{R}^{n_l\times n_l} \rightarrow \mathbb{R}^{n_l\times n_l}\) and \(\vGamma: \mathbb{R}\times\mathbb{R}^{n_\mu}\times\mathbb{R}^{n_l} \rightarrow \mathbb{R}^{n_l}\) are learnable, time-dependent corrective functions applied to the original linear operators \(\mA(t)\) and \(\vb(t)\), respectively.
The role of $\mGamma$ is to adapt the transition matrix by correcting $\mA(t)$ such that the exponential operator better approximates the true system evolution.
When \(\mA(t)\) commutes with its time integral, \(\mGamma\) effectively captures the integral value in equation \eqref{eq:lin_sys.transmat.commute}.
Similarly, \(\vGamma\) adjusts the forcing term \(\vb(t)\) to ensure its contribution is faithfully represented. Under the commutativity assumption, \(\vGamma\) effectively approximates the integral expression of the inhomogeneous solution as given in equation~\eqref{eq:lin_sys.action.const}.
In the simplest formulation, these corrections are applied via the Hadamard product, a choice justified by the inherently component-wise nature of tensor integration, which holds irrespective of dimensionality.

The use of matrix exponential in general LTV systems is theoretically valid only when $\mA(t)$ commutes with its integral, a condition that is difficult to prove analytically. Therefore, we propose a practical diagnostic to evaluate the suitability of this approximation. Specifically, we define a relative variation metric:
\begin{equation}\label{eq:delta_a}
    \Delta \mA (t) = 100\frac{1}{M}\sum_{k=1}^M
    \left.\left\|\frac{d\mA^{(k)}(t)}{dt}\right\|_F \,\middle/\,
    \left\|\frac{\mA^{(k)}(t)}{t}\right\|_F \right. \eqspace,
\end{equation}
where $\|\cdot\|_F$ denotes the Frobenius norm, and $M=100$ is the number of representative trajectory samples. A small value of $\Delta \mA(t)$ across all times indicates that $\mA(t)$ evolves slowly and can be treated as approximately constant, thereby justifying the use of the exponential ansatz.

\paragraph{Graph-based nonlinear corrections}
The corrective functions \(\mGamma\) and \(\vGamma\) can be designed based on the structural characteristics of the underlying physics. For thermochemical systems, the entries of the linear operators \(\mA(t)\) and \(\vb(t)\) are typically determined by reaction rate constants, which may be scaled by the components of the nonlinear state vector, \(\vq_{nl}(t)\).
Our approach focuses on applying corrections directly to the forward reaction rates, while backward rates are recovered using equilibrium constants. This ensures that the steady-state equilibrium composition, when it exists, is preserved by construction.
The proposed formulation for \(\mGamma\) and \(\vGamma\) is inspired by graph-based methods~\cite{Pepiot-Desjardins2008AnMechanisms,Lehmann2004AnSystems,Albert2002StatisticalNetworks,Kolaczyk2009StatisticalModels,Mizui2017GraphicalNetworks,Holmes2021GraphEngineering,Venturi2023AnGraphs}, where the reaction mechanism is represented as a graph. In this representation, nodes correspond to chemical species \(s \in \mathcal{S}_l\), with $\mathcal{S}_l$ denoting the set of species governed by linear dynamics, and edges denote the chemical reactions connecting them.
The objective is to compute edge-specific correction factors (i.e., scaled forward reaction rates) based on the states of connected species.
The graph-based correction procedure consists of three main stages, detailed below and illustrated with an example in section~\ref{suppl:method:graph_example} of the Supplementary Material.

\begin{figure}[htb!]
    \centering
    \includegraphics[width=0.70\textwidth]{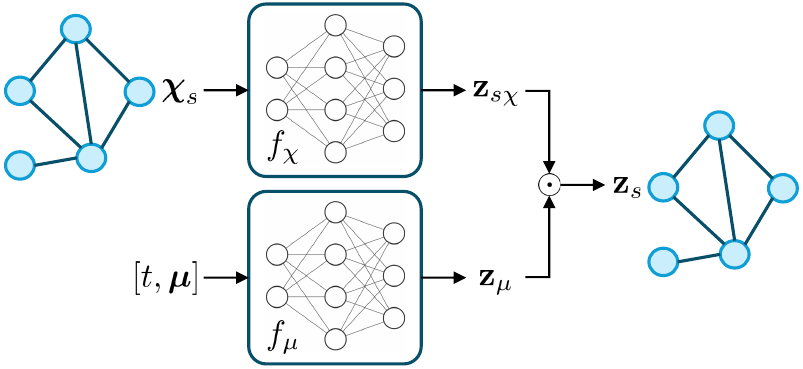}
    \caption{\textit{Nodes encoding}. Each node (species) \(s \in \mathcal{S}_l\) is encoded through a nonlinear transformation that combines two MLPs: $f_\chi$ encodes static (physical) features specific to the node, and $f_\mu$ captures time and initial condition information.}
    \label{fig:graph.encode}
\end{figure}
\begin{figure}[htb!]
    \centering
    \includegraphics[width=0.60\textwidth]{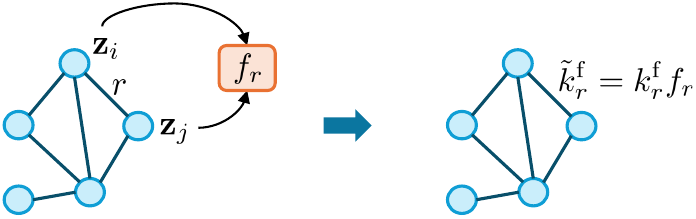}
    \caption{\textit{Message passing and edges updating.} In the message passing step (left), information from adjacent nodes is aggregated to compute the corrective factor \(f_r\) for reaction \(r\). In the edge updating step (right), the forward rate constant \(k_r^\mathrm{f}\) is scaled by \(f_r\) to obtain the corrected rate, $\tilde{k}_r^\mathrm{f}$.}
    \label{fig:graph.pass_update}
\end{figure}

\begin{enumerate}
    \item \textit{Nodes encoding}. Each node (species) $s\in\mathcal{S}_l$ is encoded using a learnable nonlinear transformation defined as follows:
    \begin{align}
        [\vz_{s\chi}, c_s, d_s] & = f_\chi(\vchi_s;\vTheta_\chi) \eqspace, \\
        \vz_{\mu} & = f_\mu(t,\vmu;\vTheta_\mu) \eqspace, \\
        \vz_s & = [d_s\vz_{\mu}\odot \vz_{s\chi}, c_s] \eqspace,
    \end{align}
    where $\vchi_s\in\mathbb{R}^{n_\chi}$ represents a set of static (physical) features corresponding to the $s$-th node.
    A multilayer perceptron (MLP) $f_\chi:\mathbb{R}^{n_\chi}\rightarrow\mathbb{R}^{p+2}$, parameterized by \(\vTheta_\chi\), encodes the static features of each node into a latent representation. Its output includes a \(p\)-dimensional embedding $\vz_{s\chi}\in\mathbb{R}^{p}$, along with two scalar parameters \(c_s\) and \(d_s\), which serve as node-specific shift and scale factors, respectively. A second MLP $f_\mu:\mathbb{R}^{1+n_\mu}\rightarrow\mathbb{R}^p$, parameterized by \(\vTheta_\mu\), maps the current time and initial conditions to a time-dependent latent vector $\vz_{\mu}\in\mathbb{R}^{p}$. The final latent state for node \(s\), denoted $\vz_s\in\mathbb{R}^{p+1}$, is constructed by element-wise multiplying \(\vz_{s\chi}\) and \(\vz_{\mu}\), scaling the result by \(d_s\), and appending the shift coefficient \(c_s\). This node encoding process is illustrated in figure \ref{fig:graph.encode}.

    \item \textit{Message passing}. For each edge (reaction) \(r\), we compute a correction factor \(f_r\) based on the latent embeddings of its adjacent nodes (see figure \ref{fig:graph.pass_update}, left):
    \begin{equation} \label{eq:edge_corr}
        f_r = \exp\left(\sum_{s\in\mathcal{S}_r\subseteq\mathcal{S}_l}\nu_s\mathbf{1}^\intercal\vz_s\right) \eqspace,
    \end{equation}
    where \(\mathcal{S}_r\) is the set of species involved in reaction \(r\), taken as a subset of the linearly modeled species \(\mathcal{S}_l\), and \(\nu_s\) denotes the stoichiometric coefficient of species \(s\) (negative for reactants).
    Equation \eqref{eq:edge_corr} mirrors the structure used in the computation of equilibrium constants~\cite{Vincenti1965IntroductionDynamics2}, with the exponential function ensuring that \(f_r\) remains strictly positive.

    \item \textit{Edges updating}. The correction factor \(f_r\) is used to scale the forward rate constant as $\tilde{k}_r^\mathrm{f} = k_r^\mathrm{f} f_r$ (see figure \ref{fig:graph.pass_update}, right). The corresponding backward rate constant is then obtained as \(\tilde{k}_r^\mathrm{b} = \tilde{k}_r^\mathrm{f} / k_r^\mathrm{eq}\), where \(k_r^\mathrm{eq}\) denotes the known equilibrium constant for reaction \(r\).
\end{enumerate}

\subsection{Model training, performance assessment, and deployment}\label{sec:method:train.perf}
\paragraph{Training}
The construction of the model is performed in three stages, following the split between nonlinear and linear state variables described in earlier sections:
\begin{enumerate}
    \item \textit{Nonlinear dynamics}. Each entry of the nonlinear state vector \(\vq_{nl}(t)\) is learned independently by a separate \textit{flexDeepONet} network, which takes as inputs the current time \(t\) and the parameter vector \(\vmu\) encoding the initial conditions.
    \item \textit{Linear dynamics}.
    The two MLPs in MENO, \(f_\chi\) and \(f_\mu\), introduced in section~\ref{sec:method:meno}, are trained to accurately model the time evolution of the linear state vector \(\vq_l(t)\), conditioned on the current time \(t\) and the initial condition vector \(\vmu\), using the neural matrix exponential integrator defined in equation~\eqref{eq:lin_sys.sol.ml}.
    During this stage, the true trajectories of \(\vq_{nl}(t)\) are provided as input.
    \item \textit{Joint fine-tuning}. The final stage involves joint fine-tuning of all pretrained components to predict the full state vector \(\vq(t;\vmu)\). This final step ensures that interactions between the nonlinear and linear components are captured coherently within the learned model.
\end{enumerate}
Note that the first two stages can be executed independently and in parallel, while the final fine-tuning stage requires joint optimization of the full architecture.
All surrogate models presented in this work were developed and trained using PyCOMET~\cite{Zanardi2025PyCOMET:Discovery}, a TensorFlow-based~\cite{TF_2016} library designed for physics-informed applications and operator learning. 

\paragraph{Performance assessment}
Model accuracy is evaluated using the mean absolute percentage error (MAPE) for each state component \(q_i\) for $i\in\{1,\dots,n\}$:
\begin{equation}\label{eq:mape}
	\mathrm{e}_i=\frac{1}{N_tN_\mu}\sum_{j=1}^{N_t}\sum_{k=1}^{N_\mu}
	\dfrac{|\,\hat{q}_i(t_j;\vmu_k) - q_i(t_j;\vmu_k)\,|}
	{|\,q_i(t_j;\vmu_k)\,|+\varepsilon} \eqspace,
\end{equation}
where \(\hat{q}_i\) and \(q_i\) denote the predicted and ground truth values, respectively. Here, \(N_t = 200\) is the number of (log-uniformly sampled) time points per trajectory, \(N_\mu=100\) is the number of test trajectories, and \(\varepsilon=10^{-8}\) prevents division by zero.

To quantify computational speedup, we compare MENO inference time against integration via the implicit backward differentiation formula (BDF) from the LSODE (``Livermore Solver for Ordinary Differential Equations'') toolbox~\cite{Radhakrishnan_LSODE_1993}.
This solver is employed by \textsc{plato}~\cite{Munafo2025PlatoPlasmas}, a physico-chemical library designed to compute thermodynamic properties and evaluate source terms associated with nonequilibrium collisional and radiative processes, for integrating zero-dimensional thermochemical reactor systems.
MENO's performance is benchmarked directly against this reference implementation.
For the solver, we fix the absolute tolerance at zero and sweep the relative tolerance from \(10^{-9}\) to \(10^{-6}\) and the scheme order from 2 to 4. The mean integration time is computed by solving 1\,000 representative trajectories up to a final time \(t_f\).
For MENO, inference is performed on a single batch of 10\,000 samples to predict \(\vq(t_f)\) for the same set of trajectories (repeated 10 times) to mimic the typical workload per core in parallel CFD, and we report the average inference time. The reported inference time is averaged per data point within the batch.
MENO's runtime is entirely evaluated in \texttt{C++} using the TF2 (``TensorFlow to Fortran'') library~\cite{Zanardi_TF2_2025}, which enables seamless integration of TensorFlow models into \texttt{C++} and \texttt{Fortran} applications.
This is achieved by leveraging the CppFlow library~\cite{cppflow_2022}, a \texttt{C++} wrapper for the TensorFlow \texttt{C} API~\cite{TF_2016}.
Details on the hardware configurations used are provided in section~\ref{suppl:method:devices} of the Supplementary Material.

\paragraph{Deployment}
Following the approach of Zanardi et al. \cite{Zanardi_SciTech_2023,Zanardi_SciTech_2024,Zanardi_SciRep_2023}, we employ Strang’s operator-splitting method \cite{Strang_SIAM_1968} to integrate the thermochemical surrogate model within CFD solvers.
In the context of standard reactive flow solvers, which govern inviscid (Euler) or viscous (Navier-Stokes) fluid systems, this method integrates the transport operator, $\boldsymbol{\mathcal{T}}\left(\mathbf{U}\right)$, and the reaction operator, $\boldsymbol{\mathcal{R}}\left(\mathbf{U}\right)$, sequentially in a symmetric fashion:
\begin{align}
	\partial_t\mathbf{U}^{(1)} & = \boldsymbol{\mathcal{T}}\left(\mathbf{U}^{(1)}\right)\eqspace, & \hspace{15mm} &
	\hspace{-5cm}\mathbf{U}^{(1)}\left(t_n\right)=\mathbf{U}_n \label{eq:split.01}\\
	\partial_t\mathbf{U}^{(2)} & = \boldsymbol{\mathcal{R}}\left(\mathbf{U}^{(2)}\right)\eqspace, & \hspace{15mm} &
    \hspace{-5cm}\mathbf{U}^{(2)}\left(t_n\right)=\mathbf{U}^{(1)}\left(t_n+\Delta t/2\right) \label{eq:split.02} \\
	\partial_t\mathbf{U}^{(3)} & = \boldsymbol{\mathcal{T}}\left(\mathbf{U}^{(3)}\right)\eqspace, & \hspace{15mm} &
	\hspace{-5cm}\mathbf{U}^{(3)}\left(t_n+\Delta t/ 2\right)=\mathbf{U}^{(2)}\left(t_n+\Delta t\right) \label{eq:split.03} \\
	\mathbf{U}_{n+1} &=\mathbf{U}^{(3)}\left(t_n+\Delta t\right)\eqspace. & & \label{eq:split.04}
\end{align}
Here, $\Delta t$ denotes the time step, and $\mathbf{U}$ represents the state vector of conservative variables, including thermochemical and hydrodynamic quantities such as species partial densities, momentum, and total energy.
The operator-splitting formulation enables seamless integration of the neural surrogate. Specifically, instead of solving the stiff reaction step \eqref{eq:split.02} via an implicit method, the trained model directly evolves the state in time.
It takes the output of the flux integration step as input and provides the updated gas state for the final stage.
All flow simulations are performed using \textsc{hegel}~\cite{Munafo2024HEGEL:Simulations}, a parallel, multi-block, structured solver for plasma hydrodynamics. The solver is coupled with \textsc{plato} to provide reference solutions and with MENO, integrated via the TF2 library, to evaluate the surrogate model.

\section{Numerical experiments}\label{sec:num_exp}
In this section, we evaluate the performance of the proposed MENO framework across three thermochemical systems of increasing complexity.
The first is the well-established POLLU problem, a benchmark in atmospheric chemistry involving 20 chemical species and 25 reactions, which models the temporal evolution of pollutant concentrations in air.
The second system is an adiabatic oxygen mixture in thermochemical nonequilibrium, involving 47 degrees of freedom that account for chemical reactions and the excitation of internal energy modes, including vibrational and rotational states.
This test case is particularly relevant in the hypersonics community~\cite{Venturi2020Data-InspiredSystem,Grover2019VibrationalCollisions}, as it models the kinetics of oxygen, one of the dominant species in Earth’s upper atmosphere, and plays a crucial role in accurately predicting thermal and chemical behavior during atmospheric reentry of space vehicles.
The third case is a collisional-radiative argon plasma model that includes excitation, ionization, and radiative energy transfer processes.
It is commonly used in CFD shock-tube simulations to study reactive plasma flows.
This test case serves as a well-established benchmark in the plasma physics community due to its representative complexity and relevance to high-temperature plasma applications~\cite{Vlcek1989AData,Bultel2002InfluenceModel2,Kapper2011IonizingStructure}.

\subsection{Example 1: POLLU problem}\label{sec:num_exp:pollu}
POLLU is an atmospheric chemistry model developed at the National Institute for Public Health and the Environment (RIVM), originally described by Verwer~\cite{Verwer1994GaussSeidelKinetics}. It models the temporal evolution of chemical species concentrations in polluted air using a stiff system of ODEs.
The state vector is defined as \(\vq = \vy \in \mathbb{R}^{20}\), where each component \(y_i\) represents the concentration of the \(i\)-th chemical species, measured in parts per million (ppm). The reaction mechanism comprises 25 chemical reactions involving 20 species.
\begin{table}[!htb]
    \centering
    \begin{tabular}{lll|lll}
    \arrayrulecolor{black}\midrule
    Index & Source term & Rate constant & Index & Source term & Rate constant \\
    \arrayrulecolor{black}\midrule
    1 & $k_1 {\color{blue}y_1}$ & $3.50$ & 14 & $k_{14} {\color{blue}y_1} {\color{blue}y_6}$ & $1.63\times 10^5$ \\
    2 & $k_2 {\color{blue}y_2} y_4$ & $2.66\times 10^2$ & 15 & $k_{15} y_3$ & $4.80\times 10^7$ \\
    3 & $k_3 y_5 {\color{blue}y_2}$ & $1.20\times 10^5$ & 16 & $k_{16} y_4$ & $3.50\times 10^{-3}$ \\
    4 & $k_4 y_7$ & $8.60\times 10^{-3}$ & 17 & $k_{17} y_4$ & $1.75\times 10^{-1}$ \\
    5 & $k_5 y_7$ & $8.20\times 10^{-3}$ & 18 & $k_{18} y_{16}$ & $1.00\times 10^9$ \\
    6 & $k_6 y_7 {\color{blue}y_6}$ & $1.50\times 10^5$ & 19 & $k_{19} y_{16}$ & $4.44\times 10^{12}$ \\
    7 & $k_7 y_9$ & $1.30\times 10^{-3}$ & 20 & $k_{20} y_{17} {\color{blue}y_6}$ & $1.24\times 10^4$ \\
    8 & $k_8 y_9 {\color{blue}y_6}$ & $2.40\times 10^5$ & 21 & $k_{21} y_{19}$ & $2.10\times 10^1$ \\
    9 & $k_9 y_{11} {\color{blue}y_2}$ & $1.65\times 10^5$ & 22 & $k_{22} y_{19}$ & $5.78\times 10^1$ \\
    10 & $k_{10} y_{11} {\color{blue}y_1}$ & $9.00\times 10^4$ & 23 & $k_{23} {\color{blue}y_1} y_4$ & $4.74\times 10^{-1}$ \\
    11 & $k_{11} y_{13}$ & $2.20\times 10^{-1}$ & 24 & $k_{24} y_{19} {\color{blue}y_1}$ & $1.78\times 10^4$ \\
    12 & $k_{12} y_{10} {\color{blue}y_2}$ & $1.20\times 10^5$ & 25 & $k_{25} y_{21}$ & $3.12\times 10^1$ \\
    13 & $k_{13} y_{14}$ & $1.88\times 10^1$ & & & \\
    \arrayrulecolor{black}\midrule
    \end{tabular}
    \caption{\textit{Reaction mechanism for the POLLU problem.} Each reaction is shown with its corresponding source term expression and rate constant \(k\). Species that contribute nonlinearly to the ODE system are highlighted in blue.}
    \label{tab:pollu}
\end{table}
Table~\ref{tab:pollu} summarizes the full set of reactions, including their source terms in the ODE system and corresponding reaction rate constants, \(k\). Further details on the model derivation and formulation are available in~\cite{Verwer1994GaussSeidelKinetics}.

Following the methodology described in section~\ref{sec:method:meno}, table~\ref{tab:pollu} identifies the species \(y_1\), \(y_2\), and \(y_6\) (highlighted in blue) as the only contributors to the system's nonlinearity. These species introduce quadratic terms into the governing ODEs, making them the sole source of nonlinear behavior.
Accordingly, we partition the full set of variables into two groups: the nonlinear state vector \(\vq_{nl} = [y_1, y_2, y_6] \in \mathbb{R}^3\), and the remaining variables, which form the linear state vector \(\vq_l \in \mathbb{R}^{17}\). Since the dimensionality of the nonlinear part is much smaller than that of the linear one, this problem is well-suited for the MENO framework.
To model the nonlinear dynamics, we employ three independent \textit{flexDeepONet}, each tasked with learning the temporal evolution of one component of \(\vq_{nl}\). The linear ODE subsystem for \(\vq_l\), conditioned on the learned nonlinear trajectories, is then solved using the neural matrix exponential integrator defined in equation~\eqref{eq:lin_sys.sol.ml}.
\begin{figure}[htb!]
\centering
\includegraphics[width=0.35\textwidth]{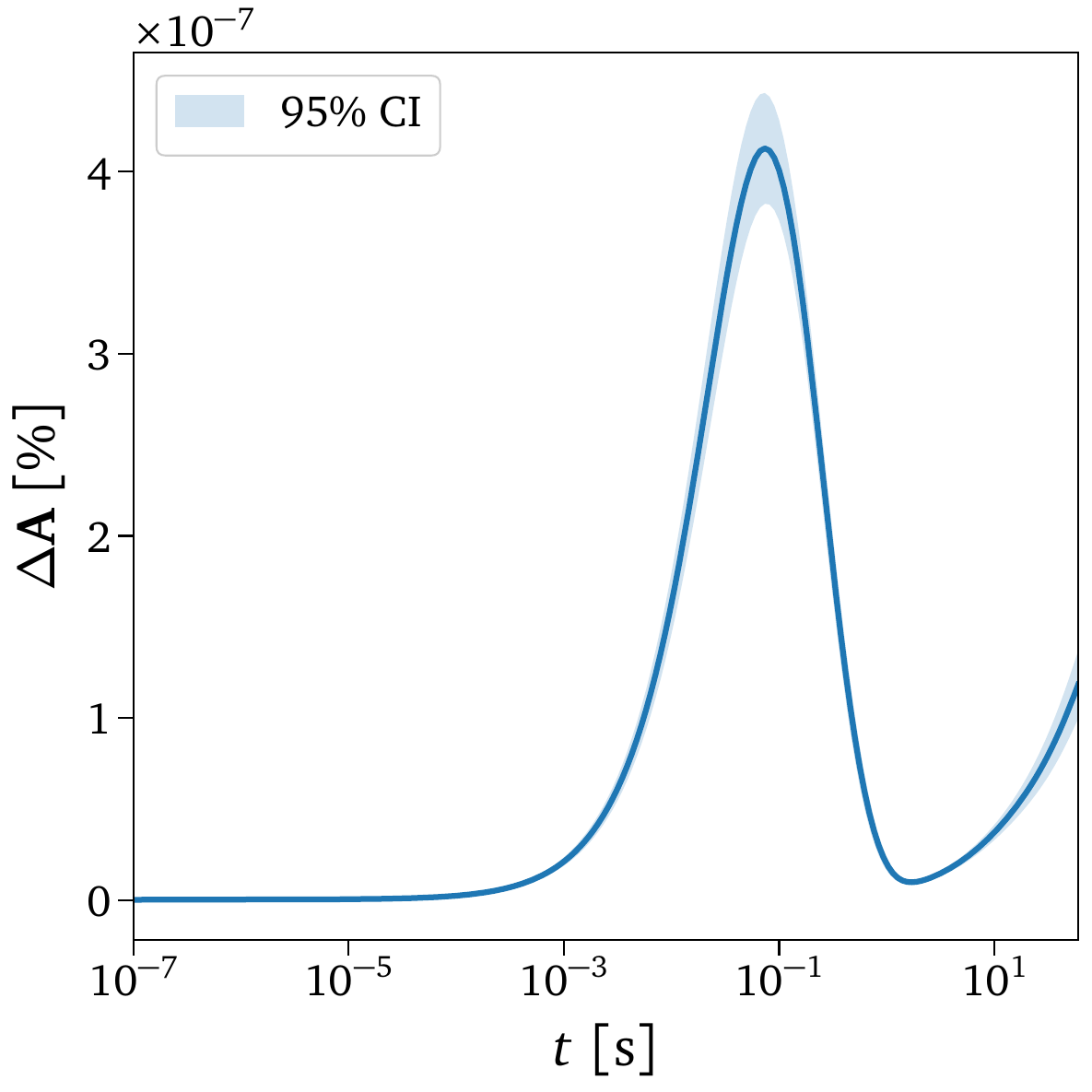}
\caption{\textit{Variability of the operator \(\mA(t)\) in the POLLU problem}. The mean variability of operator $\mA$ shown with a 95\% confidence interval, calculated using equation \eqref{eq:delta_a} across 100 test trajectories.}
\label{fig:pollu.delta_a}
\end{figure}
To validate the applicability of this formulation, we compute the variability metric \(\Delta \mA(t)\) (equation~\eqref{eq:delta_a}) over 100 test trajectories. Figure~\ref{fig:pollu.delta_a} shows the mean value of this metric, along with 95\% confidence intervals, across the time interval considered.
The consistently small values indicate that \(\mA(t)\) is approximately time-invariant, justifying the use of the matrix exponential. Consequently, most corrective learning is applied to the linear forcing term \(\vb(t)\), rather than the transition matrix.

Further details on data generation, model architecture, and training are provided in section~\ref{suppl:num_exp:pollu} of the Supplementary Material.

\begin{figure}[htb!]
    \centering
    \begin{subfigure}[htb!]{0.32\textwidth}
        \centering\hspace{1.5mm}
        \includegraphics[width=0.95\textwidth]{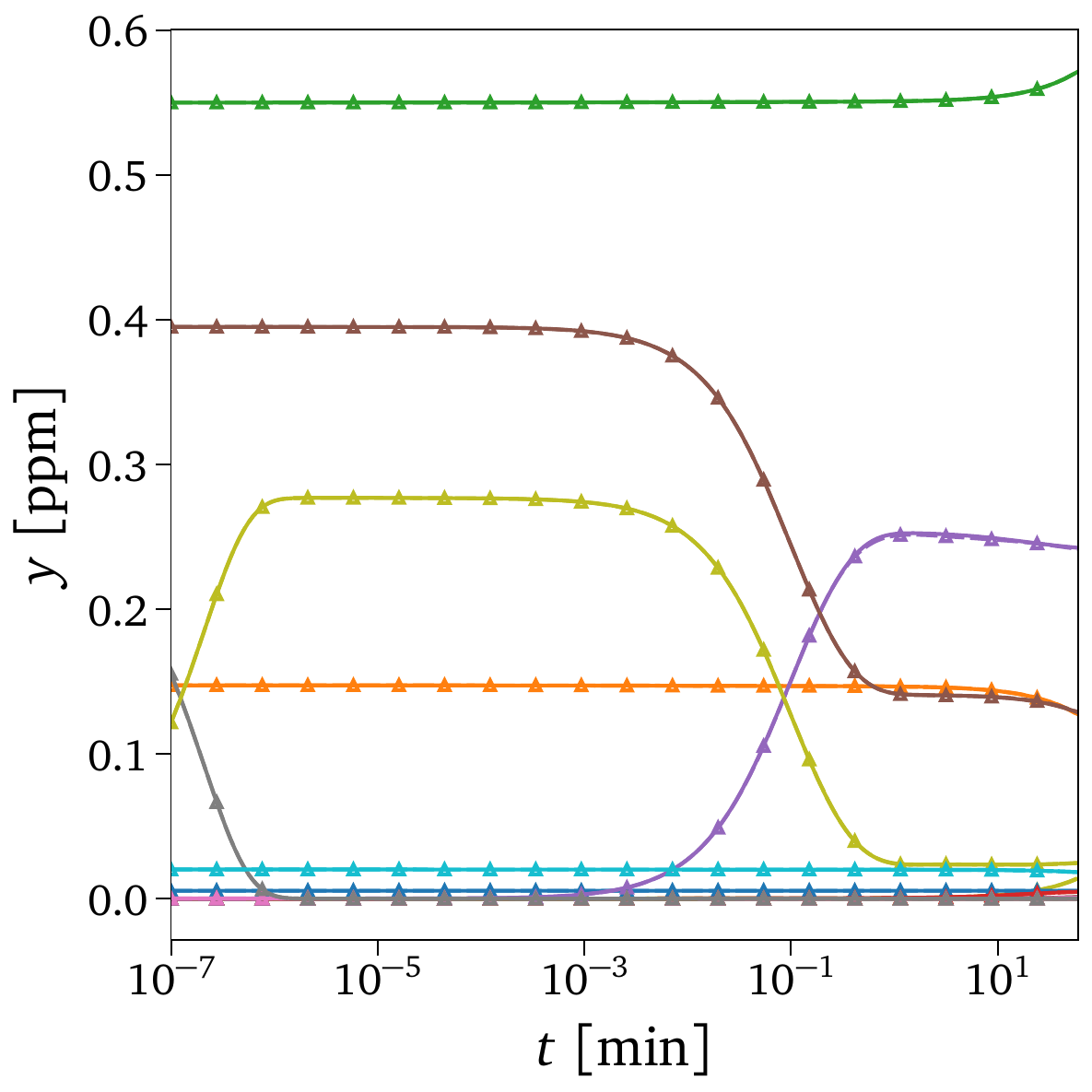}
    \end{subfigure}
    \begin{subfigure}[htb!]{0.32\textwidth}
        \centering\hspace{1.5mm}
        \includegraphics[width=0.95\textwidth]{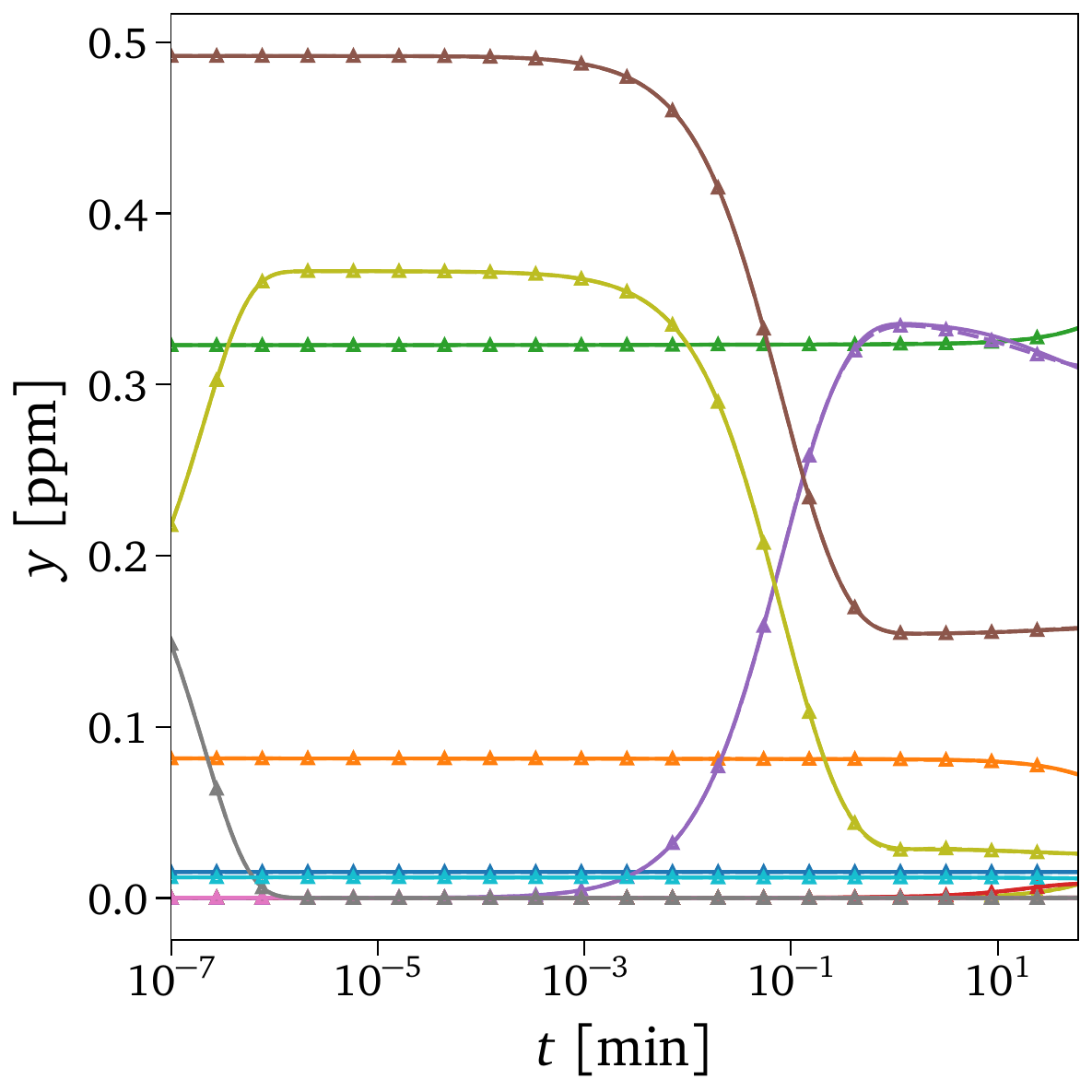}
    \end{subfigure}
    \begin{subfigure}[htb!]{0.32\textwidth}
        \centering\hspace{1.5mm}
        \includegraphics[width=0.95\textwidth]{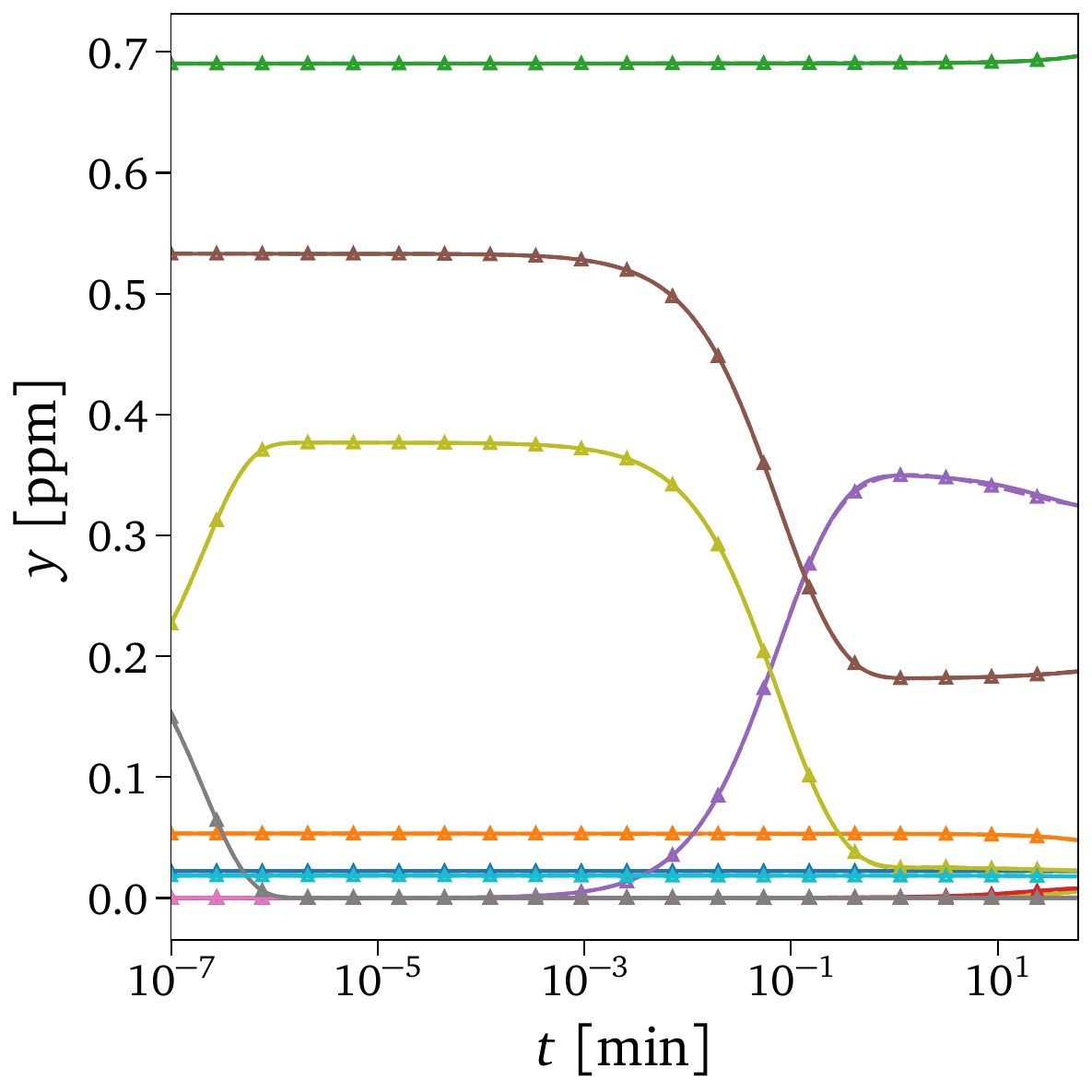}
    \end{subfigure}
    \\[2pt]
    \begin{subfigure}[htb!]{0.32\textwidth}
        \centering
        \includegraphics[width=\textwidth]{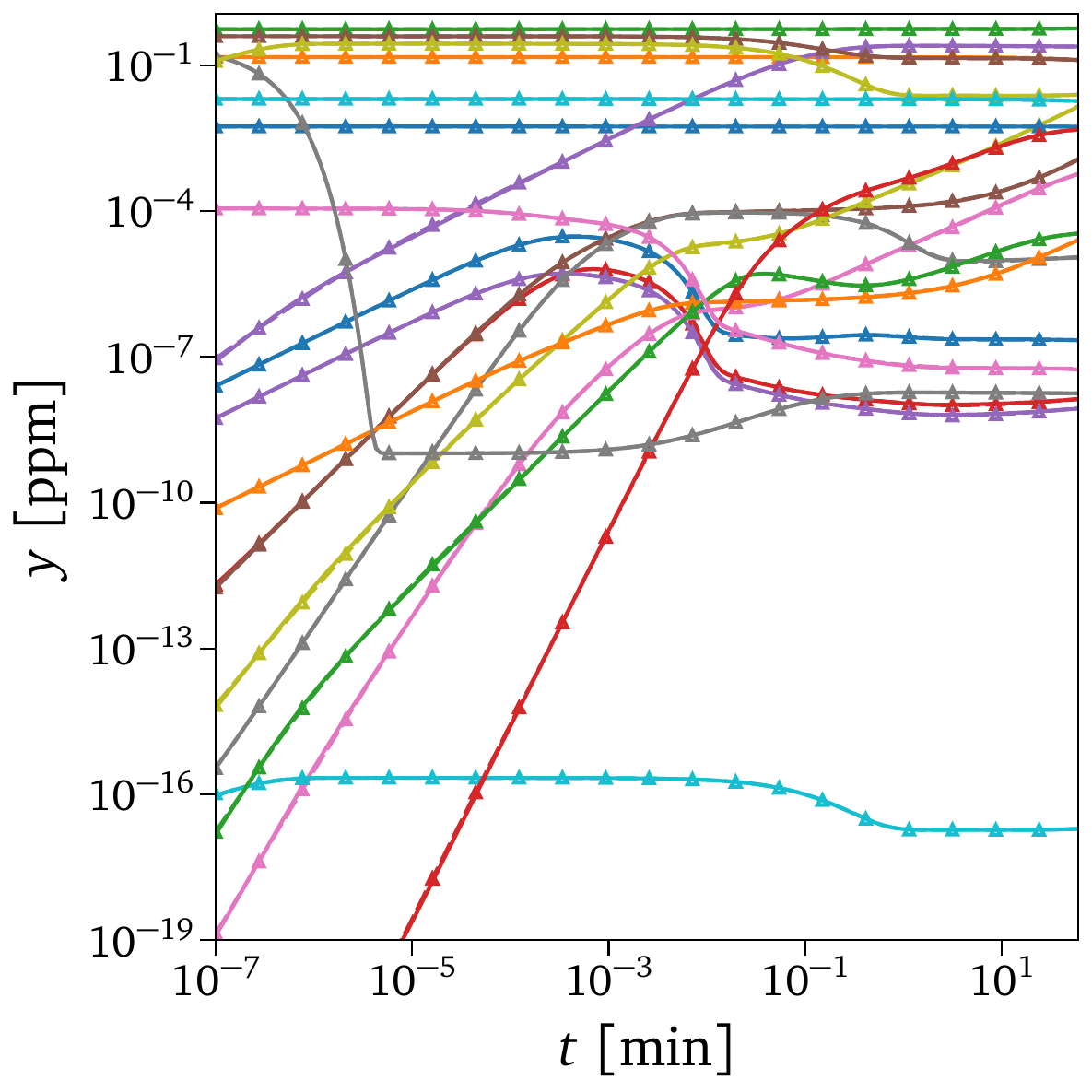}
    \end{subfigure}
    \begin{subfigure}[htb!]{0.32\textwidth}
        \centering
        \includegraphics[width=\textwidth]{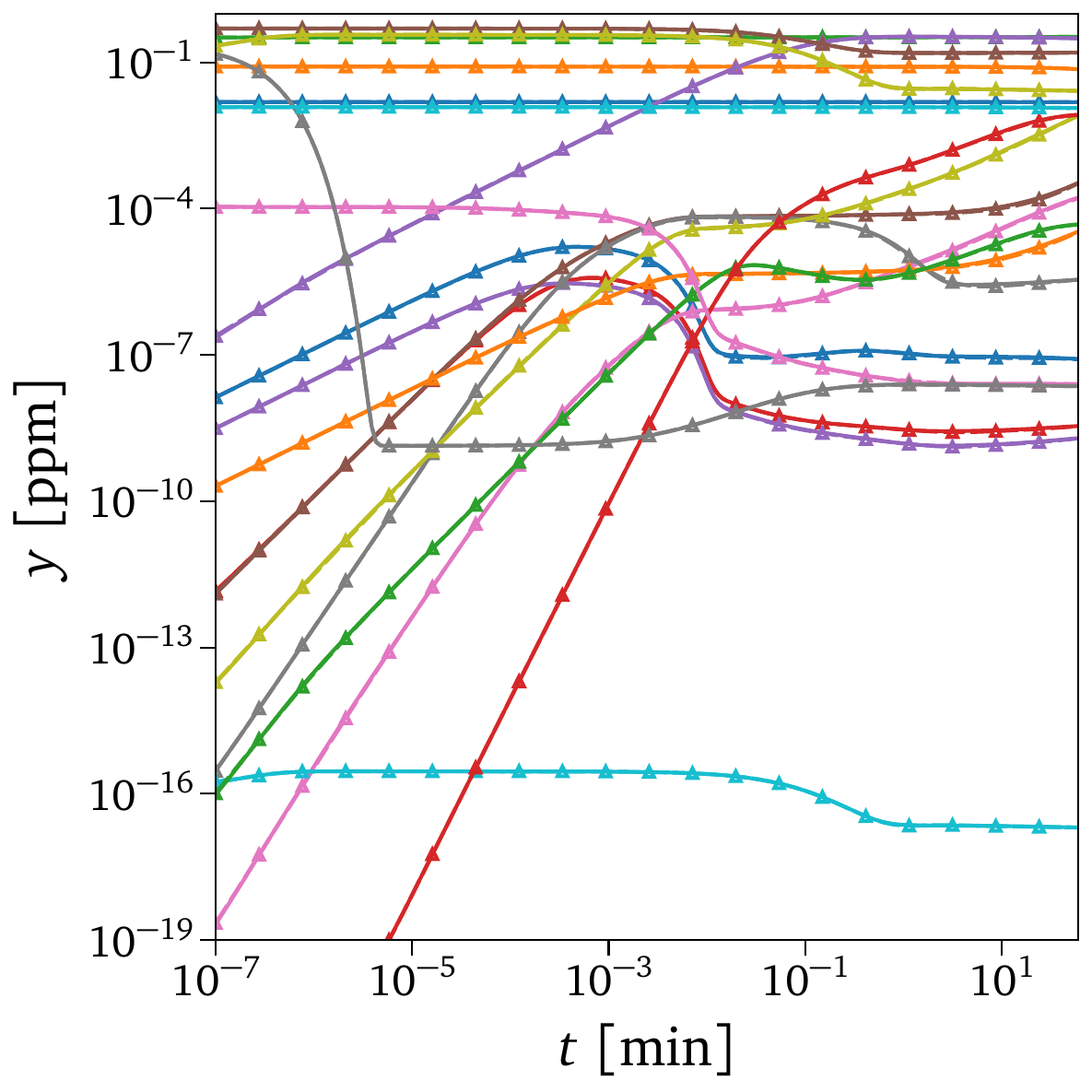}
    \end{subfigure}
    \begin{subfigure}[htb!]{0.32\textwidth}
        \centering
        \includegraphics[width=\textwidth]{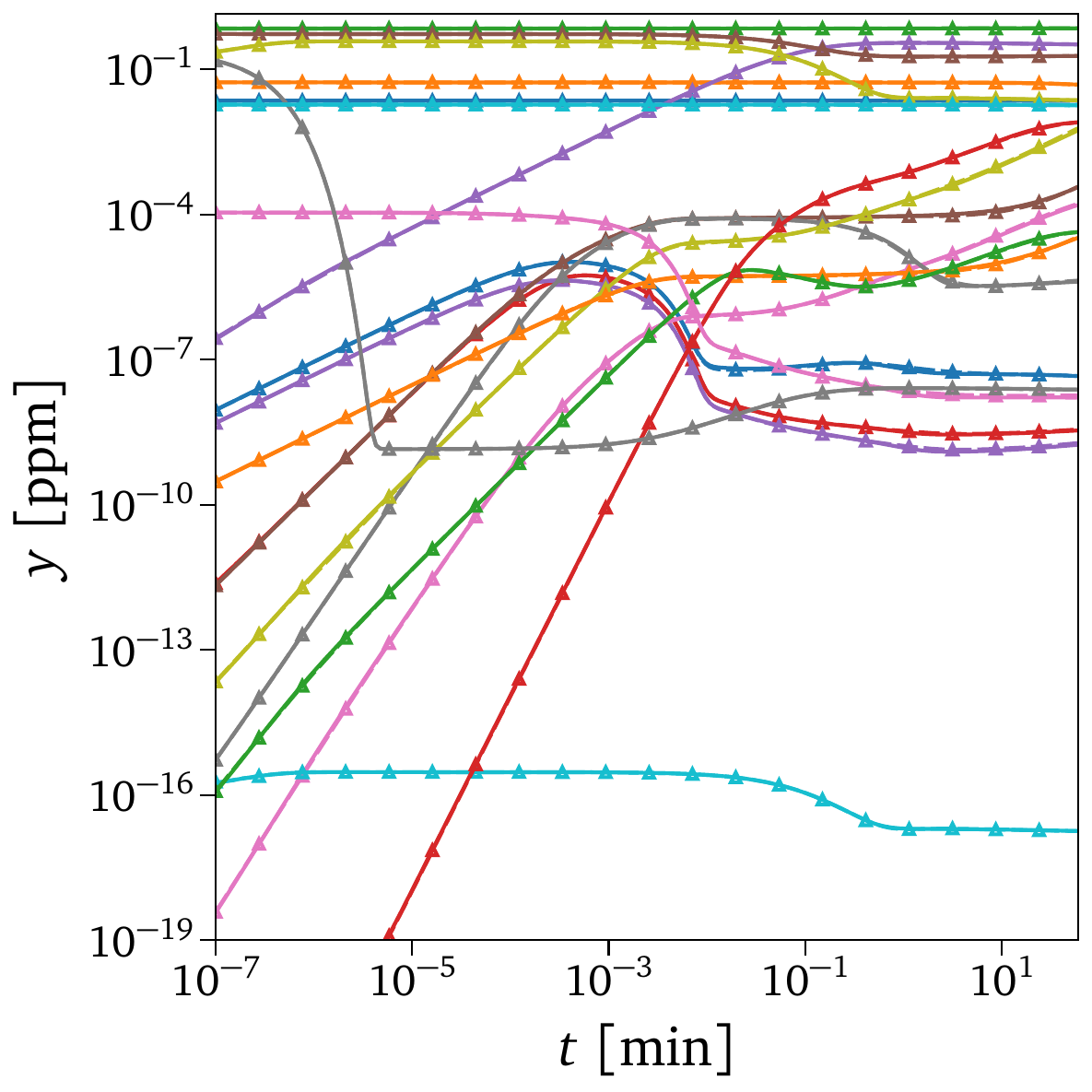}
    \end{subfigure}
    \caption{\textit{Comparison of MENO and true solutions for the POLLU problem.} Reference solutions (solid lines) are compared with predictions from the trained MENO model (dashed lines with markers) across three representative test cases (columns). Each subplot shows species concentrations over time, presented in linear scale (top row) and logarithmic scale (bottom row).}
    \label{fig:pollu.sol}
\end{figure}
Figure~\ref{fig:pollu.sol} provides a qualitative assessment of the surrogate model’s accuracy. It shows the time evolution of the 20 chemical species concentrations for three unseen test cases, arranged in columns.
Each subplot presents results in both linear (top row) and logarithmic (bottom row) scales to capture dynamics across several orders of magnitude.
The comparison reveals excellent agreement between the ground truth (solid lines) and the surrogate predictions (dashed lines with markers).
The model successfully captures concentrations spanning 19 orders of magnitude, from \(10^{-19}\) to \(1\) ppm, over the full time interval.
\begin{figure}[htb!]
\centering
\begin{subfigure}[htb!]{0.35\textwidth}
    \centering
    \caption*{\hspace{7mm}\fontfamily{mdbch}\small MENO - No corrections}
    \includegraphics[width=\textwidth]{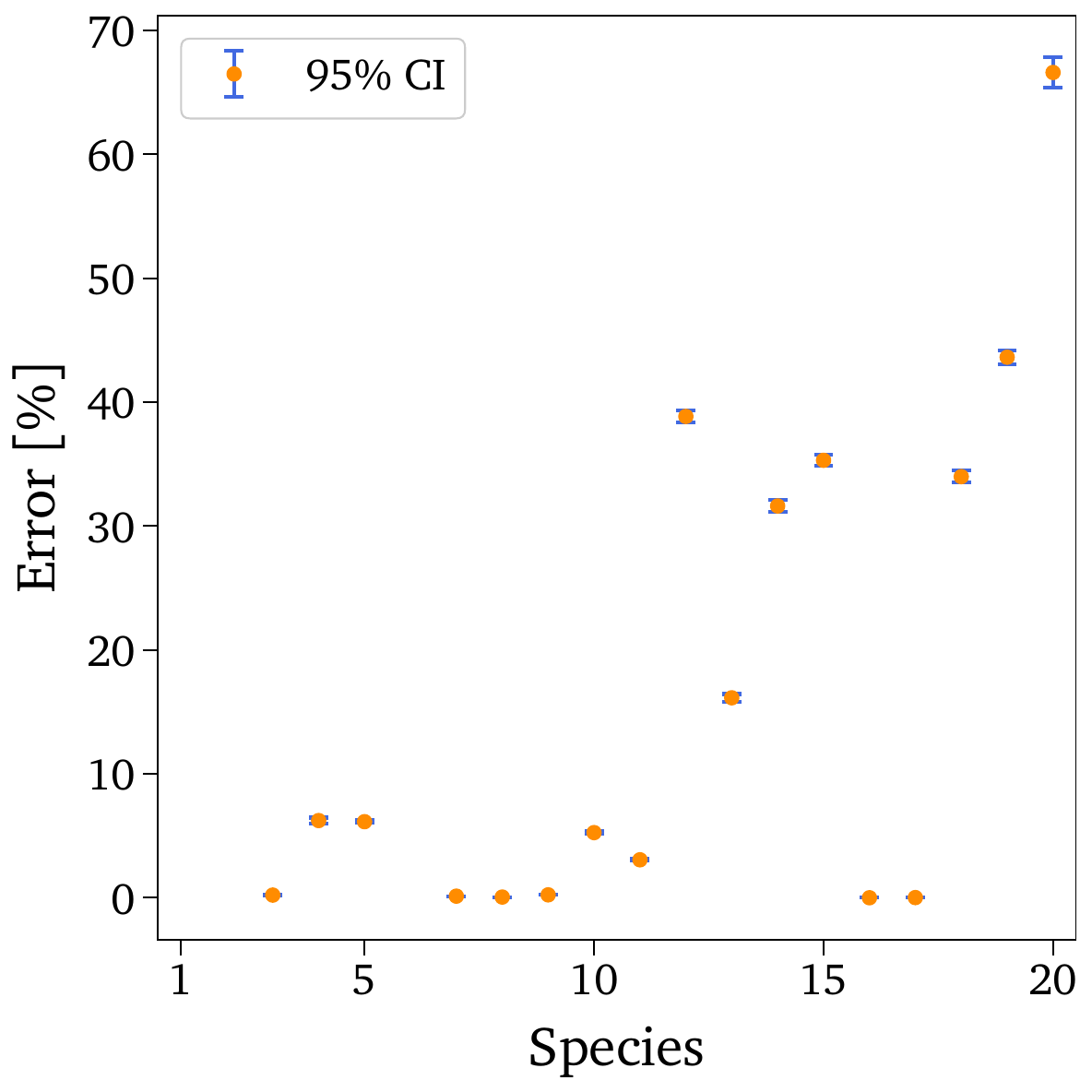}
\end{subfigure}
\hspace{1mm}
\begin{subfigure}[htb!]{0.35\textwidth}
    \centering
    \caption*{\hspace{7mm}\fontfamily{mdbch}\small MENO}
    \includegraphics[width=\textwidth]{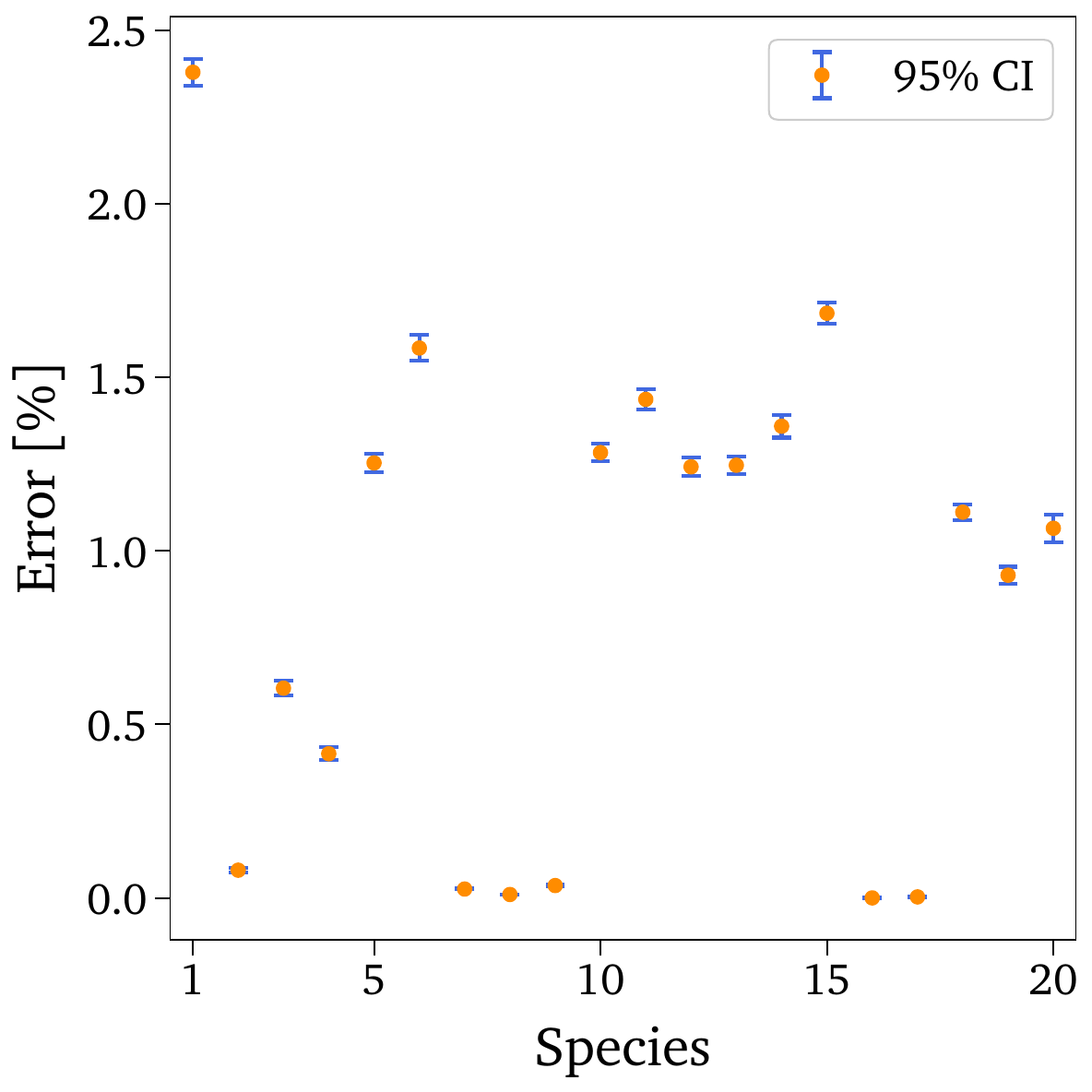}
\end{subfigure}
\caption{\textit{Test error for the POLLU problem}. Mean relative test errors with 95\% confidence intervals for the predicted species concentrations using MENO without corrective factors (left) and the fully trained MENO (right).}
\label{fig:pollu.err_lin}
\end{figure}
This qualitative agreement is supported by a quantitative evaluation based on the MAPE metric, defined in equation~\eqref{eq:mape}, and shown in figure~\ref{fig:pollu.err_lin}, along with 95\% confidence intervals for each species.
The left panel of figure~\ref{fig:pollu.err_lin} reports the MAPE for the linear state vector \(\vq_l\), predicted using the MENO framework without any corrective factors, and conditioned on the exact trajectories of the nonlinear state \(\vq_{nl}\).
This setup enables an \textit{a priori} assessment of the model’s representational capacity before any training is performed.
As shown in the figure, most species exhibit errors below 10\%, with only a few exceeding 30\%. These results suggest that the matrix exponential structure in MENO inherently captures much of the linear dynamics, even without learning-based corrections.
This built-in accuracy significantly simplifies and accelerates the second stage of training described in section~\ref{sec:method:train.perf}, especially when compared to training a generic neural network from random initialization.
This observation is further supported by results shown in figure~\ref{fig:pollu.sol_nocorr} of the Supplementary Material, where the same test cases from figure~\ref{fig:pollu.sol} are evaluated using the untrained MENO model.
Note that species \(y_1\), \(y_2\), and \(y_6\), which constitute the nonlinear state vector \(\vq_{nl}\), are omitted from the left panel of figure~\ref{fig:pollu.err_lin}, as their dynamics are modeled independently. The right panel, instead, presents the MAPE for the fully trained MENO model, including the three \textit{flexDeepONet} networks used to learn the nonlinear components.
In this case, prediction errors are consistently below 2\% for all species, with the exception of \(y_1\), which reaches approximately 2.5\%.
This slight deviation is expected, as \(y_1\) is one of the three nonlinear species modeled using a black-box neural operator, which does not incorporate the same physical priors as the linear part of MENO.
We emphasize that, in contrast to the model proposed by Goswami et al.~\cite{Goswami2024LearningOperators}, our framework maintains high accuracy throughout the entire temporal domain, whereas their discretized formulation restricts accuracy to a limited set of time points.
Furthermore, we provide a detailed evaluation of model performance using logarithmic scales in both time and concentration space, enabling a clearer assessment of accuracy across orders of magnitude.
In comparison, their results are presented exclusively in linear scale~\cite[Figure 3]{Goswami2024LearningOperators}, which may obscure the extent of modeling discrepancies and hinder a comprehensive evaluation of predictive fidelity.

\subsection{Example 2: Nonequilibrium oxygen}\label{sec:num_exp:noneq}
In this numerical experiment, we consider the rovibrational excitation and dissociation of molecular oxygen (O$_2$) colliding with atomic oxygen (O) within an adiabatic, isochoric chemical reactor.
This system is relevant for studying thermochemical nonequilibrium in high-temperature gas dynamics, particularly in the context of atmospheric reentry of hypersonic vehicles~\cite{Park_Book_1990}. Both O$_2$ and O are assumed to be in their electronic ground states.
From quantum chemistry calculations~\cite{Venturi_JCP_2020}, it is known that O$_2$ possesses 6115 distinct rovibrational energy levels, which can be arranged in ascending order of energy and indexed by a global identifier $i\in\mathcal{I}$, with $\mathcal{I}$ representing the set of all rovibrational states.
The kinetic model for the O$_2$-O system accounts for collisional excitation and dissociation processes, detailed in section~\ref{suppl:num_exp:noneq} of the Supplementary Material.
The time evolution of the system is governed by the following set of ODEs:
\begin{align}
    \frac{dn_i}{d t} =
        & \sum_{j \in \mathcal{I}} \left(-k_{i j}^{\mathrm{e}} n_in_\atom
        + k_{j i}^{\mathrm{e}} n_j n_\atom \right)
        -k_i^{\mathrm{d}}n_in_\atom + k_i^{\mathrm{r}}n_\atom^3
        \eqspace, \label{eq:noneq.mol} \\
    \frac{dn_\atom}{dt} = & - 2\sum\limits_{i\in\mathcal{I}}\frac{dn_i}{d t}
        \eqspace, \label{eq:noneq.atom}
\end{align}
where $n_i$ is the number density of the $i$-th state of O$_2$, and $n_\atom$ is the number density of atomic oxygen. The rate constants $k_{i}^{\mathrm{d}}$, $k_{i}^{\mathrm{r}}$, $k_{ij}^{\mathrm{e}}$, and $k_{ji}^{\mathrm{e}}$ (see section~\ref{suppl:num_exp:noneq}) are temperature-dependent and derived from quasi-classical trajectory (QCT) calculations~\cite{Jaffe_AIAA_2015,Venturi_JCPA_2020,Priyadarshini2023EfficientArchitectures}, fit to Arrhenius-type expressions.
The system is closed with the conservation of total mass and energy, $d\rho/dt=de/dt=0$, with the energy constraint used to determine the system temperature $T$.

Due to the large number of rovibrational states, we apply a physics-based model reduction strategy to reduce the number of variables.
Specifically, we use a coarse-grained (CG) approach~\cite{Magin2012Coarse-grainNitrogen,Panesi_Lani_PofF_2013,Munafo_PR_2014,Munafo_PF_2015,Liu_JCP_2015} that clusters the rovibrational energy levels of O$_2$ into a manageable number of pseudo-species, or bins. These bins are constructed using an energy-based grouping method~\cite{Macdonald_I_JCP_2018,Macdonald_II_JCP_2018}, and we assume a local Boltzmann distribution at the system temperature $T$ to describe the population within each bin.
The reduced model is derived by taking successive energy-weighted moments of equation~\eqref{eq:noneq.mol} over each group $g$:
\begin{equation}\label{eq:noneq.cg}
    \sum_{i\in\mathcal{I}_g\subseteq\mathcal{I}}\left(\epsilon_i\right)^m\frac{dn_i}{dt} = \Omega^g_{m}
    \eqspace,
    \quad\forall \;g\in\{1,\dots,N_g\} \eqspace.
\end{equation}
where $\epsilon_i$ is the internal energy of state $i$, $\mathcal{I}_g$ is the set of states in group $g$, $m$ is the moment order, and $\Omega^g_{m}$ is the corresponding reactive source term.
In this work, we retain only the zeroth-order moment ($m=0$), yielding a reduced system whose unknowns are the total populations of each bin.
The resulting CG model preserves the same mathematical structure as equations~\eqref{eq:noneq.mol}-\eqref{eq:noneq.atom}, with state index $i$ now referring to a bin, and rate constants describing transitions between bins rather than individual rovibrational states.
For this study, we use a 45-bin CG model, which, along with atomic oxygen and temperature, results in a system with 47 degrees of freedom. The corresponding state vector of primitive variables is defined as $\vq=[n_g,n_\atom,T]$, where $g\in\{1,\dots,45\}$ indexes the CG bins.
For additional details on the CG technique, interested readers are referred to reference \cite{Liu_JCP_2015}.

In this thermochemical system, nonlinear behavior arises primarily from a small subset of variables: the atomic number density $n_\atom$, which introduces quadratic and cubic nonlinearities, and the temperature $T$, which influences the system nonlinearly via the Arrhenius rate law governing temperature-dependent reaction rates.
Based on this observation, we define the nonlinear state vector as \(\vq_{nl} = [n_\atom, T] \in \mathbb{R}^2\), and the linear state vector as \(\vq_l= [n_g]\in \mathbb{R}^{45}\), which makes the system, due to its low-dimensional nonlinearity, well suited for the MENO framework.
We model $n_\atom$ and $T$ using two separate \textit{flexDeepONet} networks, while the linear subsystem for \(\vq_l\), conditioned on the learned nonlinear trajectories, is then solved using the MENO formulation via equation~\eqref{eq:lin_sys.sol.ml}.
\begin{figure}[htb!]
    \centering
    \includegraphics[width=0.35\textwidth]{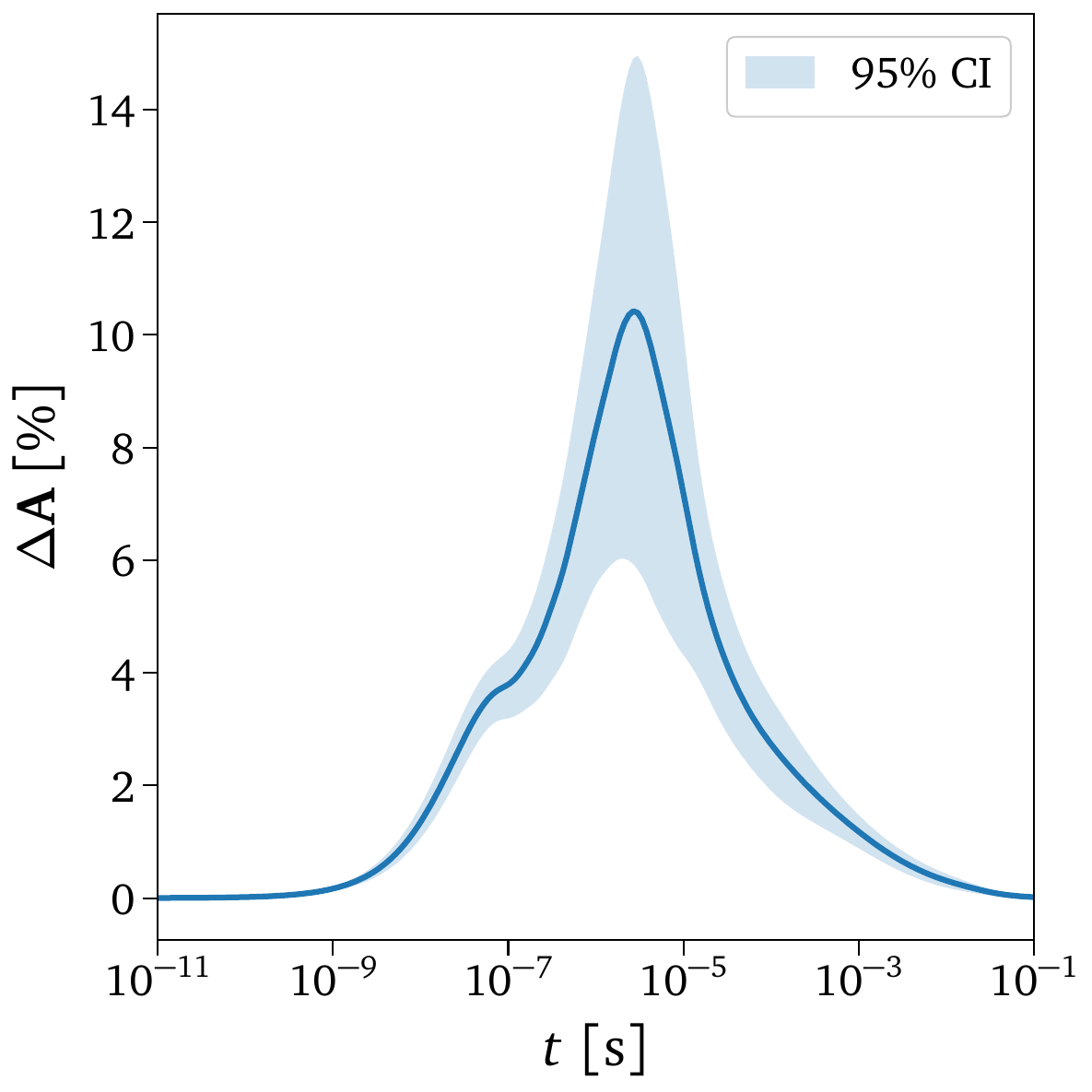}
    \caption{\textit{Variability of the operator \(\mA(t)\) in the nonequilibrium oxygen problem}. The mean variability of operator $\mA$ shown with a 95\% confidence interval, calculated using equation \eqref{eq:delta_a} across 100 test trajectories.}
    \label{fig:noneq.delta_a}
\end{figure}
To assess the validity of the matrix exponential formulation, we compute the variability metric $\Delta \mA(t)$, defined in equation~\eqref{eq:delta_a}, across 100 representative trajectories.
Figure~\ref{fig:noneq.delta_a} presents the mean value of $\Delta \mA(t)$, along with 95\% confidence intervals, over the full simulation time span. The maximum observed variability is approximately 10\%, which is sufficiently small to justify the use of the matrix exponential formulation in equation \eqref{eq:lin_sys.sol.ml}.
Beyond its numerical role, the behavior of $\Delta \mA(t)$ also carries physical significance. For very early times ($t < 10^{-7}$ s), only thermal excitation processes are active, and the system behaves linearly with $T(t) \approx T(0)$ and $n_\atom(t) \approx n_\atom(0)$. At the other end of the spectrum, for $t > 10^{-4}$ s, the system approaches equilibrium, and the state becomes effectively constant. In the intermediate range ($10^{-7} < t < 10^{-4}$ s), dissociation dominates the dynamics, resulting in notable changes in both $T(t)$ and $n_\atom(t)$. Interestingly, these variations tend to counterbalance each other in their effect on $\mA(t)$, leading to relatively modest fluctuations in the operator. This behavior is consistent with the quasi-steady-state (QSS) assumption: during this phase, internal energy exchange is effectively frozen, with thermal excitation suppressed and chemical processes (specifically dissociation) taking precedence. In this regime, energy is transferred from the translational energy pool (manifested as a reduction in temperature $T$) to the chemical energy pool (reflected in the increasing concentration $n_\atom$), enabling bond-breaking in molecular oxygen.

Further details on data generation, model architecture, and training are provided in section~\ref{suppl:num_exp:noneq} of the Supplementary Material.

\begin{figure}[htb!]
    \centering
    \begin{subfigure}[htb!]{0.32\textwidth}
        \centering
        \includegraphics[width=\textwidth]{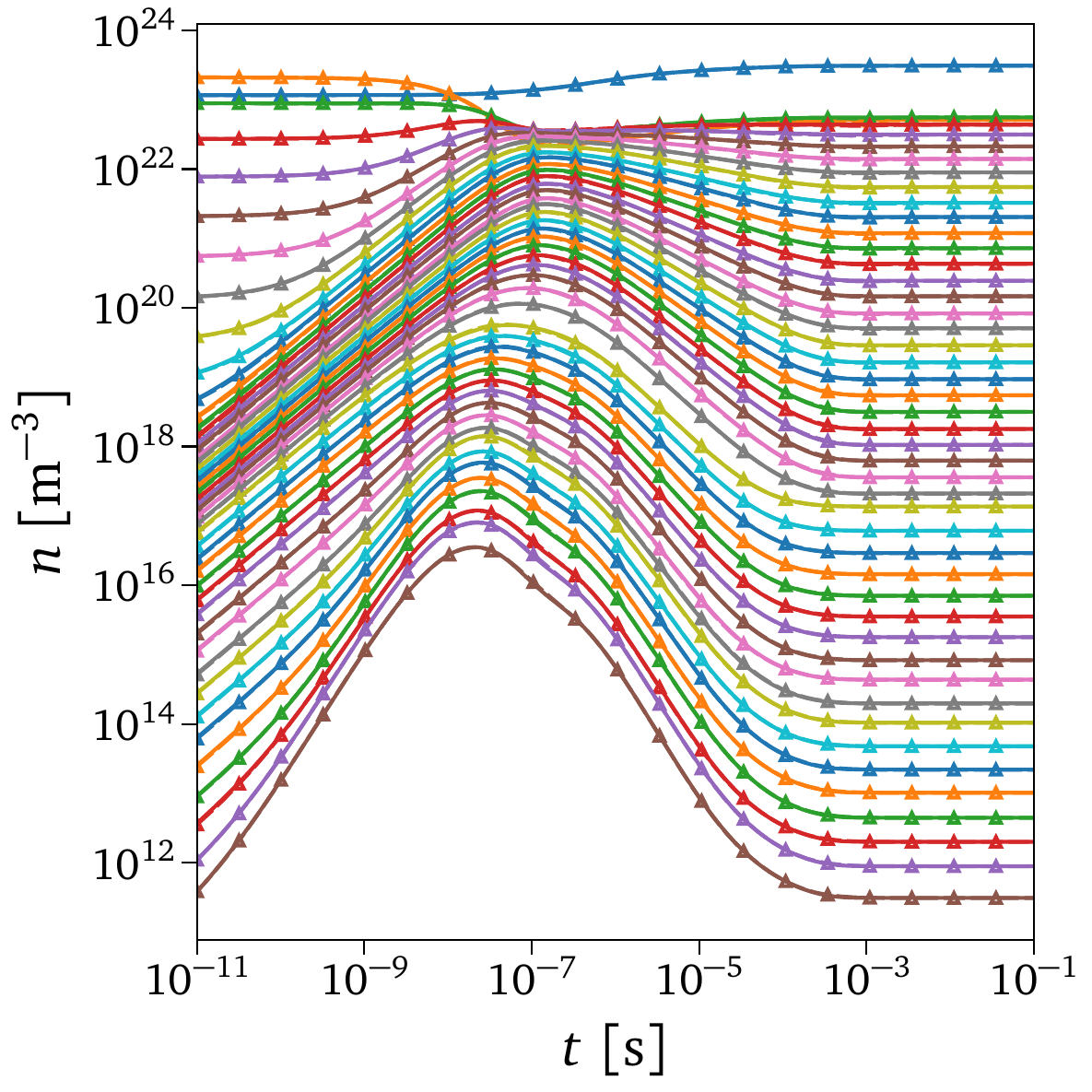}
    \end{subfigure}
    \begin{subfigure}[htb!]{0.32\textwidth}
        \centering
        \includegraphics[width=\textwidth]{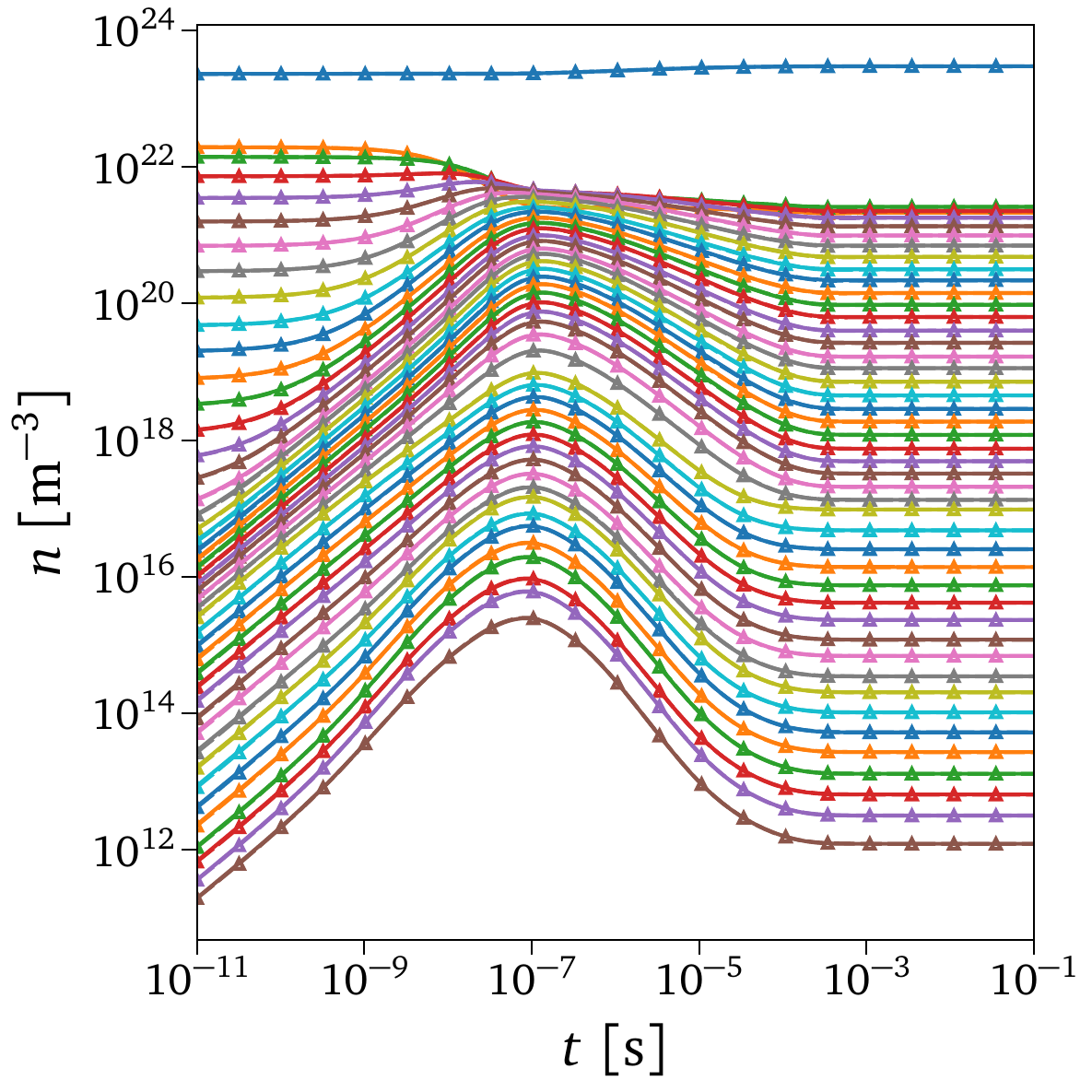}
    \end{subfigure}
    \\
    \begin{subfigure}[htb!]{0.32\textwidth}
        \centering
        \includegraphics[width=\textwidth]{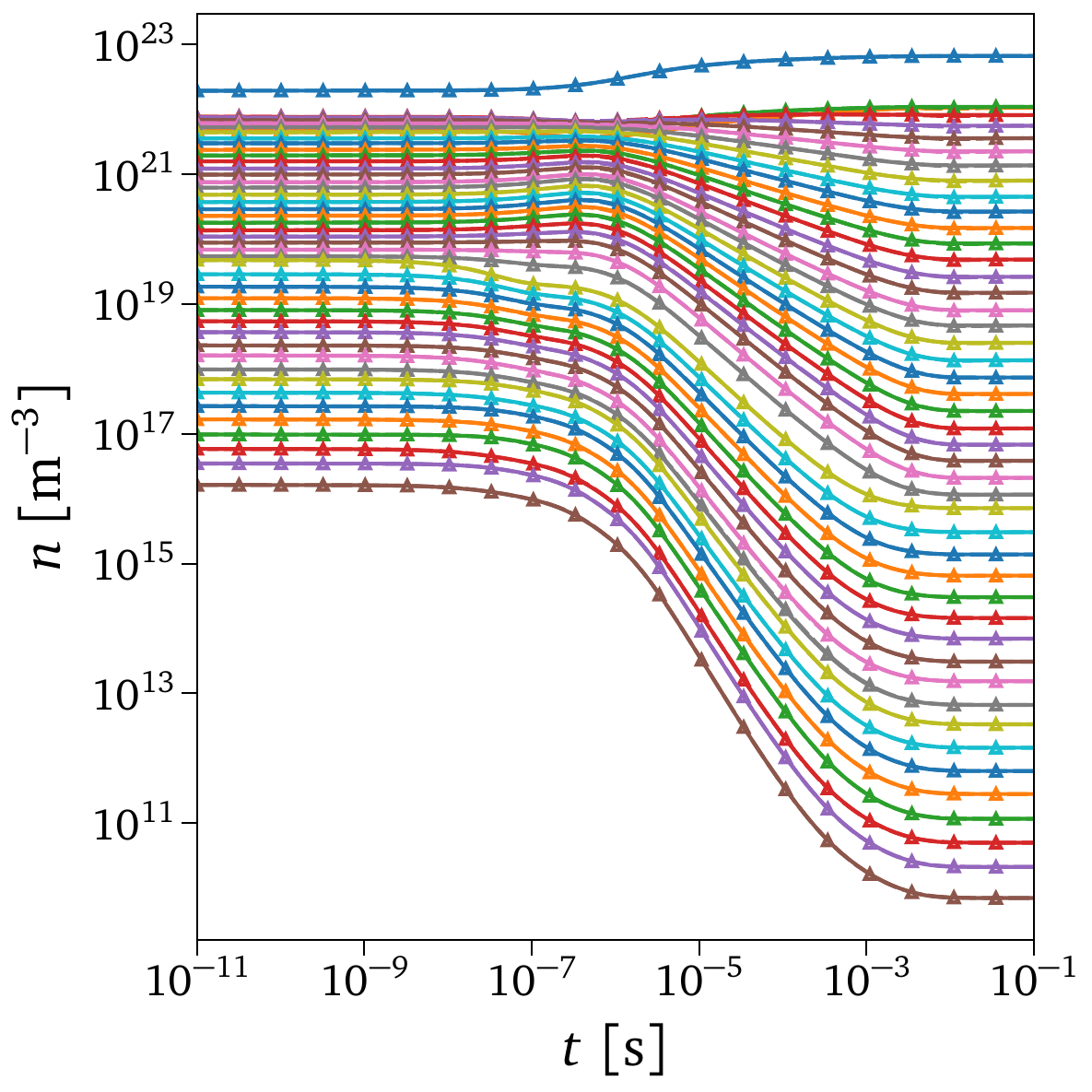}
    \end{subfigure}
    \begin{subfigure}[htb!]{0.32\textwidth}
        \centering
        \includegraphics[width=\textwidth]{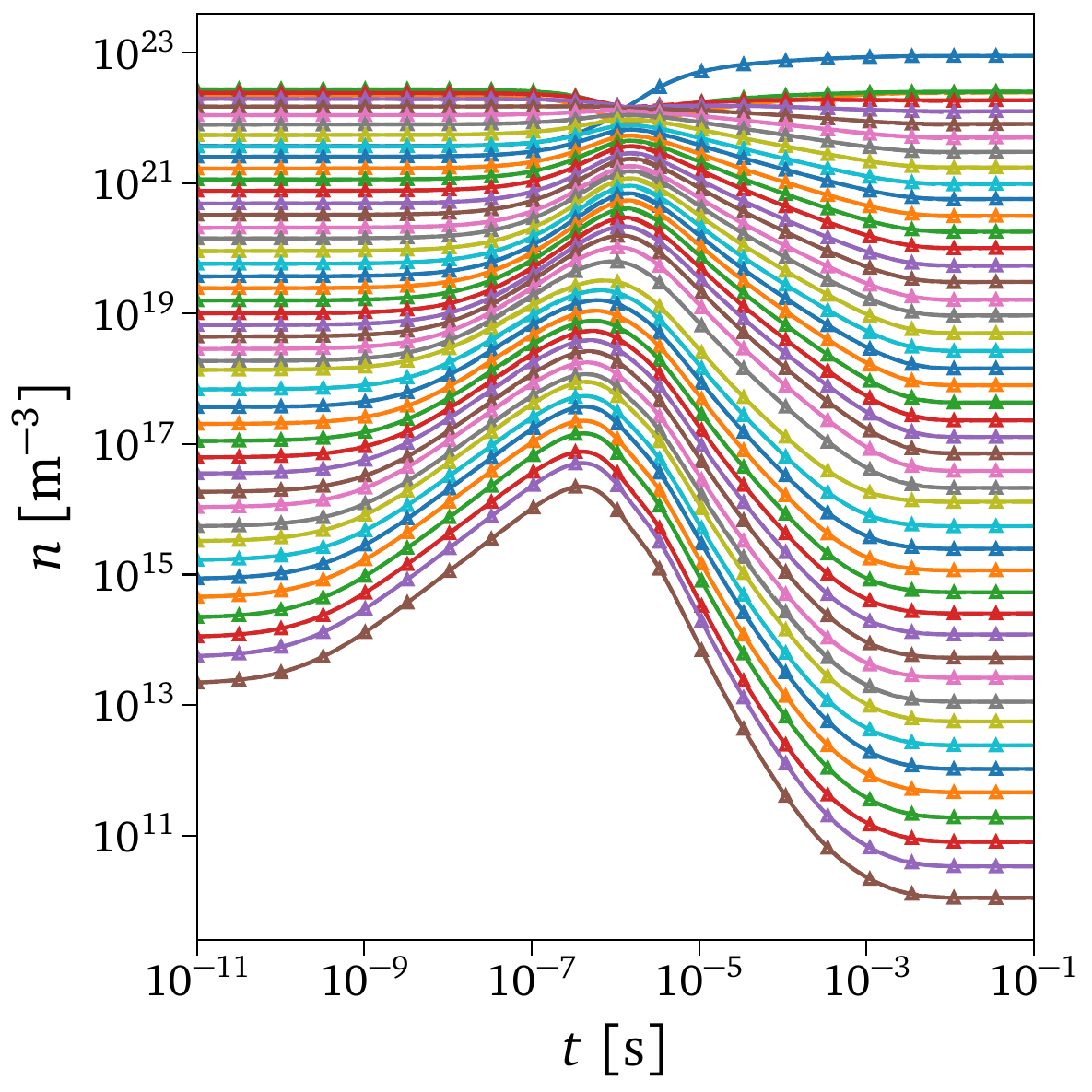}
    \end{subfigure}
    \caption{\textit{Comparison of MENO and true solutions for the nonequilibrium oxygen mixture}. The reference solutions (solid lines) are compared against predictions from the trained MENO (dashed lines with markers) for four different test cases. Results are shown for species number densities.}
    \label{fig:noneq.nd}
\end{figure}
Figure~\ref{fig:noneq.nd} presents a qualitative evaluation of the surrogate model’s performance by illustrating the time evolution of number densities for the mixture pseudo-species across four unseen test cases.
Corresponding temperature profiles for the same cases are provided in figure~\ref{fig:noneq.temp} in the Supplementary Material.
Across all scenarios, the predicted trajectories (dashed lines with markers) closely match the ground truth solutions (solid lines), accurately capturing both the species concentrations and temperature evolution.
Notably, the model successfully resolves concentrations spanning 14 orders of magnitude, from \(10^{10}\) to \(10^{24}\) m$^{-3}$, throughout the full time domain.

\begin{figure}[htb!]
\centering
\begin{subfigure}[htb!]{0.32\textwidth}
    \caption*{\hspace{6mm}\fontfamily{mdbch}\small MENO - No corrections}
    \includegraphics[width=\textwidth]{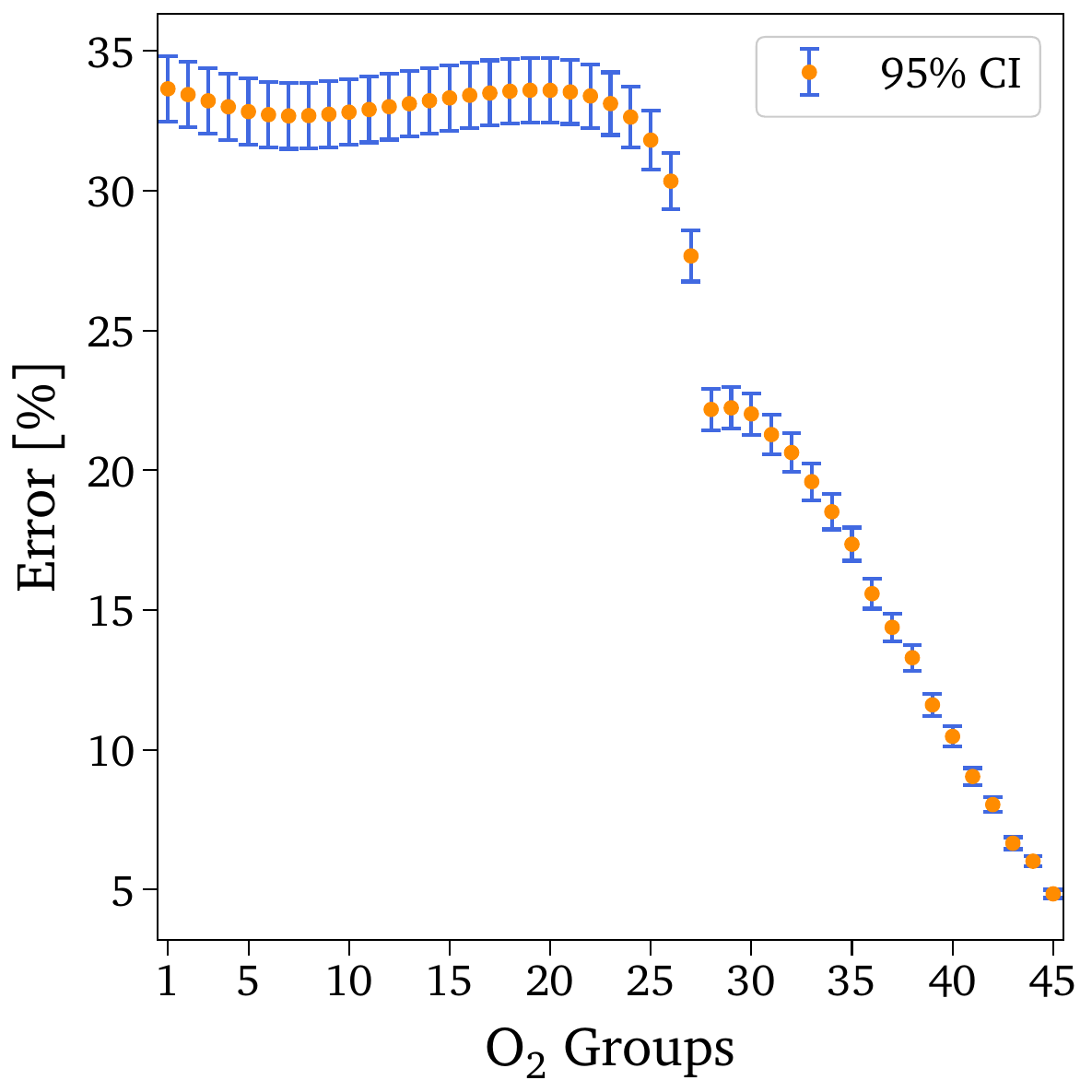}
\end{subfigure}
\begin{subfigure}[htb!]{0.32\textwidth}
    \caption*{\hspace{6mm}\fontfamily{mdbch}\small MENO}
    \includegraphics[width=\textwidth]{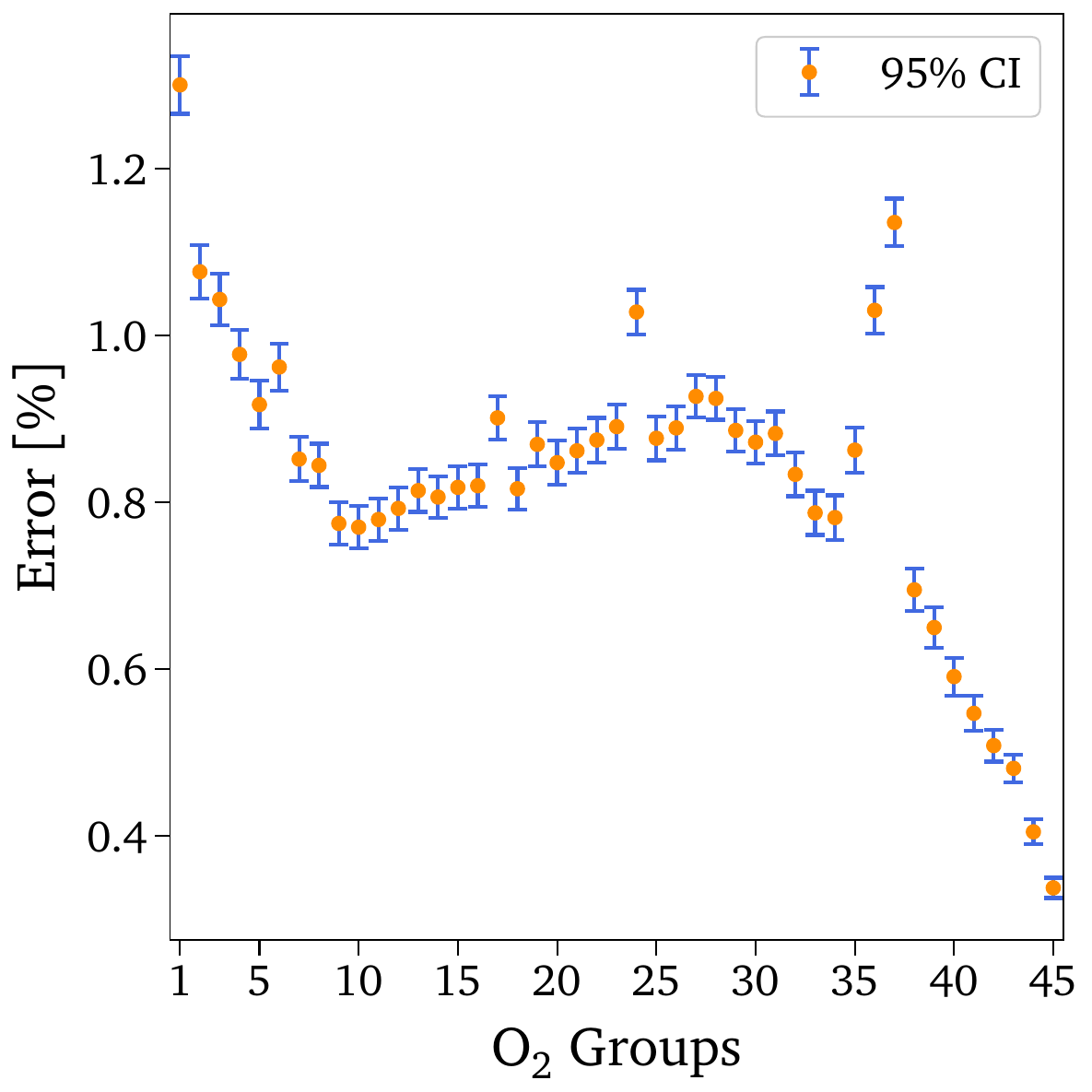}
\end{subfigure}
\begin{subfigure}[htb!]{0.32\textwidth}
    \caption*{\hspace{6mm}\fontfamily{mdbch}\small Vanilla DeepONet}
    \includegraphics[width=\textwidth]{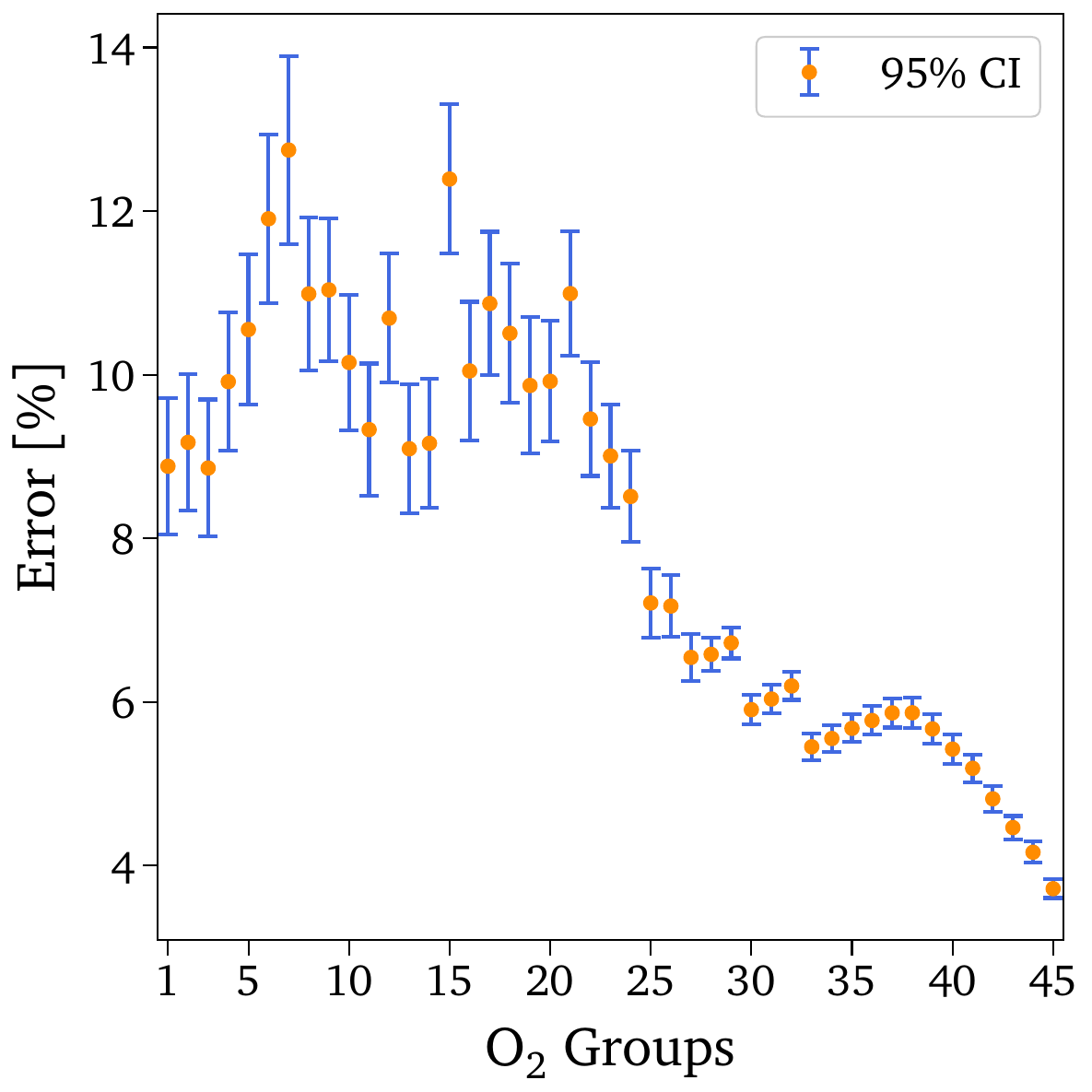}
\end{subfigure}
\caption{\textit{Test error for the nonequilibrium oxygen mixture}. Mean relative test errors with 95\% confidence intervals for the predicted molar concentrations of O$_2$ groups. Results are shown for MENO without corrective factors (left), fully trained MENO (center), and the vanilla DeepONet (right).}
\label{fig:noneq.err_lin}
\end{figure}
This qualitative agreement is further supported by a quantitative analysis using the MAPE metric, defined in equation~\eqref{eq:mape}.
For the nonlinear state variables, specifically the temperature and the molar fraction of O, the surrogate model achieves high predictive accuracy, with mean relative test errors of 0.079\% for $T$ and 0.171\% for O concentration.
The corresponding errors for the linear variables, namely the molar concentrations of the O$_2$ groups, are shown in figure~\ref{fig:noneq.err_lin}.
To assess the inherent representational capability of the MENO framework, the left panel of figure~\ref{fig:pollu.err_lin} shows the MAPE for $\vq_l$ using an untrained model (i.e., with corrective factors disabled), conditioned on the exact nonlinear trajectories $\vq_{nl}$.
This provides an \textit{a priori} assessment of the matrix exponential integrator’s expressiveness. Even without learning-based corrections, the maximum error remains below 35\%, indicating that the MENO architecture captures a substantial portion of the linear dynamics by design, which facilitates the second training stage described in section~\ref{sec:method:train.perf}.
Additional support for this conclusion is shown in figure~\ref{fig:noneq.nd_nocorr} (Supplementary Material), where the same test cases from figure~\ref{fig:noneq.nd} are evaluated using the untrained model.
In contrast, the center panel of figure~\ref{fig:pollu.err_lin} shows MAPE results after full training of MENO, where the nonlinear variables are predicted by two separate \textit{flexDeepONet} models.
In this case, errors drop to around 1\% across all species. For comparison, the right panel shows results from a vanilla DeepONet~\cite{Lu_NMI_2021}, which yields significantly larger errors. 
To ensure a fair comparison, the vanilla DeepONet was configured with a number of parameters such that its inference time remained comparable to that of the MENO framework across varying final integration times $t_f$ (see figure~\ref{fig:noneq.don_vs_meno.time}, Supplementary Material).

\begin{figure}[htb!]
    \centering
    \begin{subfigure}[htb!]{0.35\textwidth}
        \centering
        \caption*{\hspace{7mm}\fontfamily{mdbch}\normalsize CPU}
        \includegraphics[width=\textwidth]{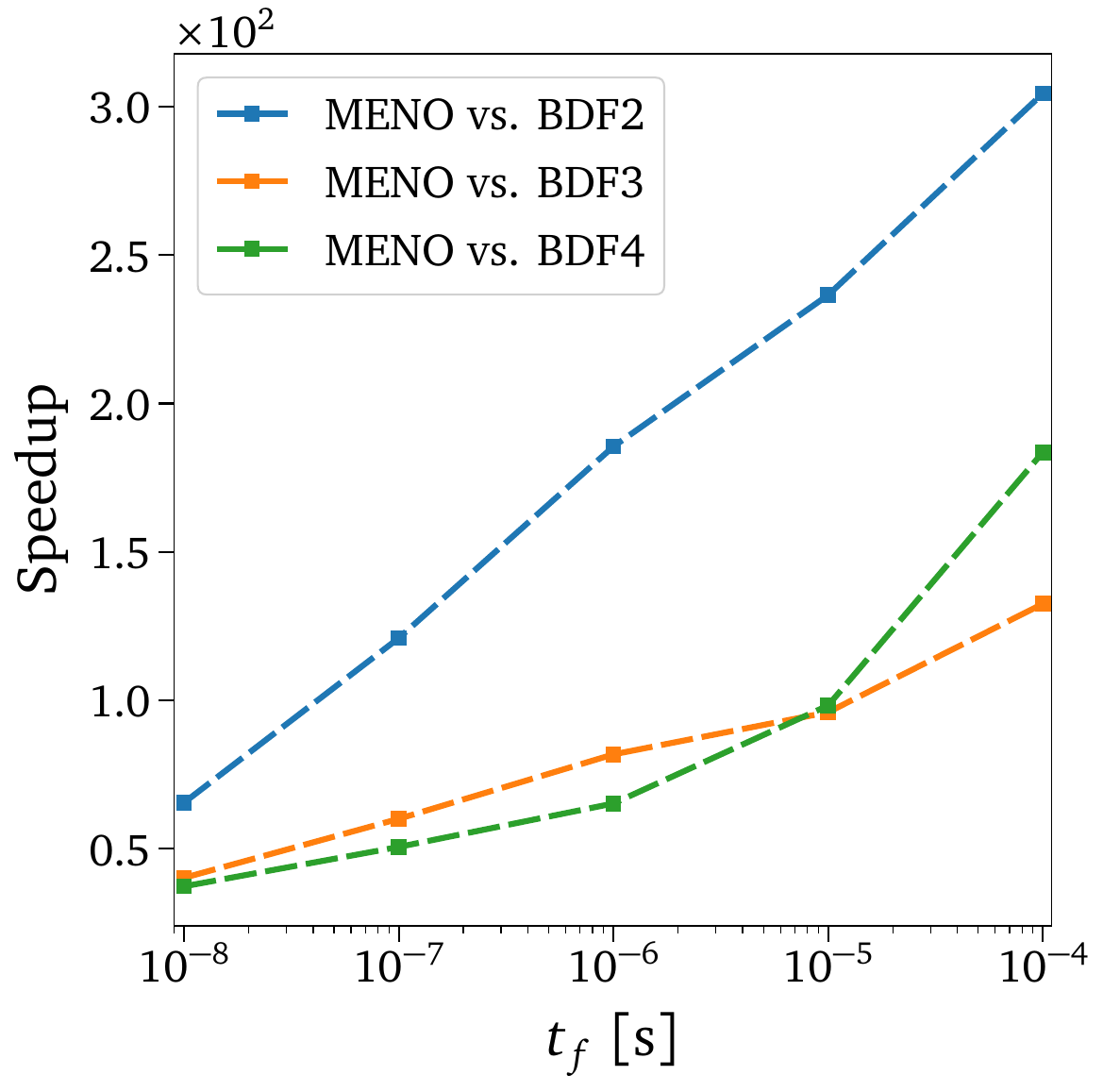}
    \end{subfigure}
    \begin{subfigure}[htb!]{0.35\textwidth}
        \centering
        \caption*{\hspace{7mm}\fontfamily{mdbch}\normalsize GPU}
        \includegraphics[width=\textwidth]{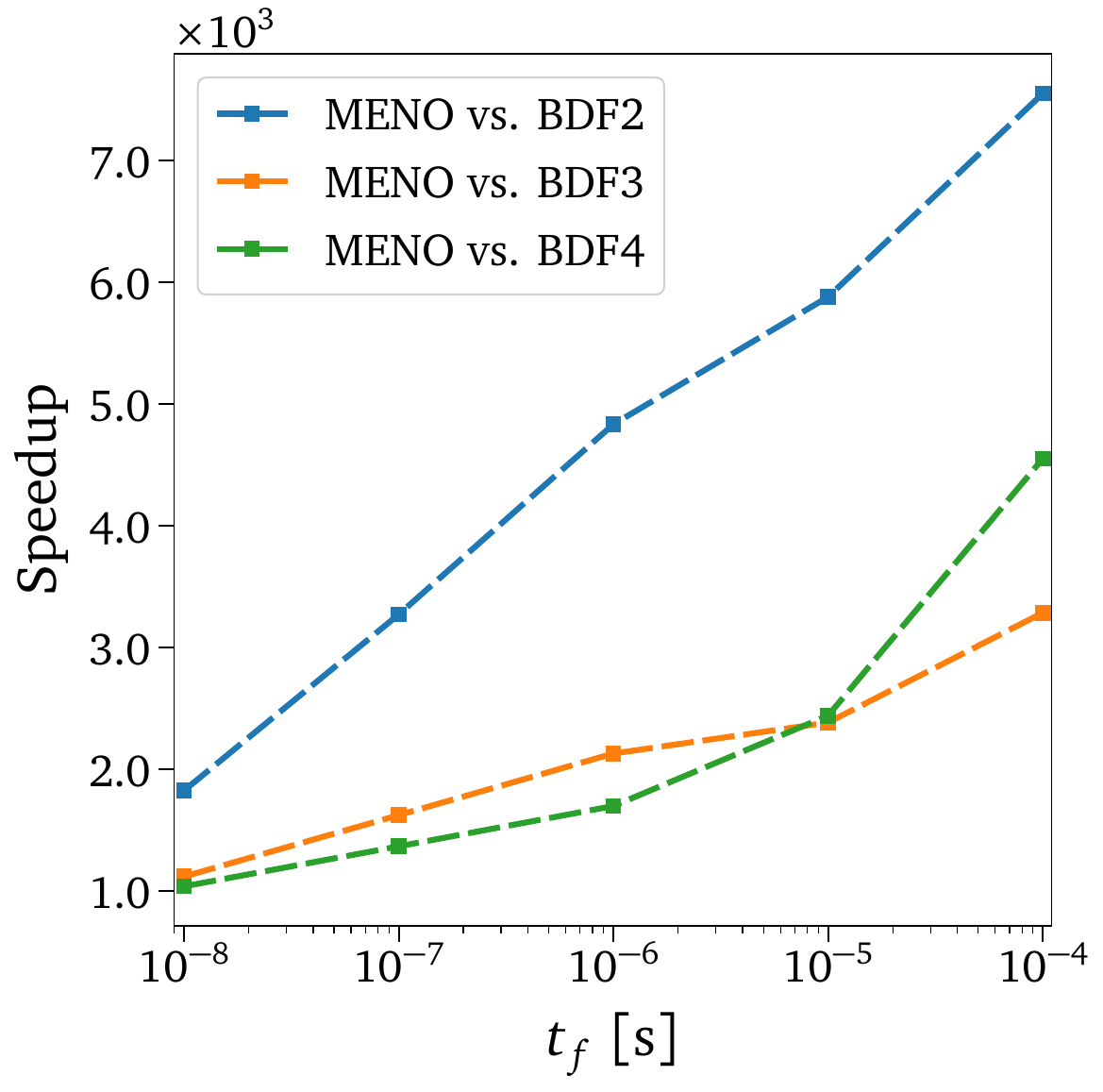}
    \end{subfigure}
    \caption{\textit{Computational speedup for the nonequilibrium oxygen mixture}. Computational speedup achieved by executing MENO on a single CPU core (left panel) and a single GPU (right panel). In both the figures, MENO is compared to the BDF implicit scheme run on a single CPU core. The analysis encompasses various scheme orders (BDF2 to BDF4) and different final integration times, $t_f$, with a fixed relative tolerance of $10^{-7}$ for the BDF scheme.}
    \label{fig:noneq.speedup}
\end{figure}
To quantify the computational advantage of MENO, we measure the speedup relative to a conventional implicit integration method. Following the methodology described in section~\ref{sec:method:train.perf}, figure~\ref{fig:noneq.speedup} reports the speedup of the fully trained MENO model compared to the BDF method.
The analysis is conducted across varying final integration times $t_f$ and BDF scheme orders (2-4), with fixed solver tolerances: absolute tolerance set to zero and relative tolerance set to $10^{-7}$.
While higher-order BDF methods typically yield faster integration due to fewer time steps, they demand more memory and can suffer instability in highly oscillatory regimes due to the limited stability region of orders above two~\cite{Enright1975ComparingO.D.E:s}.
Detailed runtime measurements for different BDF tolerances and orders on both CPU and GPU are summarized in tables~\ref{table:noneq.speedup.cpu}-\ref{table:noneq.speedup.gpu} (Supplementary Material), which reveal a near-linear increase in speedup with $t_f$.
This trend is consistent with expectations: as $t_f$ increases, the BDF solver requires more integration steps, while MENO evaluates the entire trajectory in a fixed number of operations.
Although computing the matrix exponential in MENO depends on the norm of the exponent ($\hmA(t_f),t_f$), as evaluated through Padé approximants, its sensitivity to $t_f$ remains comparatively low.
For a representative case in hypersonic flow applications, where a typical physical integration time is on the order of $t_f=10^{-6}$ seconds, the speedup achieved by MENO is substantial: 185$\times$ on a CPU and 4\,835$\times$ on a GPU relative to BDF2. These results highlight MENO’s significant potential for real-time or large-scale reactive flow simulations.

\subsection{Example 3: Collisional-radiative argon plasma}\label{sec:num_exp:plasma_cr}
In this example, we develop a surrogate model for a weakly ionized argon plasma using a detailed collisional-radiative (CR) framework, where each electronic state is treated as a separate pseudo-species.
This non-Boltzmann description of the atomic state distribution function (ASDF) enables high-fidelity simulation of nonequilibrium plasmas.
The CR mechanism used in this work was originally developed by Vl{\^c}ek~\cite{Vlcek1989AData} and Bultel~\cite{Bultel2002InfluenceModel2}, and has been validated against experimental data from the UTIAS (University of Toronto, Institute for Aerospace Studies) shock tube~\cite{Glass1991OverWaves} by Kapper and Cambier~\cite{Kapper2011IonizingStructure,Kapper2011IonizingEffects}.
The plasma consists of free electrons (e$^-$), neutral argon (Ar), and singly ionized argon (Ar$^+$), collectively represented by the set $\mathcal{S}=\{\mathrm{e}^-,\mathrm{Ar},\mathrm{Ar}^+\}$.
The electronic structure is resolved into 31 states for Ar and 2 for Ar$^+$, yielding 34 pseudo-species.
The CR model accounts for five types of fundamental processes~\cite{Vlcek1989AData,Bultel2002InfluenceModel2,Kapper2011IonizingStructure,Kapper2011IonizingEffects}, including
\begin{inparaenum}[(i)]
    \item excitation and ionization by electron impact, 
    \item excitation and ionization by heavy-particle impact,
    \item spontaneous emission/absorption (bound-bound),
    \item photo-ionization/radiative recombination (free-bound and bound-free),
    \item and Bremsstrahlung radiation (free-free).
\end{inparaenum}
The plasma flow is described using a single-fluid formulation, in which all species share a common velocity field, and thermal nonequilibrium is captured by distinguishing between electron and heavy-particle temperatures, $T_\mathrm{e}$ and $T_\mathrm{h}$, respectively.
Neglecting viscous and species diffusion effects, the evolution of the system is thus governed by the two-temperature Euler equations~\cite{Kapper2011IonizingEffects}.
Further details on the CR model and governing equations are provided in section~\ref{suppl:num_exp:plasma_cr} of the Supplementary Material.
For a comprehensive description of the reaction rate coefficients, collisional-radiative source terms, and thermodynamic closures, the reader is referred to the methodology developed by Kapper and Cambier~\cite{Kapper2011IonizingStructure,Kapper2011IonizingEffects}.

\subsubsection{Zero-dimensional surrogate model}\label{sec:num_exp:plasma_cr:0d}
To construct a surrogate model for the zero-dimensional (0D) thermochemical system governed by equations (52)-(54) in~\cite{Kapper2011IonizingStructure}, we first generate a physically relevant dataset for training and testing. This is accomplished by simulating a representative one-dimensional (1D) unsteady shock-tube problem (see section~\ref{sec:num_exp:plasma_cr:1d}) and collecting the primitive thermochemical variables $\vq = [\rho_s^i, T_\mathrm{h}, T_\mathrm{e}] \in \mathbb{R}^{36}$, where $\rho_s^i$ denotes the partial mass density of state $i$ for species $s\in\mathcal{S}$.
Due to the large volume of data points collected across space and time, we apply data compression and kernel density estimation techniques to perform informed subsampling, thereby obtaining a representative and tractable dataset. Further details on the data generation and preprocessing procedures can be found in section~\ref{suppl:num_exp:plasma_cr} of the Supplementary Material.

According to the 0D governing equations (52)-(54) in~\cite{Kapper2011IonizingStructure}, only the concentrations of e$^-$ and Ar(1), and the two temperatures \(T_\mathrm{h},T_\mathrm{e}\) introduce nonlinear terms. 
These four variables constitute the nonlinear state vector $\vq_{nl} \in \mathbb{R}^{4}$ and are each modeled using a separate \textit{flexDeepONet}.
The remaining 32 pseudo-species, which contribute linearly to the system dynamics, form the linear state vector $\vq_l \in \mathbb{R}^{32}$ and are integrated using the neural matrix exponential formulation described in equation~\eqref{eq:lin_sys.sol.ml}.
However, for this system, directly solving the linear system \(\hat{\mA}\vy = \hat{\vb}\) (see equation~\eqref{eq:lin_sys.sol.ml}) can lead to numerical instabilities, particularly when high-energy states attain very low values.
To eliminate the need for solving this linear system, we modify the formulation by enforcing \(\vb = \mathbf{0}\). This is done by expanding the linear state vector to include Ar(1), redefining it as $\vq_l \in \mathbb{R}^{33}$.
Although Ar(1) is still predicted by one of the \textit{flexDeepONet} models as part of \(\vq_{nl}\), its predicted value is used solely to parametrize the linear operators.
The actual concentration of Ar(1) is then recovered via the matrix exponential solution in equation~\eqref{eq:lin_sys.sol.ml}, ensuring numerical stability.
Additionally, to maintain quasi-neutrality of the plasma, we enforce the charge conservation condition $n_\mathrm{e}=n_{\mathrm{Ar}^+}$, thereby recomputing the electron density at each time step based on the Ar$^+$ concentration.
Although this results in redundant computation of both e$^-$ and Ar(1) concentrations, it incurs minimal overhead while enhancing numerical stability and ensuring consistency with physical conservation laws.
As in the previous test cases, the variability metric \(\Delta \mA(t)\), defined in equation~\eqref{eq:delta_a} and shown in figure~\ref{fig:plasma_cr.0d.delta_a} of the Supplementary Material, remains consistently low, supporting the applicability of the matrix exponential formulation to the argon plasma problem.

\begin{figure}[htb!]
\centering
\begin{subfigure}[htb!]{0.35\textwidth}
    \centering
    \caption*{\hspace{5mm}\fontfamily{mdbch}\small MENO - No corrections}
    \includegraphics[width=\textwidth]{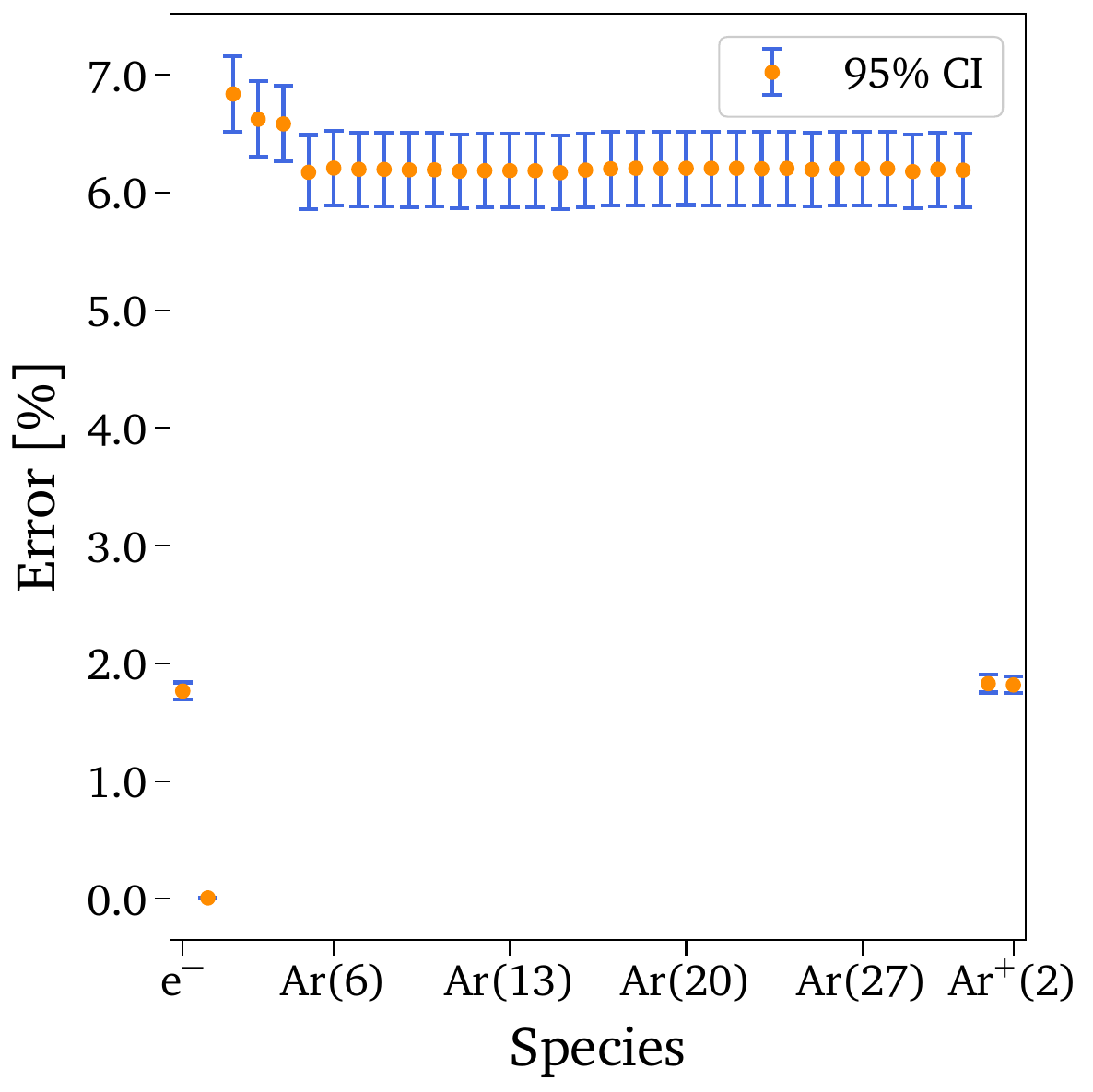}
\end{subfigure}
\begin{subfigure}[htb!]{0.35\textwidth}
    \centering
    \caption*{\hspace{5mm}\fontfamily{mdbch}\small MENO}
    \includegraphics[width=\textwidth]{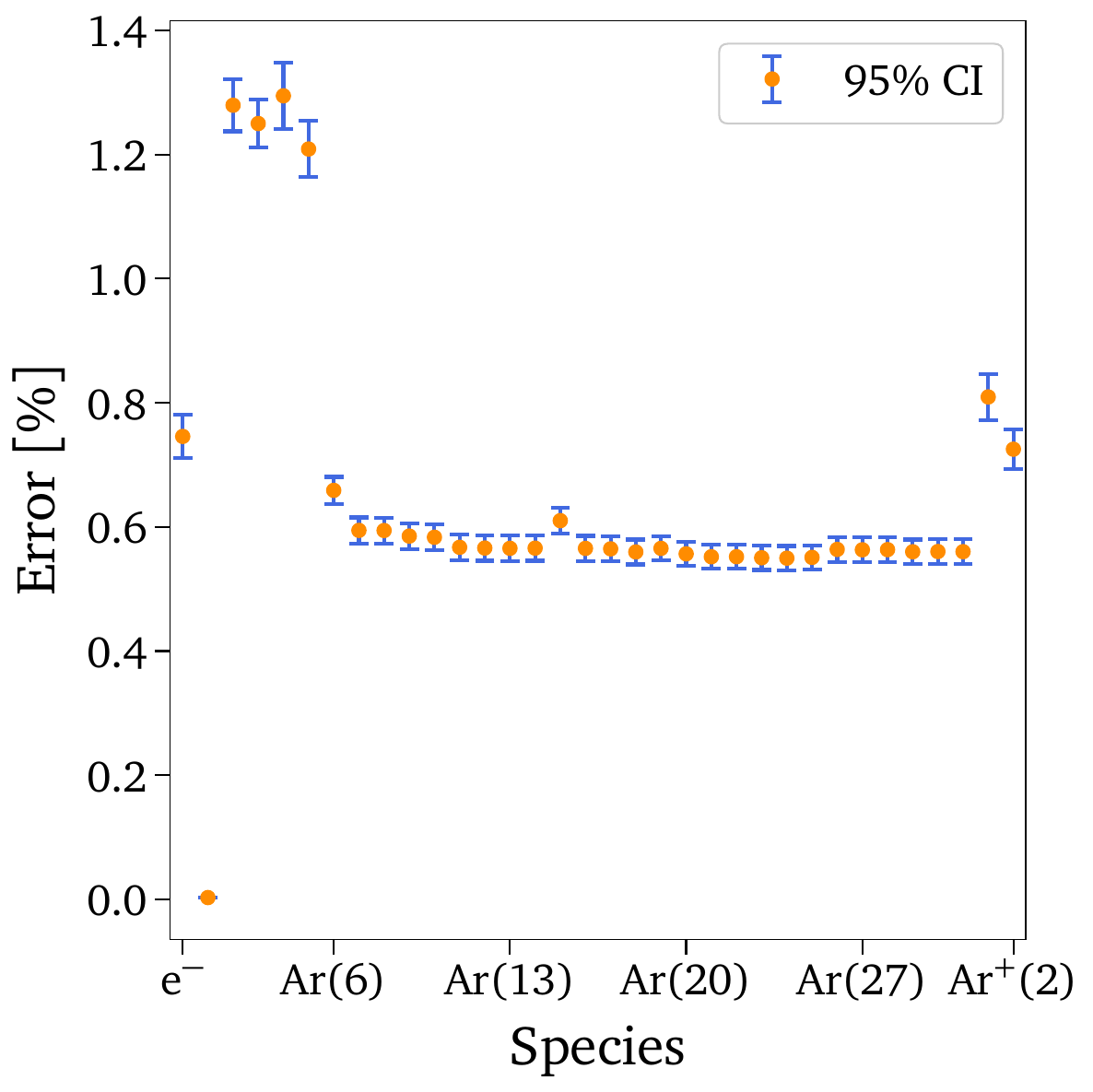}
\end{subfigure}
\caption{\textit{Test error for the argon plasma problem}. Mean relative test errors with 95\% confidence intervals for the predicted species concentrations using MENO without corrective factors (left) and the fully trained MENO (right).}
\label{fig:plasma_cr.0d.err_lin}
\end{figure}
To assess the accuracy of the 0D surrogate model, we report quantitative results based on the MAPE metric.
The errors for the two temperatures are remarkably low: approximately 0.006\% for the heavy-particle temperature \(T_\mathrm{h}\) and 0.095\% for the electron temperature \(T_\mathrm{e}\). 
Figure~\ref{fig:plasma_cr.0d.err_lin} shows the MAPE values for the species mass fractions.
The left panel presents the results using the MENO framework without corrective factors and with exact \(\vq_{nl}\) trajectories provided.
Even in this configuration, the model achieves strong performance with maximum errors around 7\%, providing an excellent baseline for training.
The right panel displays the results from the fully trained MENO model, where the prediction errors drop significantly to around 1\% across all species, confirming the effectiveness of the correction mechanisms and joint training procedure.
Additional qualitative comparisons for three unseen test cases are provided in figure~\ref{fig:plasma_cr.sol} in the Supplementary Material, further demonstrating the surrogate model’s accuracy.
\begin{figure}[htb!]
    \centering
    \begin{subfigure}[htb!]{0.35\textwidth}
        \centering
        \caption*{\hspace{7mm}\fontfamily{mdbch}\normalsize CPU}
        \includegraphics[width=\textwidth]{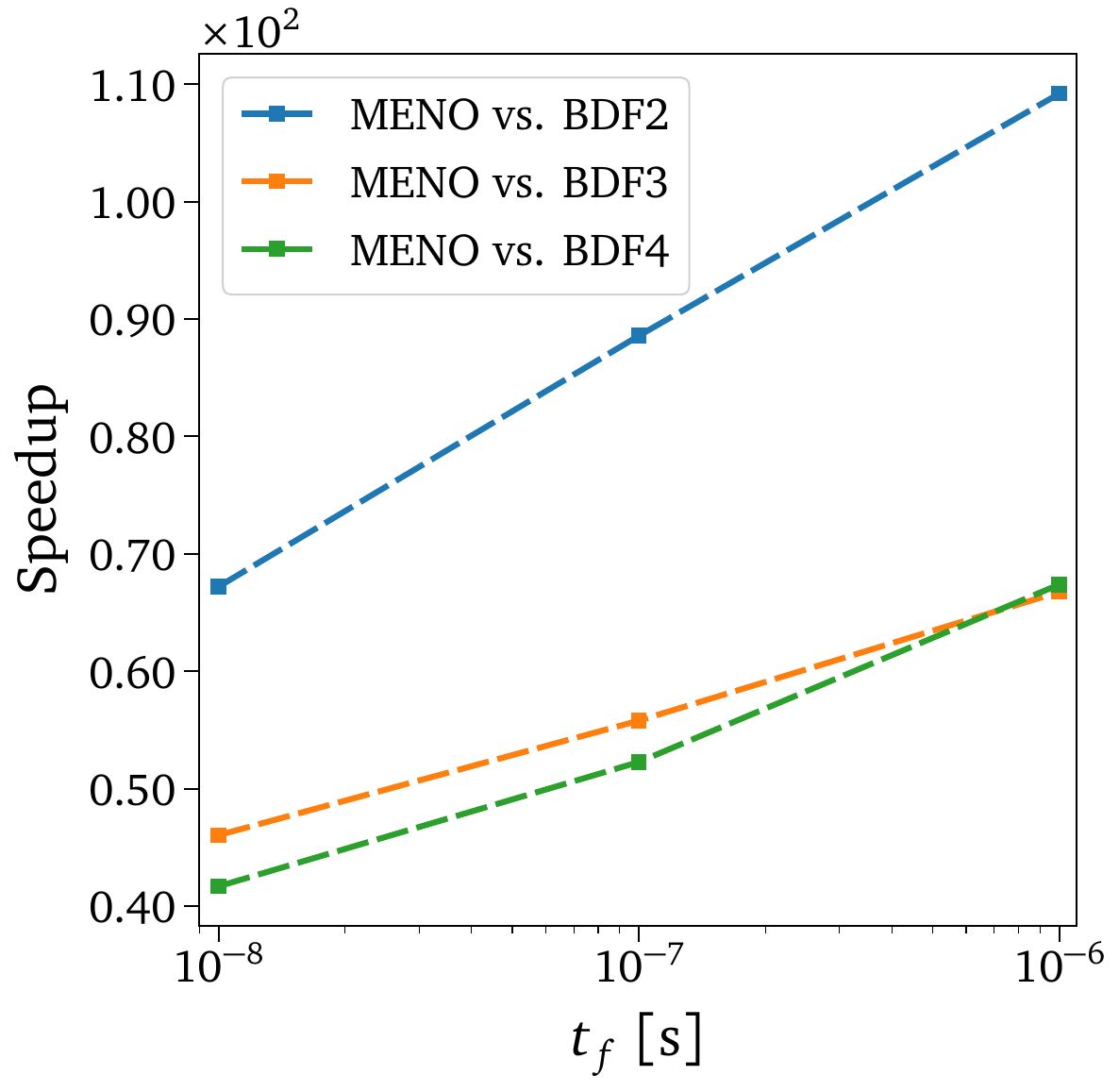}
    \end{subfigure}
    \begin{subfigure}[htb!]{0.35\textwidth}
        \centering
        \caption*{\hspace{7mm}\fontfamily{mdbch}\normalsize GPU}
        \includegraphics[width=\textwidth]{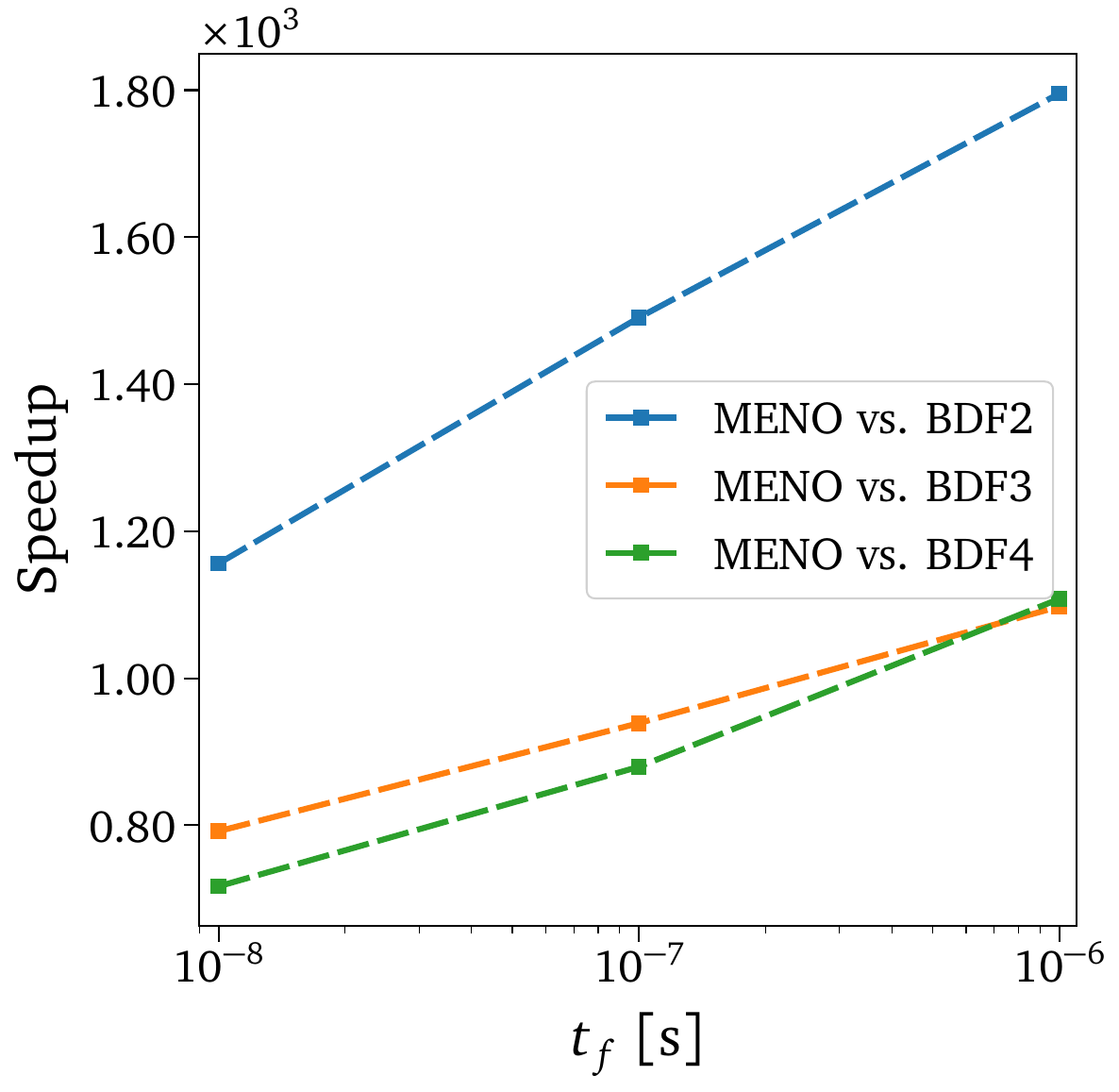}
    \end{subfigure}
    \caption{\textit{Computational speedup for the argon plasma problem}. Computational speedup achieved by executing MENO on a single CPU core (left panel) and a single GPU (right panel), compared to the BDF implicit scheme run on a single CPU core. The analysis encompasses various scheme orders (BDF2 to BDF4) and different final integration times, $t_f$, with a fixed relative tolerance of $10^{-7}$ for the BDF scheme.}
    \label{fig:plasma_cr.speedup}
\end{figure}
As the final performance metric for the zero-dimensional argon plasma system, we assess computational speedup. Following the same procedure used in the previous example, we compare the fully trained MENO model against the backward differentiation formula (BDF) solver across a range of final integration times \(t_f\) and BDF scheme orders (2 to 4), using a fixed relative tolerance of \(10^{-7}\).
As illustrated in figure~\ref{fig:plasma_cr.speedup}, the MENO framework consistently outperforms the traditional BDF solver, with computational speedup increasing almost linearly with the final integration time \(t_f\) on both CPU and GPU architectures. Detailed runtime measurements supporting this trend are provided in tables~\ref{table:plasma_cr.0d.speedup.cpu}-\ref{table:plasma_cr.0d.speedup.gpu} (Supplementary Material).
For the plasma flow applications considered in the next sections, where a typical integration time is \(t_f = 10^{-7}\) seconds, MENO achieves substantial acceleration: 88$\times$ speedup on CPU and 1\,490$\times$ on GPU, relative to BDF2.

\subsubsection{One-dimensional simulations}\label{sec:num_exp:plasma_cr:1d}
\begin{figure}[!htb]
    \centering
    \includegraphics[width=0.74\textwidth]{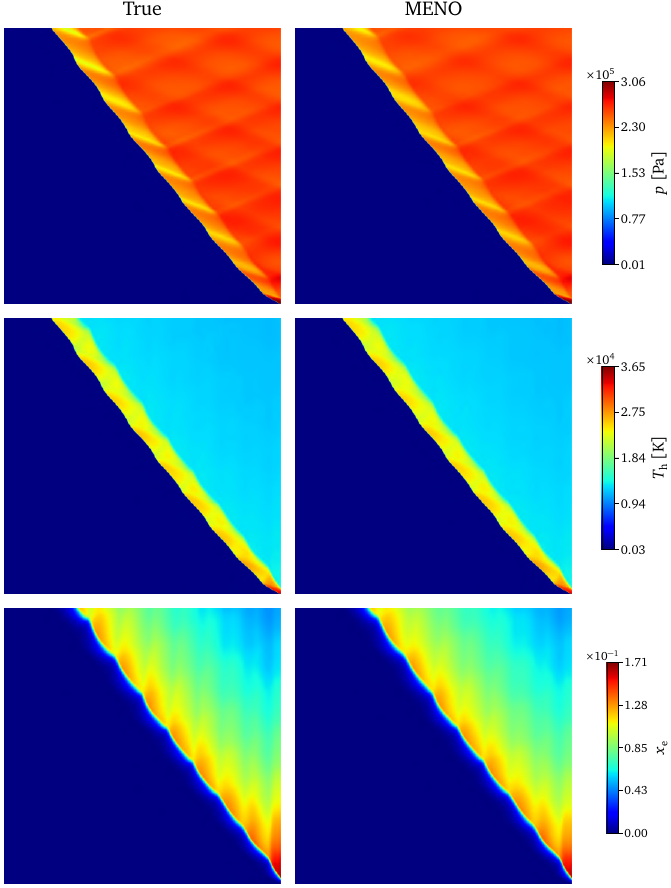}
    \caption{\textit{MENO vs. true solutions: space-time fields from 1D simulation.} Comparison between MENO-based predictions, coupled with a CFD solver, and ground-truth results for pressure $p$, heavy-particle temperature $T_\mathrm{h}$, and electron molar fraction $x_{\mathrm{e}}$ in space-time ($x$-$y$ plane) diagrams.}
    \label{fig:plasma_cr.1d.timespace}
\end{figure}
We assess the performance of the MENO-based thermochemical surrogate in a canonical 1D shock-tube configuration, where coupling between hydrodynamics and CR kinetics governs the plasma dynamics.
This test case is particularly well-suited for evaluating the surrogate’s ability to capture unsteady, nonlinear flow phenomena driven predominantly by thermochemical effects.
Details on the numerical discretization, time integration, and coupling strategy between MENO and the CFD solver are provided in section~\ref{suppl:num_exp:plasma_cr} of the Supplementary Material.
The setup follows the shock-reflection experiments by Kapper and Cambier~\cite{Kapper2011IonizingEffects}.
In this configuration, a supersonic argon stream impinges on a perfectly reflecting wall, generating a strong shock that propagates upstream.
The freestream conditions are: pressure $p_\infty = 685.2$~Pa, $T_\infty = 293.6$~K, and velocity $u_\infty = 4\,535$~m/s, the same conditions used to generate the 0D training data (see section~\ref{sec:num_exp:plasma_cr:0d}).
The domain spans \(x \in [0, 0.2]\) m with grid spacing \(\Delta x = 3\times 10^{-4}\) m. The simulation runs over \(t \in [0, 2.5 \times 10^{-4}]\) s with time step \(\Delta t=2\times 10^{-7}\) s.
Supersonic inflow is applied at the left boundary, and a reflecting wall is imposed on the right.
Additional simulations at freestream velocities $[3\,000, 4\,000, 5\,000, 6\,000]$ m/s are also performed to test the model’s capabilities to extrapolate at regimes unseen during training.

Figure~\ref{fig:plasma_cr.1d.timespace} shows space-time ($x$-$y$ plane) diagrams of pressure \(p\), heavy-particle temperature \(T_\mathrm{h}\), and electron molar fraction \(x_\mathrm{e}\), comparing the reference solution with the predictions obtained from the MENO-based surrogate model coupled to the CFD solver.
The shock front is visible as the leftmost propagating feature, followed by the electron avalanche, defined by the first peak in electron density. These structures travel from the bottom-right to the top-left of each panel, with the induction region lying between them.
The dominant physical mechanism consists of a repeating feedback loop: the electron avalanche emits a pressure wave that reflects off the shock, modifies its strength, and produces an entropy wave.
This entropy wave enhances excitation and ionization, triggering the next avalanche and restarting the cycle. These mechanisms were analyzed in detail by Kapper and Cambier~\cite{Kapper2011IonizingEffects}.
The MENO-based solution accurately reproduces this oscillatory behavior, including the amplitude and phase of the fluctuations, thus preserving the core physical dynamics.
\begin{figure}[!htb]
    \centering
    \includegraphics[width=0.96\textwidth]{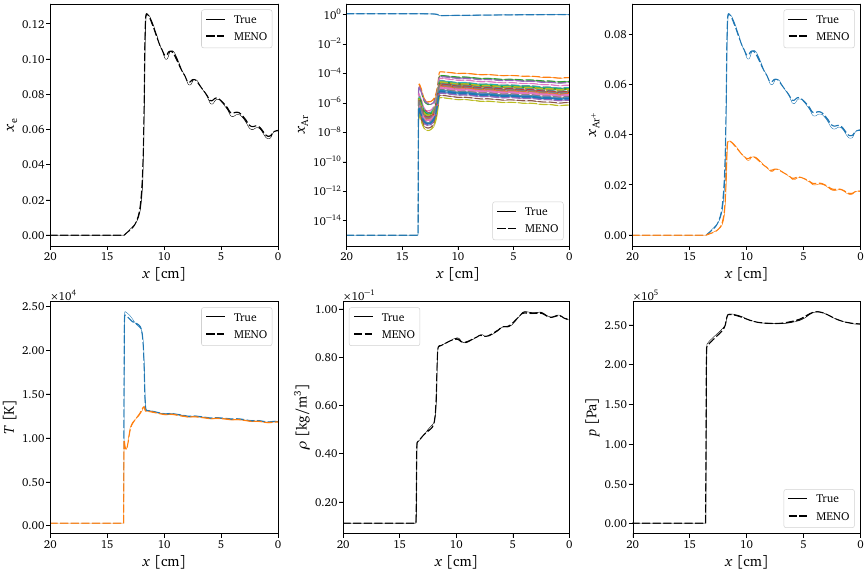}
    \caption{\textit{MENO vs. true solutions: single-time snapshots from 1D simulation.} Comparison between MENO-based predictions, coupled with a CFD solver, and ground-truth results at \(t = 2 \times 10^{-4}\,\mathrm{s}\) for the 1D simulation shown in figure~\ref{fig:plasma_cr.1d.timespace}. Top row: species molar fractions of e$^-$, Ar, and Ar$^+$. Bottom row: heavy-particle temperature $T_{\mathrm{h}}$ (blue), electron temperature $T_{\mathrm{e}}$ (orange), density $\rho$, and pressure $p$.}
    \label{fig:plasma_cr.1d.snapshots}
\end{figure}
Figure~\ref{fig:plasma_cr.1d.snapshots} provides spatial snapshots at \(t = 2 \times 10^{-4}\) s for several key quantities.
The top row shows molar fractions of e\(^-\), Ar, and Ar\(^+\), while the bottom row displays heavy-particle temperature \(T_\mathrm{h}\), electron temperature \(T_\mathrm{e}\), density \(\rho\), and pressure \(p\).
MENO predictions match the true solution with high fidelity, correctly resolving sharp transitions and fine-scale oscillations.

\begin{figure}[htb!]
\centering
\begin{subfigure}[htb!]{0.35\textwidth}
    \includegraphics[width=\textwidth]{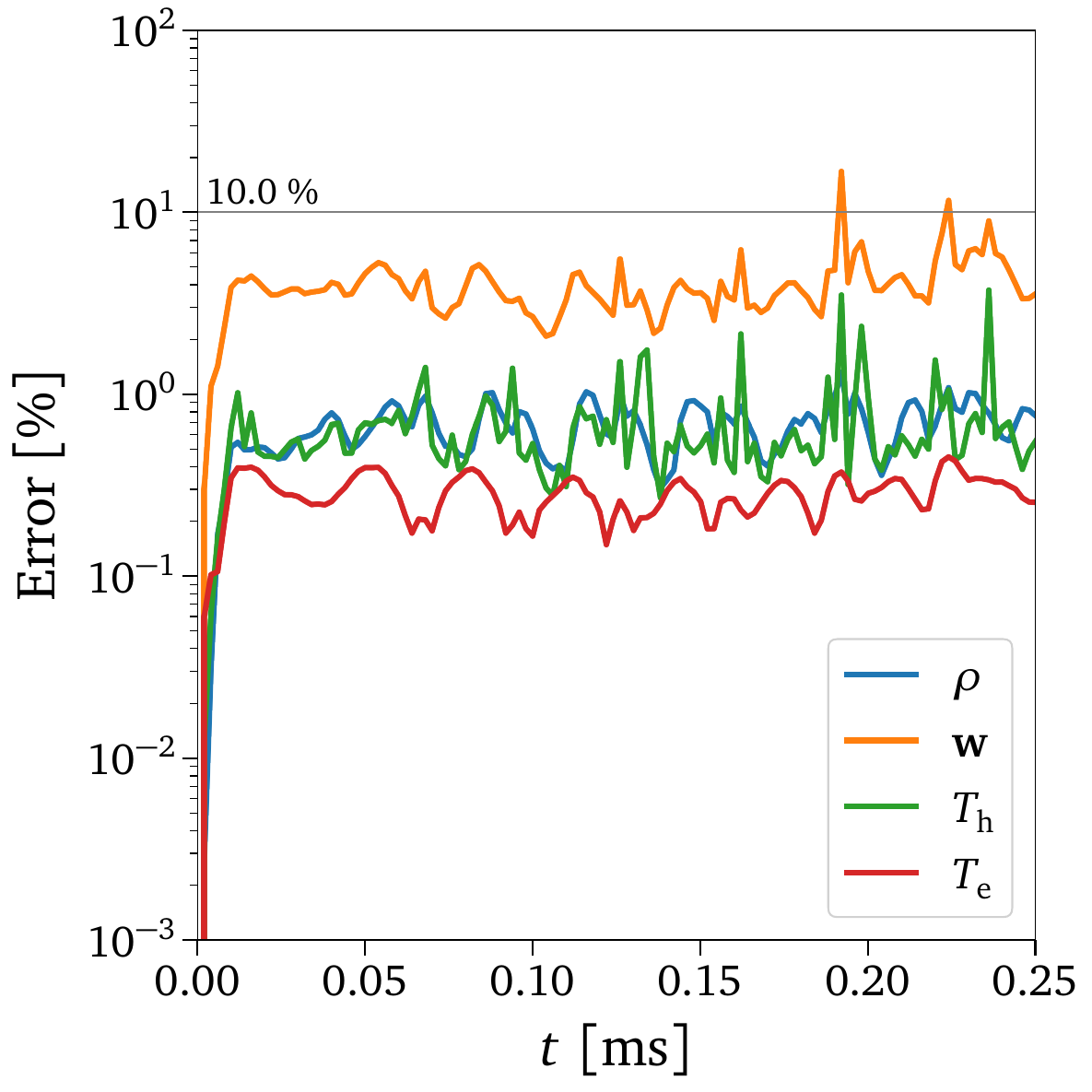}
\end{subfigure}
\begin{subfigure}[htb!]{0.35\textwidth}
    \includegraphics[width=\textwidth]{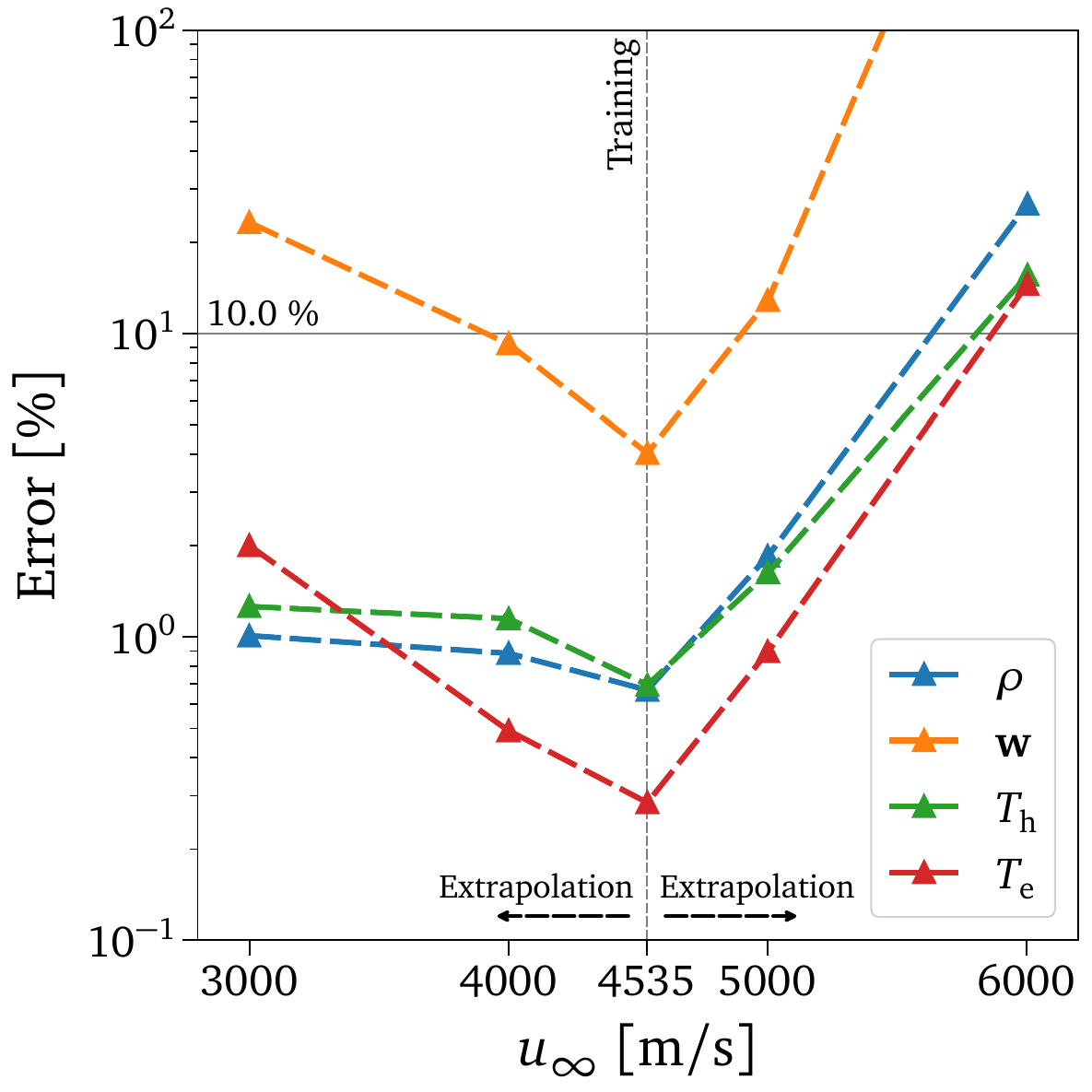}
\end{subfigure}
\caption{\textit{Error evolution of MENO in 1D simulation.} Left panel: Time history of the spatially averaged relative error for key quantities of interest at a freestream velocity of $u_\infty = 4535$ m/s. Right panel: Space-time-averaged relative error for the same quantities across a range of extrapolated freestream velocities. Note: Training was performed only at 4535 m/s; all other velocities represent extrapolation scenarios.}
\label{fig:plasma_cr.1d.err_evol}
\end{figure}
Quantitative accuracy is summarized in figure~\ref{fig:plasma_cr.1d.err_evol}.
The left panel shows the evolution of space-averaged relative errors for a set of thermochemical variables: the mixture density \(\rho\), species mass fractions vector \(\vw\in\mathbb{R}^{34}\) (averaged across all species), and the heavy-particle and electron temperatures.
To ensure the error metric reflects meaningful dynamics, points within the freestream region (where values remain constant) are excluded from the computation to prevent artificial deflation of the error.
MENO maintains errors below 1\% for macroscopic quantities (density and temperatures) and around 4\% for detailed species distribution functions, confirming excellent agreement.
The right panel reports the average space-time MAPE for different $u_\infty$, revealing the surrogate’s extrapolation capabilities.
For speeds below the training regime ($<4535$ m/s), the model retains acceptable accuracy, suggesting that training on a strong shock implicitly covers milder regimes.
However, extrapolation to higher velocities (e.g., 6000 m/s) results in a marked increase in error, indicating limitations when moving far beyond the learned solution space.

\subsubsection{Two-dimensional simulations}\label{sec:num_exp:plasma_cr:2d}
\begin{figure}[!htb]
    \centering
    \includegraphics[width=0.98\textwidth]{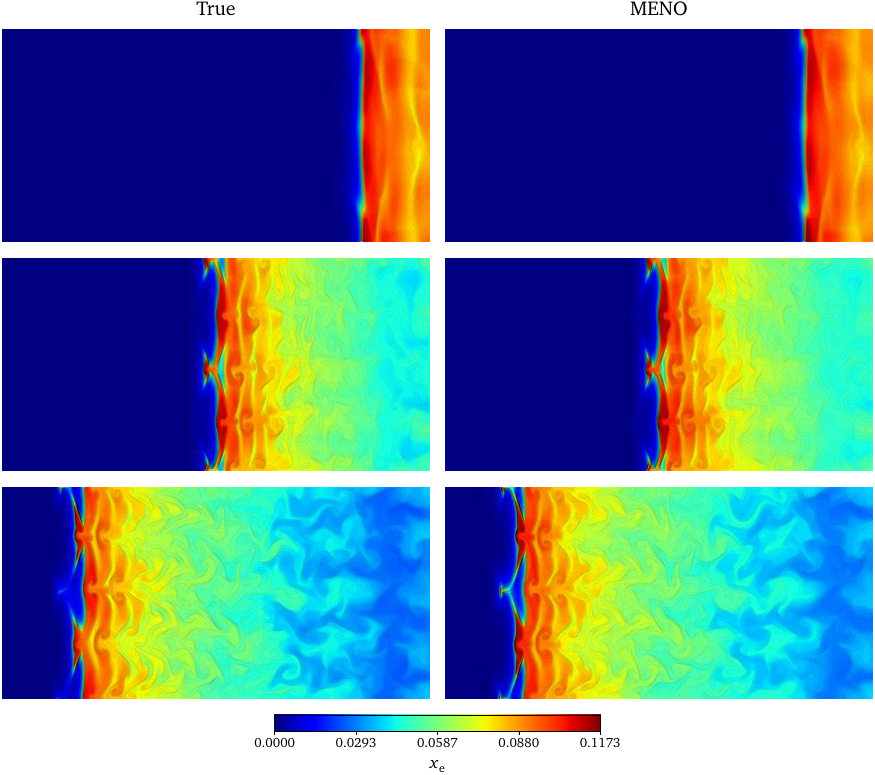}
    \vspace{-2mm}
    \caption{\textit{MENO vs. true solutions: Electron molar fraction snapshots from 2D simulation.} Comparison between MENO-based predictions, coupled with a CFD solver, and ground-truth results for the electron molar fraction \(x_{\mathrm{e}}\) at three time instants: \(t_i = [1,\,3,\,5] \times 10^{-4}\)~s.}
    \label{fig:2d.xe}
\end{figure}
Building on the insights gained from the 1D simulations, we extend our analysis to a more complex two-dimensional (2D) shock-reflection problem to evaluate the generalization capability of the MENO surrogate model.
In the 2D setting, disturbances are no longer confined to the longitudinal direction, allowing for the development of transverse instabilities and rich wave interactions.
This presents a rigorous test for the MENO surrogate and its ability to capture multidimensional plasma flow features.
Notably, this case is fully extrapolatory, as no data from 2D simulations were used during training.

The computational domain spans $(x, y) \in [0, 0.36] \times [0, 0.18]$~m and is discretized uniformly with grid spacing $\Delta x = \Delta y = 3 \times 10^{-4}$~m.
Time integration is performed over the interval $t \in [0, 5] \times 10^{-4}$~s with a fixed time step $\Delta t = 2 \times 10^{-7}$~s.
As in the 1D setup, a high-speed argon flow impinges on a reflecting wall to generate a strong shock propagating upstream.
The freestream conditions are maintained at $p_\infty = 685.2$~Pa and $T_\infty = 293.6$~K, while the velocity is varied over extrapolative values $u_\infty = [4\,000,4\,250,4\,535,4\,750]$~m/s.
These values mostly cover the lower-speed regime, which exhibited more robust surrogate performance in the 1D case.
To trigger transverse instabilities and enhance multidimensional flow features, a weak Gaussian pressure perturbation is introduced near the right wall, as described in section~\ref{suppl:num_exp:plasma_cr} of the Supplementary Material.
This perturbation accelerates the development of instabilities while remaining small enough to preserve the underlying flow physics.

Figure~\ref{fig:2d.xe} shows snapshots of the electron molar fraction $x_\mathrm{e}$ at three representative times, $t_i=[1,3,5]\times 10^{-4}$~s, for the fully extrapolatory case with $u_\infty = 4\,250$~m/s.
The true solution (left) and the MENO prediction (right) are compared side by side.
Initially, weak transverse waves form and progressively evolve into organized structures composed of incident and reflected shocks, Mach stems, and triple points.
These features form a repeating cellular pattern that resembles detonation cells observed in chemically reactive flows~\cite{Kapper2011IonizingEffects}.
The MENO surrogate successfully captures the overall dynamics, including the formation of ionization cells, their vorticity patterns, and the timing of key events.
Additional field comparisons provided in the Supplementary Material (figures \ref{fig:2d.rho}–\ref{fig:2d.Te}) further confirm the model's accuracy across a range of thermochemical variables.

\begin{figure}[htb!]
    \centering
    \includegraphics[width=0.35\textwidth]{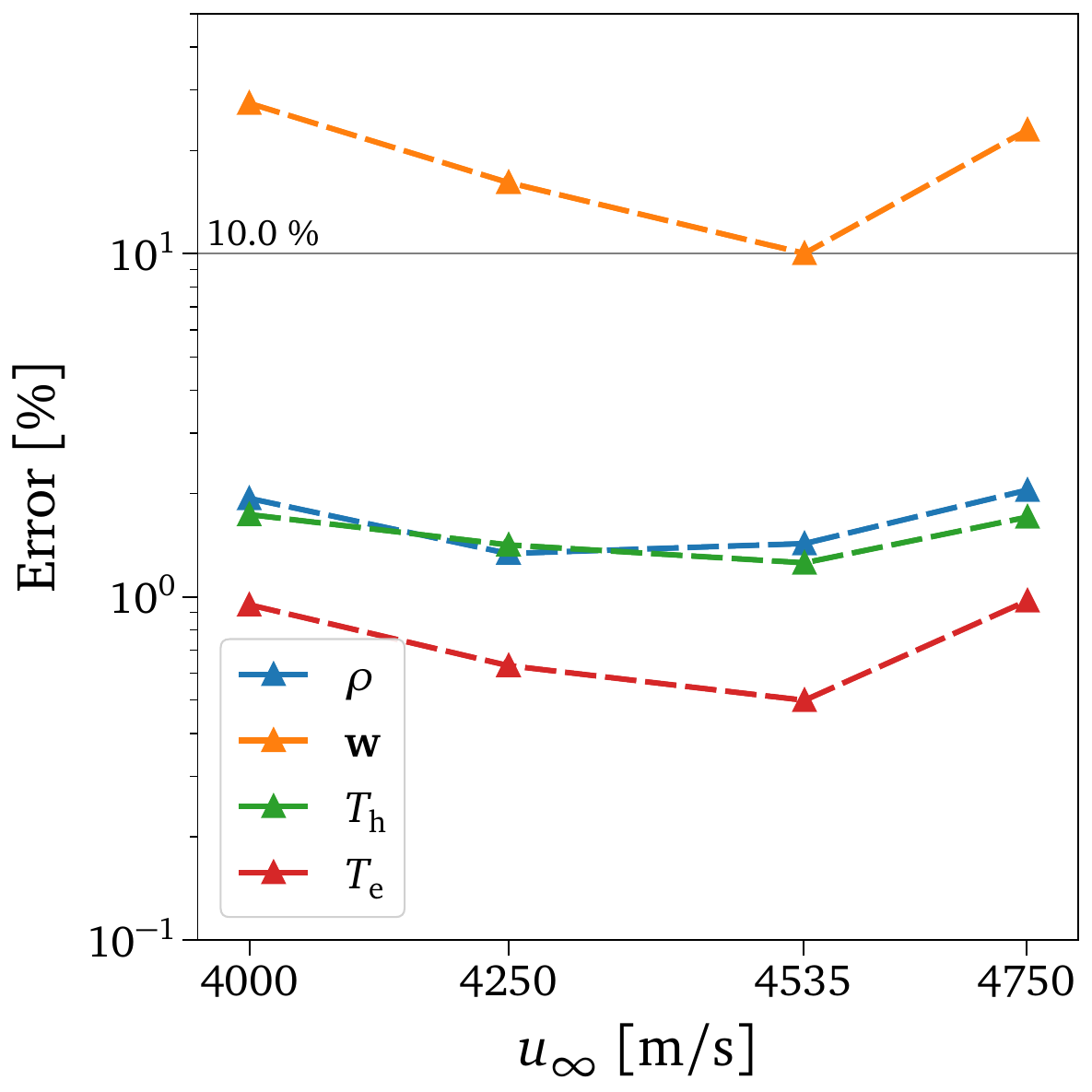}
    \caption{\textit{MENO error in 2D simulations.} Space-time-averaged relative error for key quantities of interest across a range of extrapolated freestream velocities.}
    \label{fig:2d.err_evol}
\end{figure}
A quantitative performance evaluation is presented in figure~\ref{fig:2d.err_evol}, which reports the space-time-averaged relative errors across freestream velocities $u_\infty$. 
While the error levels remain acceptable, consistent with the qualitative agreement observed in figure~\ref{fig:2d.xe}, the 2D case exhibits errors approximately twice as large as those in the 1D simulations.
This increase reflects the added complexity introduced by the extra spatial dimension, including richer flow structures and a broader range of thermochemical states.
Additionally, accumulated errors over time may contribute to the observed increase, with a slight lag in the shock front position likely being the primary source of discrepancy.

\section{Conclusions}\label{sec:conclusions}
In this work, we introduced MENO, a ``Matrix Exponential–based Neural Operator'' designed to build fast, accurate, and physically consistent surrogates for stiff thermochemical kinetic systems.
The key idea behind MENO is to exploit the inherent structure of the governing equations by embedding them directly into the model architecture.
This is accomplished by decomposing the system into a low-dimensional nonlinear component, modeled using standard neural operators such as \textit{flexDeepONet}~\cite{Venturi_CMAME_2023}, and a high-dimensional linear subsystem, which is integrated using a novel, learnable neural matrix exponential formulation.
While evaluated on thermochemical systems, MENO is broadly applicable to any differential system exhibiting this sparse nonlinear structure.

Across three challenging case studies, including the atmospheric chemistry POLLU problem, a nonequilibrium oxygen mixture, and a collisional-radiative argon plasma applied in 1D and 2D shock-tube simulations, we demonstrated that MENO delivers robust and accurate predictions.
The approach captured system behavior across many orders of magnitude in species concentrations and reproduced complex, unsteady flow dynamics in CFD settings.
Quantitatively, across the three test cases considered, the model yielded mean relative errors below 2\% in its trained 0D configuration and maintained good accuracy in fully extrapolatory multidimensional CFD scenarios.
Furthermore, MENO delivered substantial computational gains, achieving up to 4\,800$\times$ speedup on GPU and 185$\times$ on CPU over conventional implicit ODE solvers, making it particularly well-suited for real-time and large-scale simulations.

A key strength of MENO is its strong physical grounding. Unlike traditional black-box or physics-informed neural operators, where physics enters the model only through soft constraints in the loss function, MENO embeds the system of equations directly into its computational structure.
This ensures that properties such as equilibrium steady states are satisfied by construction, not merely approximated. As a result, MENO offers superior robustness, generalization, and interpretability, while also simplifying training.
However, this physically grounded design introduces some trade-offs. MENO is an intrusive method: it requires access to the system of equations and their structural decomposition, which may limit its applicability in settings where models are not fully known or easily accessible.
Moreover, MENO’s efficiency is currently maximized when the number of nonlinear variables is small relative to the linear ones, as this minimizes reliance on pure data-driven learning for the nonlinear dynamics.
Extending MENO to handle systems with more complex or higher-dimensional nonlinearities, while preserving its physical grounding, is a promising direction for future work.

Overall, MENO offers a compelling framework that combines the rigor of physics-based modeling with the flexibility of machine learning.
By moving beyond black-box surrogates to physics-structured neural architectures, MENO opens a path toward fast, reliable, and generalizable modeling tools for complex reactive systems in science and engineering.

\section*{Acknowledgements}
I. Zanardi and M. Panesi were supported by the Vannevar Bush Faculty Fellowship OUSD(RE) Grant N00014-21-1-295. The views and conclusions expressed in this work are those of the authors and do not necessarily reflect the official policies or endorsements, either expressed or implied, of the United States Government.

\putbib  
\end{bibunit}


\clearpage

\renewcommand*{\thetable}{S\arabic{table}}
\renewcommand*{\thefigure}{S\arabic{figure}}
\renewcommand*{\thesection}{S.\arabic{section}}
\renewcommand*{\thesubsection}{\thesection.\arabic{subsection}}
\renewcommand*{\thesubsubsection}{\thesubsection.\arabic{subsubsection}}
\renewcommand*{\theequation}{\thesection.\arabic{equation}}

\setcounter{section}{0}
\setcounter{page}{1}
\setcounter{figure}{0}
\setcounter{equation}{0}


\makeatletter
\let\@title\@empty
\makeatother

\begin{bibunit}  

\title{Supplementary Material to \\ ``\textit{\fulltitle}''}


\begin{abstract}
    
\end{abstract}
\begin{keyword}
Scientific machine learning \sep Neural operator \sep Stiff systems \sep Thermochemistry
\end{keyword}

\maketitle

\clearpage
\section{Methodology}\label{suppl:method}

\subsection{Commutativity conditions for simplifying matrix exponential solutions}\label{suppl:method:matexp_commute}
If $\mA(t)$ in equations \eqref{eq:lin_sys} satisfies the commutation condition
\begin{equation}\label{eq:lin_sys.commute}
    \mA(t) \int_{t_0}^t \mA(\tau) d \tau
    = \int_{t_0}^t \mA(\tau) d \tau \,\mA(t) \eqspace,
\end{equation}
then the transition matrix $\phib\left(t, t_0\right)$ can be expressed as
\begin{equation}\label{eq:lin_sys.transmat.commute}
    \phib\left(t, t_0\right) = e^{\int_{t_0}^t \mA(\tau) d \tau}
    = \sum_{k=}^{\infty} \frac{1}{k!}\left[\int_{t_0}^t \mA(\tau) d \tau\right]^k \eqspace.
\end{equation}
The commutability of \(\mA(t)\) with its integral is generally difficult to prove but holds under specific conditions, such as:
\begin{enumerate}
    \item $\mA(t)=\mA$ is a constant matrix.
    \item $\mA(t)=\beta(t) \mA$ where $\beta : \mathbb{R} \rightarrow \mathbb{R}$ is a scalar function and $\mA$ is a constant matrix.
    \item $\mA(t)=\sum_{i=1}^m \beta_i(t) \mA_i$ where $\beta_i : \mathbb{R} \rightarrow \mathbb{R}$ are all scalar functions and $\mA_i$'s are all constant matrices that commute with each other, i.e., $\mA_i \mA_j=\mA_j \mA_i, \;\forall i, j \in\{1,2, \ldots, m\}$.
    \item There exists a factorization $\mA(t)=\mT \mD(t) \mT^{-1}$ where $\mD(t)=\operatorname{diag}\left(\lambda_1(t), \ldots, \lambda_n(t)\right)$.
\end{enumerate}

\subsection{Graph-based nonlinear corrections: Illustrative example}\label{suppl:method:graph_example}
Let us consider a simple chemical system comprising four species, \(\mathcal{S} = \{A, B, C, D\}\), governed by two reversible reactions, $\mathcal{R}=\{ABC, BD\}$:
\begin{align}
    \label{eq:reac.abc}
    \ce{ $A$ + $B$ <=>[$k_\abc^\mathrm{f}$][$k_\abc^\mathrm{b}$] $C$ }\eqspace, \\
    \label{eq:reac.bd}
    \ce{ $B$ <=>[$k_\bd^\mathrm{f}$][$k_\bd^\mathrm{b}$] $D$ }\eqspace.
\end{align}
The time evolution of the system is described by the following ODEs:
\begin{equation}\label{eq:reac.simple.dyn}
    \begin{aligned}
        \frac{dn_A}{dt} & = -k_\abc^\mathrm{f}n_An_B + k_\abc^\mathrm{b}n_C \eqspace, \\
        \frac{dn_B}{dt} & = -k_\abc^\mathrm{f}n_An_B + k_\abc^\mathrm{b}n_C 
            - k_\bd^\mathrm{f}n_B + k_\bd^\mathrm{b}n_D\eqspace, \\
        \frac{dn_C}{dt} & = k_\abc^\mathrm{f}n_An_B - k_\abc^\mathrm{b}n_C \eqspace, \\
        \frac{dn_D}{dt} & = k_\bd^\mathrm{f}n_B - k_\bd^\mathrm{b}n_D \eqspace ,
    \end{aligned}
\end{equation}
where \(n_A, n_B, n_C,\) and \(n_D\) denote the number densities of the respective species, and $k^\mathrm{f}_{(\cdot)}$, $k^\mathrm{b}_{(\cdot)}$ are the forward and backward rate constants.
In this example, the nonlinear species set is \(\mathcal{S}_{nl} = \{B\}\), while the linear species are \(\mathcal{S}_l = \{A, C, D\}\). Accordingly, we model the evolution of \(n_B\) separately and define the state vectors as \(\vq_{nl} = [n_B]\) and \(\vq_l = [n_A, n_C, n_D]\).
The corresponding linear operators in equation~\eqref{eq:meno:ltv} take the form:
\begin{equation}\label{eq:lin_ops.example}
    \mA(t) = \left[\begin{matrix}
        -k_\abc^\mathrm{f} n_B(t) & \phantom{-}k_\abc^\mathrm{b} & 0 \\
        \phantom{-}k_\abc^\mathrm{f}n_B(t) & - k_\abc^\mathrm{b} & 0 \\
        0 & 0 & - k_\bd^\mathrm{b}
    \end{matrix}\right]\eqspace,
    \quad
    \vb(t) = \left[\begin{matrix}
        0 \\
        0 \\
        k_\bd^\mathrm{f}n_B(t)
    \end{matrix}\right]\eqspace.
\end{equation}
We compute a corrective factor for each reaction involving species in the linear set $\mathcal{S}_l$, specifically in this example, both reactions \eqref{eq:reac.abc} and \eqref{eq:reac.bd}.
After obtaining the latent representations $\vz_s$ for each species $s\in\mathcal{S}_l$, the corresponding edge-wise correction factors are evaluated using equation~\eqref{eq:edge_corr} as follows:
\begin{align}
    f_\abc & = \exp\left(\mathbf{1}^\intercal\vz_C - \mathbf{1}^\intercal\vz_A\right) \eqspace, \label{eq:corr.abc} \\
    f_\bd  & = \exp\left(\mathbf{1}^\intercal\vz_D\right) \eqspace. \label{eq:corr.bd}
\end{align}
This construction is inspired by equilibrium thermochemistry~\cite{Vincenti1965IntroductionDynamics2}. For example, at equilibrium, reaction~\eqref{eq:reac.abc} satisfies:
\begin{equation}
    k_\abc^\mathrm{f}n_An_B = k_\abc^\mathrm{b}n_C \eqspace,
\end{equation}
which yields the equilibrium constant:
\begin{equation} \label{eq:eq_const.exp}
    k_\abc^\mathrm{eq}=\frac{k_\abc^\mathrm{f}}{k_\abc^\mathrm{b}} = \frac{n_C}{n_An_B}
        = \exp\left[\log(n_C)-\log(n_A)-\log(n_B)\right] \eqspace.
\end{equation}
In contrast to~\eqref{eq:eq_const.exp}, the correction in~\eqref{eq:corr.abc} omits species $B$ because \(B \in \mathcal{S}_{nl}\), and its dynamics is modeled independently. Therefore, the node embeddings used in each correction term are derived exclusively from the reaction graph restricted to species in \(\mathcal{S}_l\), as defined by the edge adjacency matrix.

\subsection{Physical devices}\label{suppl:method:devices}
To assess computational speedup, we perform benchmarks on two hardware configurations:
\begin{itemize}
    \item \textit{CPU}. Tests were conducted on a dual-socket system equipped with Intel\textsuperscript{\textregistered} Xeon\textsuperscript{\textregistered} Platinum 8276 processors, providing a total of 112 logical cores (2$\times$28 cores with hyper-threading) and a base clock frequency of 2.20 GHz (up to 4.00 GHz with turbo boost).
    \item \textit{GPU}. Benchmarks were performed on an NVIDIA RTX A6000 with 48 GB of VRAM.
\end{itemize}


\section{Results}\label{suppl:num_exp}

\subsection{Example 1: POLLU problem}\label{suppl:num_exp:pollu}

\paragraph{Data generation}
To train the surrogate model, we aim to learn the integral solution across a range of initial conditions and over a specified time interval. The initial conditions are generated by sampling the concentrations of a selected subset of species, as listed in table~\ref{table:pollu.ics}.
These are the only species with non-zero values in the canonical initial condition proposed by Verwer~\cite{Verwer1994GaussSeidelKinetics} (with the addition of $y_{16}$ species); all other species are initialized to zero.
Sampling is performed using a Latin Hypercube Sampling (LHS) strategy within the parameter bounds provided in table~\ref{table:pollu.ics}. This procedure yields 1\,000 distinct initial conditions for training and validation, and an additional 100 for testing.
The training time domain spans the interval \([10^{-7}, 60]\) minutes, consistent with Verwer’s original setup. For each trajectory, time points are selected adaptively: 256 points are sampled for training trajectories and 64 for validation, with denser sampling in regions exhibiting rapid temporal variation, as determined by the magnitude of the time derivatives. This strategy ensures adequate resolution of stiff dynamics during training.
\begin{table}[!htb]
    \centering
    \begin{tabular}{c|ccccccc}
        \arrayrulecolor{black}\cmidrule{2-8}
        & $y_2$ & $y_4$ & $y_7$ & $y_8$ & $y_9$ & $y_{16}$ & $y_{17}$ \\
        \arrayrulecolor{black}\midrule
        Min & 0.1 & 0.02 & 0.05 & 0.15 & 0.005 & 0.05 & 0.0035 \\
        Max & 0.8 & 0.16 & 0.40 & 1.20 & 0.040 & 0.40 & 0.0280 \\
        \arrayrulecolor{black}\midrule
    \end{tabular}
    \caption{\textit{Initial conditions bounds for the POLLU problem}. Sampling bounds for each parameter used to generate the training and testing trajectories.}
    \label{table:pollu.ics}
\end{table}

\paragraph{Architecture}
To improve training stability and model accuracy, we apply a series of input and output transformations.
The time input $t$ is log-transformed to better capture dynamics across multiple temporal scales, while the vectors $\vmu$ and $\vchi_s$ are standardized to have zero mean and unit variance.
On the output side, the predicted species concentrations corresponding to $\vq_{nl}$ undergo an exponential transformation followed by inverse standard normalization, effectively reversing the preprocessing applied during training.
Here, $\vmu$ encodes the initial conditions for the training trajectories and consists of a subset of species concentrations, as listed in table~\ref{table:pollu.ics}.
Each species-specific feature vector $\vchi_s$ is defined as $\ve_s$, the $s$-th standard basis vector (i.e., a one-hot vector) in $\mathbb{R}^{17}$.
This encoding is used due to the lack of additional physical descriptors for each species, allowing us to differentiate among them solely through positional identity.
A detailed breakdown of the model architecture used for the POLLU problem, including all MENO components, is provided in table~\ref{table:pollu.arch}.
\begin{table}[!htb]
	\centering
	\begin{tabular}{c|c|ccccc}
        \arrayrulecolor{black}\cmidrule{3-7}
        \multicolumn{2}{c|}{} & Architecture & Parameters & Blocks & Layers & Activations \\
        \arrayrulecolor{black}\midrule
        \multirow{6}*{$\vq_{nl}\in\mathbb{R}^3$} & \multirow{3}*{$y_1$, $y_6$} & \multirow{3}*{\textit{flexDeepONet}} 
            & \multirow{3}*{24\,743} & Branch & [128, 64, 32]
            & \multirow{3}*{tanh$\times$2 + linear} \\
        & & & & Trunk & [128, 64, 32] & \\
        & & & & Pre-Net & [64, 32, 1] & \\
        \arrayrulecolor{gray}\cmidrule{2-7}
        & \multirow{3}*{$y_2$} & \multirow{3}*{\textit{flexDeepONet}} 
            & \multirow{3}*{8\,583} & Branch & [64, 32, 16]
            & \multirow{3}*{tanh$\times$2 + linear} \\
        & & & & Trunk & [64, 32, 16] & \\
        & & & & Pre-Net & [64, 32, 1] & \\
        \arrayrulecolor{gray}\midrule
        \multirow{2}*{$\vq_l\in\mathbb{R}^{17}$} & $f_\chi$ & \multirow{2}*{\textit{Feedforward}}
            & \multirow{2}*{7\,013} & \multirow{2}*{-} & [64, 32, 18]
            & \multirow{2}*{tanh$\times$2 + linear} \\
        & $f_\mu$ & & & & [64, 32, 16] & \\
        \arrayrulecolor{black}\midrule
        \multicolumn{2}{c|}{} & MENO & 65\,073 & \multicolumn{3}{|c}{} \\
        \arrayrulecolor{black}\cmidrule{3-4}
	\end{tabular}
	\caption{\textit{Neural network architecture of MENO components for the POLLU problem.} The total number of parameters is reported as computed by the TensorFlow library~\cite{TF_2016}.}
	\label{table:pollu.arch}
\end{table}

\paragraph{Training}
The model is trained using the three-stage strategy outlined in section~\ref{sec:method:train.perf}, combining modular training of nonlinear and linear components followed by joint fine-tuning.
All training is performed using mini-batch stochastic gradient descent, with the Adam optimizer and an exponential learning rate scheduler that decays the learning rate from an initial value of \(10^{-3}\).
To promote generalization and prevent overfitting, several regularization strategies are employed:
\begin{itemize}
    \item \textit{Weight decay}. Both \(L_1\) and \(L_2\) penalties are applied with a coefficient of \(10^{-4}\).
    \item \textit{Dropout}. A dropout rate of \(10^{-3}\) is applied to each hidden layer during training.
    \item \textit{Noise injection}. Standard Gaussian noise with a standard deviation of 0.02 is added to both the inputs and outputs during training.
\end{itemize}
Each of the three training stages uses tailored hyperparameters:
\begin{enumerate}
    \item \textit{Nonlinear dynamics}. The \textit{flexDeepONet} models for the nonlinear state variables are trained for 20\,000 epochs using a batch size of 5\,120. The loss function is the mean absolute logarithmic percentage error (MALPE), a logarithmic variant of equation~\eqref{eq:mape}, where \(\log q_i\) replaces \(q_i\).
    \item \textit{Linear dynamics}. The networks \(f_\chi\) and \(f_\mu\), responsible for modeling the linear state dynamics, are trained for 10\,000 epochs using the same batch size of 5\,120. The loss function for this stage is the standard mean absolute percentage error (MAPE) as defined in equation~\eqref{eq:mape}.
    \item \textit{Joint fine-tuning}. The full MENO model is fine-tuned for 1\,000 epochs using the MALPE loss with a reduced initial learning rate of \(10^{-5}\), ensuring stable convergence as all components are jointly optimized.
\end{enumerate}

\begin{figure}[htb!]
    \centering
    \begin{subfigure}[htb!]{0.32\textwidth}
        \centering\hspace{1.5mm}
        \includegraphics[width=0.95\textwidth]{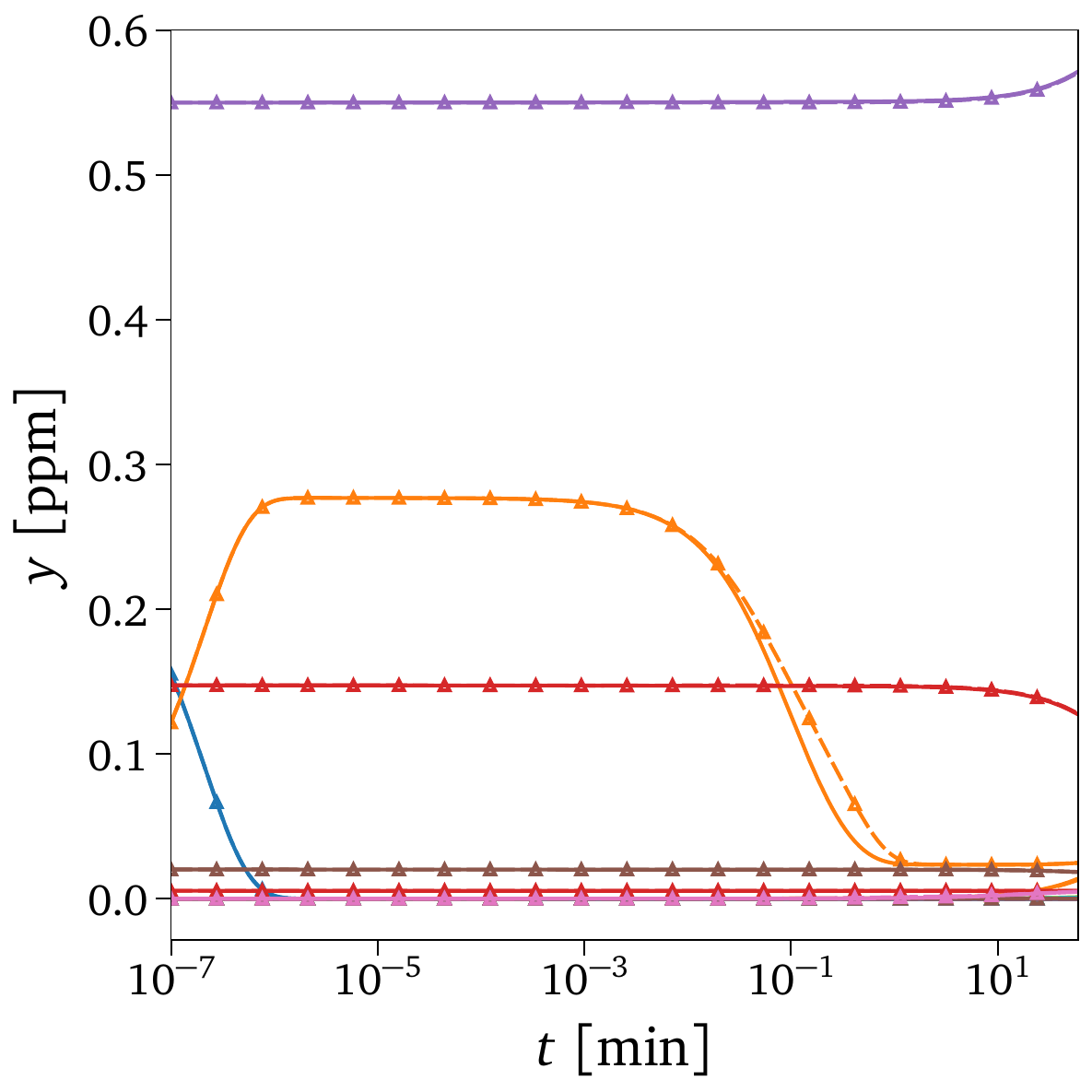}
    \end{subfigure}
    \begin{subfigure}[htb!]{0.32\textwidth}
        \centering\hspace{0.75mm}
        \includegraphics[width=0.95\textwidth]{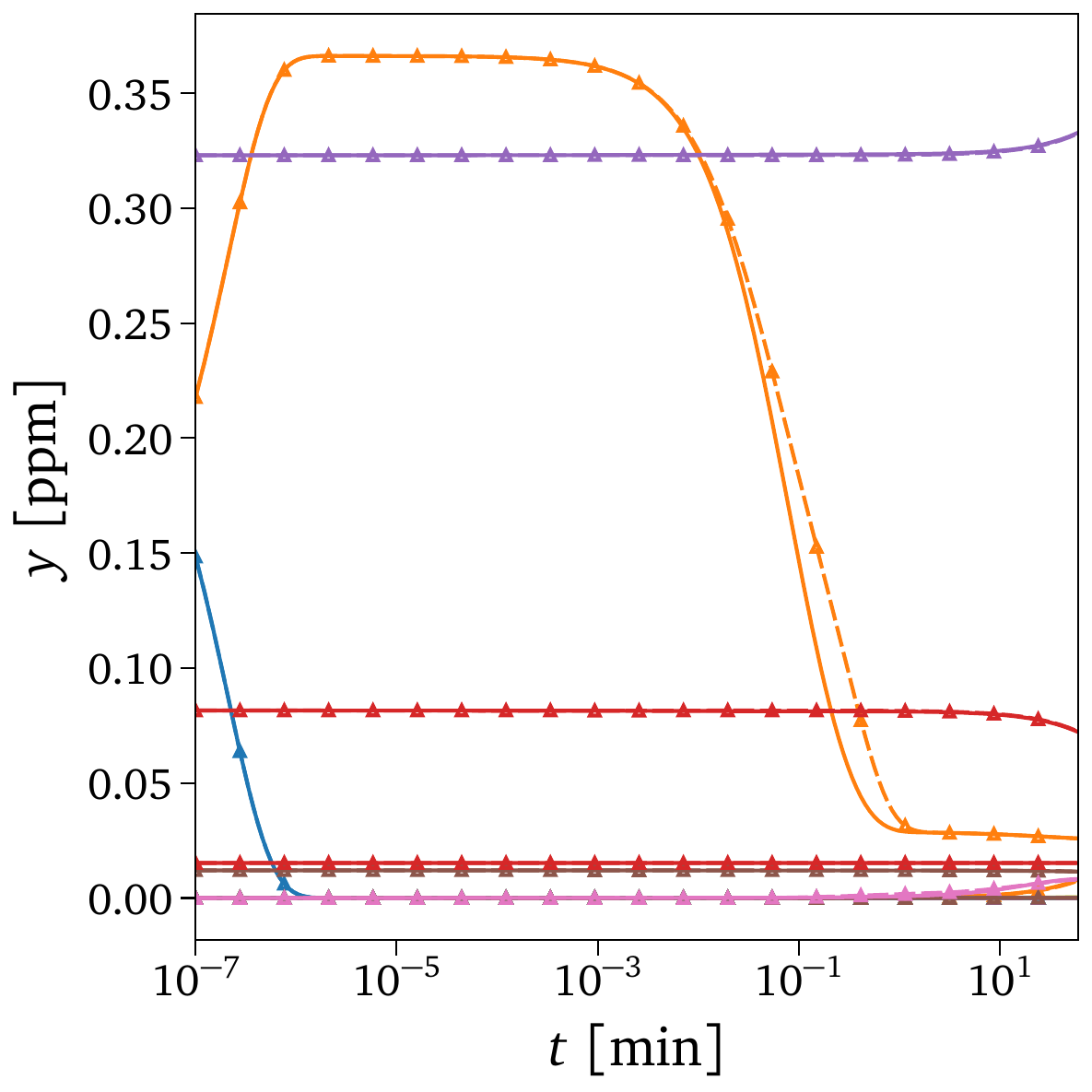}
    \end{subfigure}
    \begin{subfigure}[htb!]{0.32\textwidth}
        \centering\hspace{1.5mm}
        \includegraphics[width=0.95\textwidth]{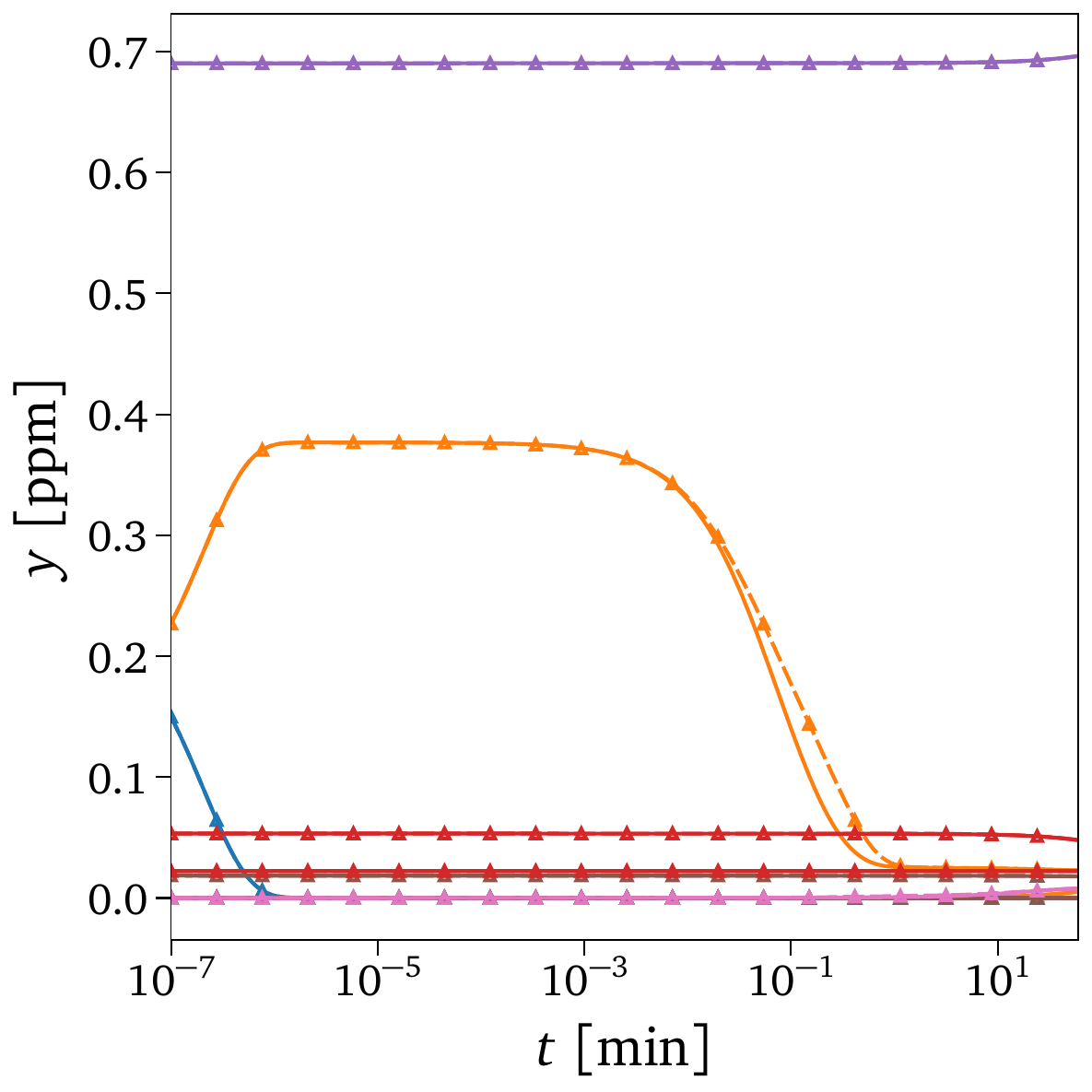}
    \end{subfigure}
    \\[2pt]
    \begin{subfigure}[htb!]{0.32\textwidth}
        \centering
        \includegraphics[width=\textwidth]{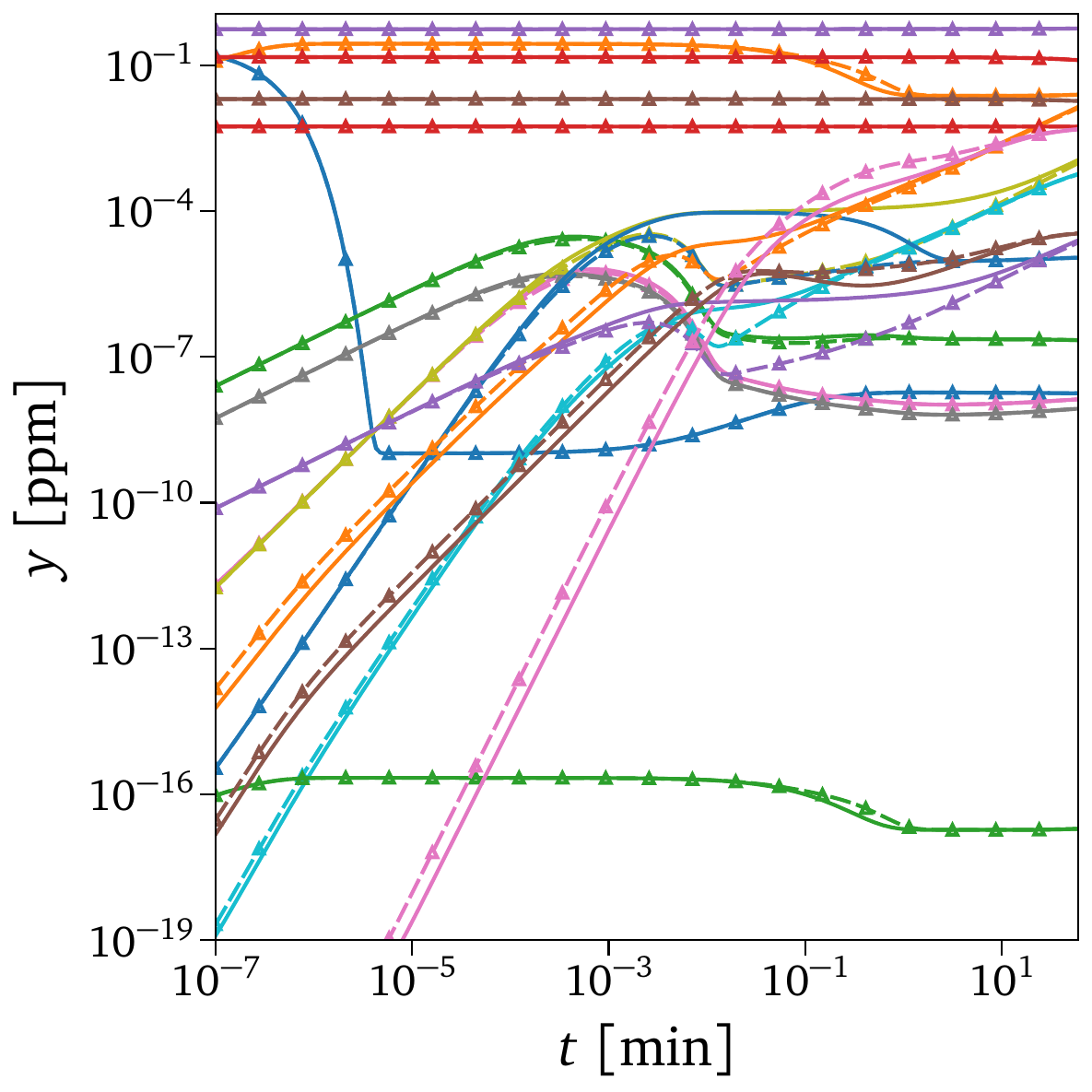}
    \end{subfigure}
    \begin{subfigure}[htb!]{0.32\textwidth}
        \centering
        \includegraphics[width=\textwidth]{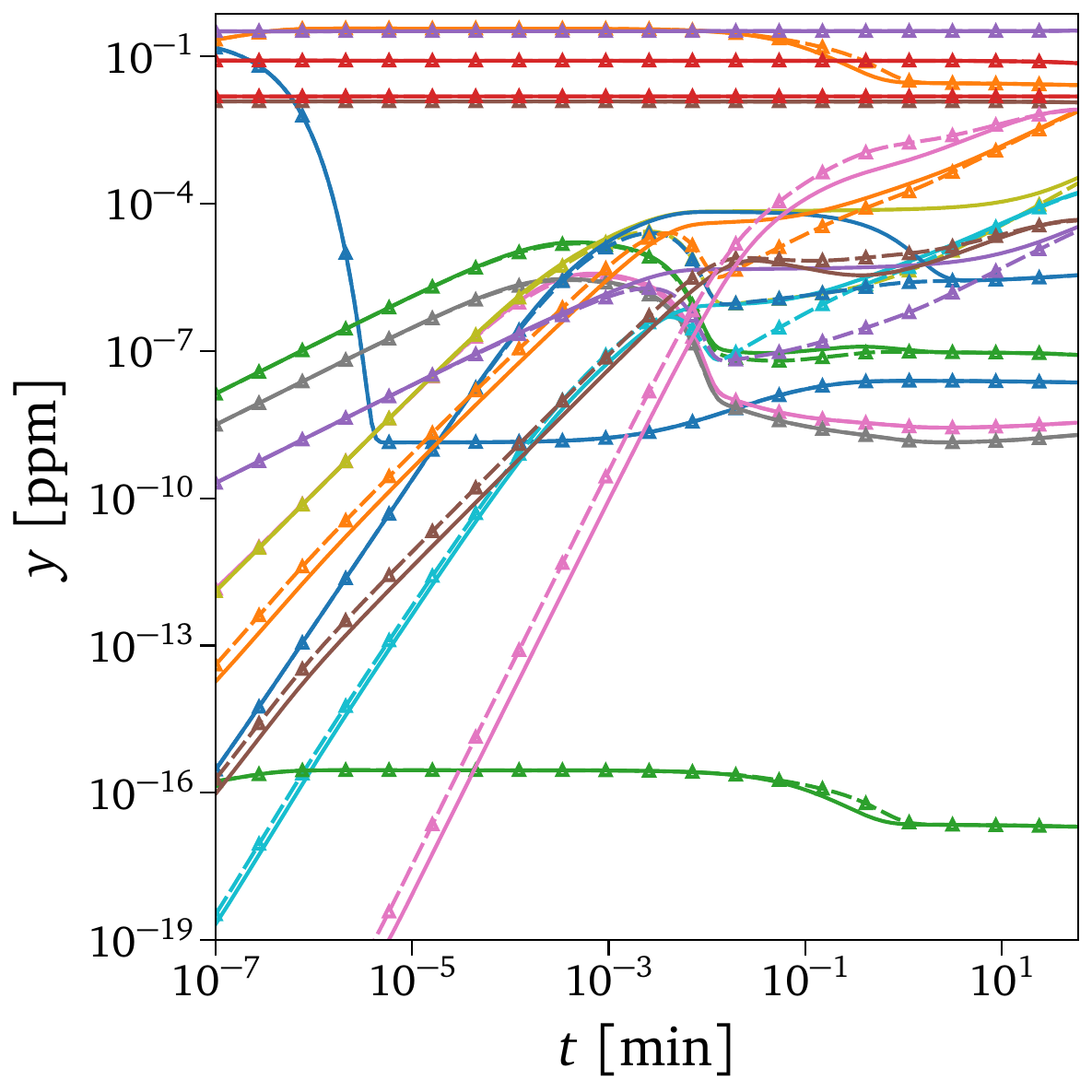}
    \end{subfigure}
    \begin{subfigure}[htb!]{0.32\textwidth}
        \centering
        \includegraphics[width=\textwidth]{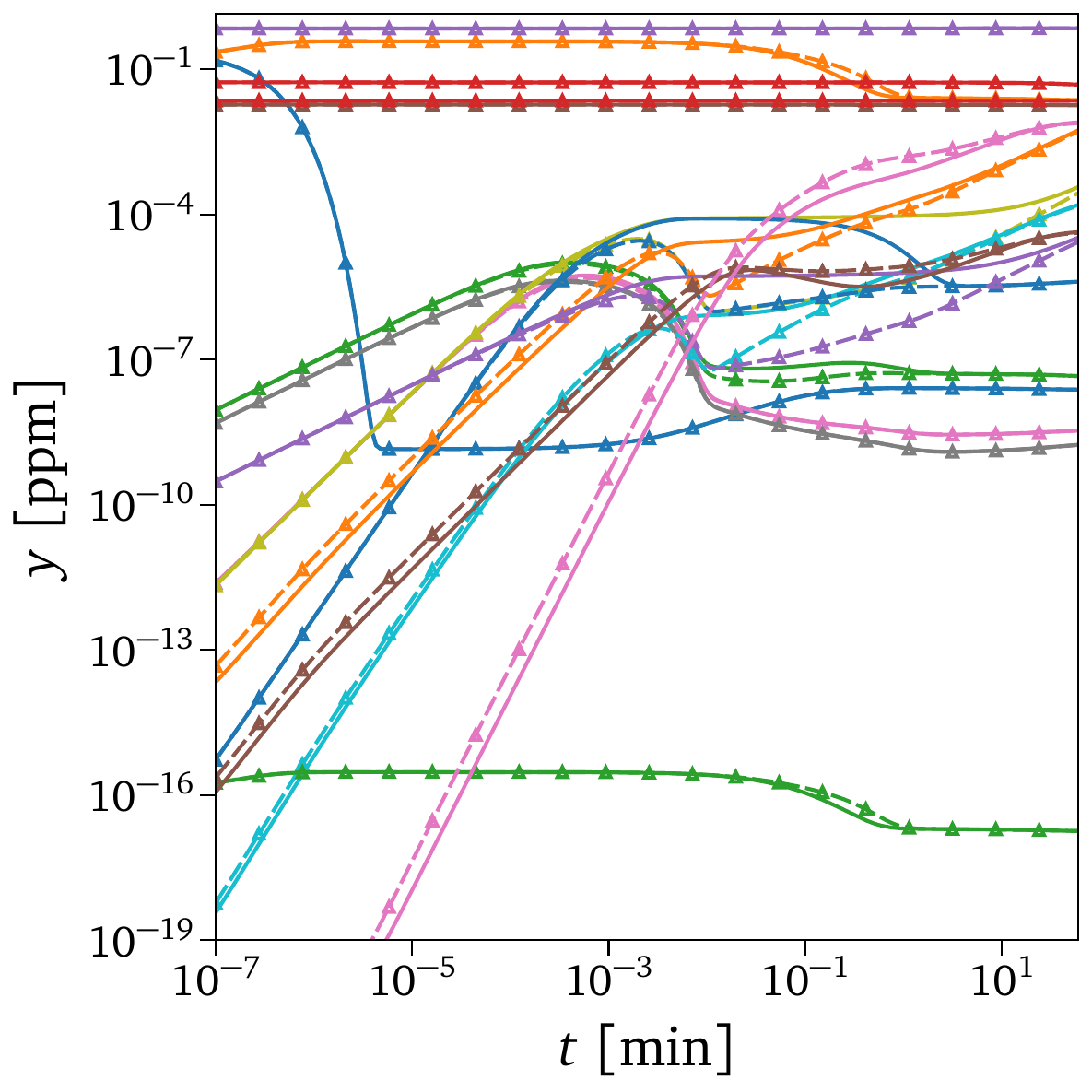}
    \end{subfigure}
    \caption{\textit{Comparison of MENO (no corrections) and true solutions for the POLLU problem.} Reference solutions (solid lines) are compared with predictions from the uncorrected MENO model (dashed lines with markers) across three representative test cases (columns). Each subplot shows species concentrations over time, presented in linear scale (top row) and logarithmic scale (bottom row).}
    \label{fig:pollu.sol_nocorr}
\end{figure}
\paragraph{Accuracy}
Figure~\ref{fig:pollu.sol_nocorr} depicts the time evolution of the 17 chemical species concentrations contained in the linear state vector $\vq_l$, across the same three representative test cases shown in figure~\ref{fig:pollu.sol}, with each column corresponding to one case.
Each subplot displays the predicted and reference concentrations using both linear (top row) and logarithmic (bottom row) scales, enabling visualization of species dynamics over several orders of magnitude.
In this comparison, predictions from the uncorrected MENO model are evaluated against the reference solutions to assess baseline performance prior to applying graph-based corrections.

\paragraph{Speedup}
For this problem, we did not evaluate the actual computational speedup achieved by the surrogate model.
This decision was made because the baseline ODE solver used for comparison was implemented in Python rather than in a compiled library such as LSODE.
As a result, any runtime measurements would be skewed by Python-specific overheads, such as interpreter latency, memory management, and the lack of low-level optimization, which do not reflect the true numerical efficiency of the method.

\subsection{Example 2: Nonequilibrium oxygen}\label{suppl:num_exp:noneq}

\paragraph{Kinetic model}
The O$_2$-O system includes two types of processes:
\begin{enumerate}[i.]
    \item collisional dissociation,
    \begin{equation}\label{eq:coll.atom.diss}
        \ce{
            O_2$(i)$ + O
            <=>[$k_{i}^{\mathrm{d}}$][$k_{i}^{\mathrm{r}}$]
            O + O + O
        }\eqspace,
    \end{equation}
    \item collisional excitation, including both inelastic (nonreactive) and exchange processes,
    \begin{equation}\label{eq:coll.atom.excit}
        \ce{
            O_2$(i)$ + O
            <=>[$k_{ij}^{\mathrm{e}}$][$k_{ji}^{\mathrm{e}}$]
            O_2$(j)$ + O
        }\eqspace.
    \end{equation}
\end{enumerate}
Here, $k_{i}^{\mathrm{d}}$, $k_{i}^{\mathrm{r}}$, $k_{ij}^{\mathrm{e}}$, and $k_{ji}^{\mathrm{e}}$ are temperature-dependent rate constants, obtained by fitting quasi-classical trajectory~\cite{Jaffe_AIAA_2015,Venturi_JCPA_2020,Priyadarshini2023EfficientArchitectures} data to Arrhenius-type expressions.

\paragraph{Data generation}
We aim to develop a surrogate model for the zero-dimensional thermochemical system described by equations~\eqref{eq:noneq.mol}-\eqref{eq:noneq.atom} when subjected to sudden heating, which mimics shock conditions typical of atmospheric reentry flows.
The initial state of the reactor is characterized by four key parameters: the initial pressure $p_0$, the initial molar fraction of atomic oxygen $x_{\atom_0}$, the initial system temperature $T_0$, and the initial internal temperature $T_{\mathrm{int}_0}$, which defines a Boltzmann distribution over the internal energy bins of molecular oxygen.
For training and testing trajectories, these parameters are sampled using the LHS method within the bounds listed in table~\ref{table:noneq.ics}.
This procedure yields 1\,000 distinct initial conditions for training and validation, and an additional 100 for testing.
The training time domain spans the interval \([10^{-11}, 10^{-1}]\) seconds, which spans the relevant excitation and dissociation processes of the nonequilibrium problem under investigation.
Temporal sampling of the training points is carried out following the procedure described in the \textit{Data generation} paragraph of section~\ref{suppl:num_exp:pollu}.
\begin{table}[!htb]
	\centering
	\begin{tabular}{c|cccc}
		\toprule
		& $p_0$ [Pa] & $x_{\atom_0}$ & $T_{\mathrm{int}_0}$ [K] & $T_0$ [K] \\
		\midrule
		Min & 1\,000 & 0 & 1\,000 & 8\,000 \\
		Max & 10\,000 & 0.95 & 8\,000 & 15\,000 \\
		\bottomrule
	\end{tabular}
    \caption{\textit{Initial conditions bounds for the nonequilibrium oxygen problem}. Sampling bounds for each parameter used to generate the training and testing trajectories.}
    \label{table:noneq.ics}
\end{table}

\paragraph{Architecture}
To enhance training stability and improve predictive accuracy, we apply several transformations to both inputs and outputs of the model.

The time variable $t$ is log-scaled and normalized using the initial pressure of atomic oxygen, $p_{\atom_0}$, computed from the initial condition vector $\vmu$.
Specifically, we define $\hat{t}_{\ln}=\ln(t\,p_{\atom_0})$, which helps align trajectories by mitigating the time shifts introduced by varying species concentrations.
Since collision frequency scales with density, this transformation reduces translational symmetries in the dataset and improves learning by promoting better temporal alignment across samples.
The input vectors $\vmu$ and $\vchi_s$ are both log-transformed and then standardized to have zero mean and unit variance.
The parametric initial condition vector is defined as $\vmu=[n_{\atom_0},n_0,e_{\mol_0},T_0]$, where $n_{\atom_0}$ is the initial number density of atomic oxygen, $n_0$ is the total initial number density, and $e_{\mol_0}$ is the initial internal energy of molecular oxygen.
Each species-specific feature vector $\vchi_s$, for group $s=g\in\{1,...,45\}$, is defined as $\vchi_s=[e^\mathrm{vib}_s,e^\mathrm{rot}_s]$, where $e^\mathrm{vib}_s$ and $e^\mathrm{rot}_s$ are the mean vibrational and rotational energies of group $s$, respectively.

On the output side, the nonlinear variables $n_\atom$ and $T$ from the state vector $\vq_{nl}$ are first passed through a \textit{sigmoid} function and then linearly rescaled using their respective equilibrium values. These equilibrium values are computed based on mass and energy conservation given the initial condition vector $\vmu$.

\begin{table}[!htb]
	\centering
	\begin{tabular}{c|c|ccccc}
        \arrayrulecolor{black}\cmidrule{3-7}
        \multicolumn{2}{c|}{} & Architecture & Parameters & Blocks & Layers & Activations \\
        \arrayrulecolor{black}\midrule
        \multirow{3}*{$\vq_{nl}\in\mathbb{R}^2$} & \multirow{3}*{$n_\atom$, $T$} & \multirow{3}*{\textit{flexDeepONet}} 
            & \multirow{3}*{30\,727} & Branch & [128, 64, 32]
            & \multirow{3}*{tanh$\times$2 + linear} \\
        & & & & Trunk & [128, 64, 32] & \\
        & & & & Pre-Net & [128, 64, 1] & \\
        \arrayrulecolor{gray}\midrule
        \multirow{2}*{$\vq_l\in\mathbb{R}^{45}$} & $f_\chi$ & \multirow{2}*{\textit{Feedforward}}
            & \multirow{2}*{2\,997} & \multirow{2}*{-} & [32, 32, 10]
            & \multirow{2}*{tanh$\times$2 + linear} \\
        & $f_\mu$ & & & & [32, 32, 8] & \\
        \arrayrulecolor{black}\midrule
        \multicolumn{2}{c|}{} & MENO & 64\,445 & \multicolumn{3}{|c}{} \\
        \arrayrulecolor{black}\cmidrule{3-4}
	\end{tabular}
	\caption{\textit{Neural network architecture of MENO components for the nonequilibrium oxygen problem.} The total number of parameters is reported as computed by the TensorFlow library~\cite{TF_2016}.}
	\label{table:noneq.arch}
\end{table}
\begin{table}[!htb]
	\centering
	\begin{tabular}{c|ccccc}
        \arrayrulecolor{black}\cmidrule{2-6}
        & Architecture & Parameters & Blocks & Layers & Activations \\
        \arrayrulecolor{black}\midrule
        \multirow{3}*{$T\in\mathbb{R}$} & \multirow{3}*{\textit{flexDeepONet}} 
            & \multirow{3}*{30\,727} & Branch & [128, 64, 32]
            & \multirow{3}*{tanh$\times$2 + linear} \\
        & & & Trunk & [128, 64, 32] & \\
        & & & Pre-Net & [128, 64, 1] & \\
        \arrayrulecolor{gray}\midrule
        \multirow{2}*{$\vn\in\mathbb{R}^{46}$} & \multirow{2}*{\textit{DeepONet}} 
            & \multirow{2}*{8\,453\,315} & Branch & [1280, 1280, 1280, 16$\times$46]
            & \multirow{2}*{tanh$\times$2 + linear} \\
        & & & Trunk & [1280, 1280, 1280, 16$\times$46] & \\
        \arrayrulecolor{black}\midrule
	\end{tabular}
	\caption{\textit{Vanilla DeepONet architecture for the nonequilibrium oxygen problem.}}
	\label{table:noneq.arch_don}
\end{table}
A detailed summary of the model architecture used for the nonequilibrium oxygen problem, including all components of the MENO framework, is presented in table~\ref{table:noneq.arch}.
The architecture of the vanilla DeepONet~\cite{Lu_NMI_2021} used for comparison is provided in table~\ref{table:noneq.arch_don}.
In the DeepONet configuration, temperature $T$ is modeled using the same \textit{flexDeepONet} architecture as in MENO, while species composition $\vn\in\mathbb{R}^{46}$ is predicted via a single DeepONet model. In this setup, the branch net takes the transformed initial condition $\vmu$, while the trunk net receives the transformed time input.
A \textit{softmax} activation is applied to the output to predict mass fractions (ensuring mass conservation), which are then converted to number densities for comparison.

\paragraph{Training}
The model is trained using the three-stage strategy described in section~\ref{sec:method:train.perf}, consisting of modular training of the nonlinear and linear components, followed by a final joint fine-tuning stage.
Optimization is performed using mini-batch stochastic gradient descent with the Adam optimizer. An exponential learning rate scheduler is employed to decay the learning rate from an initial value of \(10^{-3}\), promoting stable convergence.
To enhance generalization and mitigate overfitting, the following regularization techniques are employed:
\begin{itemize}
    \item \textit{Weight decay}. \(L_1\) and \(L_2\) penalties are used during training, with a regularization coefficient of \(10^{-3}\) for the nonlinear state variables $\vq_{nl}$, and \(10^{-4}\) for both the linear variables $\vq_{l}$ and during joint fine-tuning.
    \item \textit{Dropout}. A dropout rate of \(10^{-2}\) is applied to each hidden layer during training of $\vq_{nl}$, and \(10^{-3}\) during training of $\vq_{l}$ and the fine-tuning stage.
    \item \textit{Noise injection}. Gaussian noise is added to both inputs and outputs during training, with standard deviations of 0.04 for $\vq_{nl}$ and 0.02 for $\vq_{l}$ and the fine-tuning step.
\end{itemize}
Each training stage employs specific hyperparameters suited to its task:
\begin{enumerate}
    \item \textit{Nonlinear dynamics}. The \textit{flexDeepONet} models for the nonlinear state variables are trained independently for 20\,000 epochs with a batch size of 5\,120. The loss function used is MAPE (equation~\eqref{eq:mape}) for the atomic oxygen number density $n_\atom$, and the mean absolute error (MAE) for temperature $T$.
    \item \textit{Linear dynamics}. The networks \(f_\chi\) and \(f_\mu\) are trained for 2\,000 epochs with a batch size of 2\,560 using the MAPE loss.
    \item \textit{Joint fine-tuning}. In the final stage, the entire MENO architecture is fine-tuned for 1\,000 epochs using a reduced initial learning rate of \(10^{-5}\) and the MALPE loss, ensuring stable convergence as all components are jointly optimized.
\end{enumerate}

\begin{figure}[!htb]
    \centering
    \begin{subfigure}[htb!]{0.32\textwidth}
        \centering
        \includegraphics[width=\textwidth]{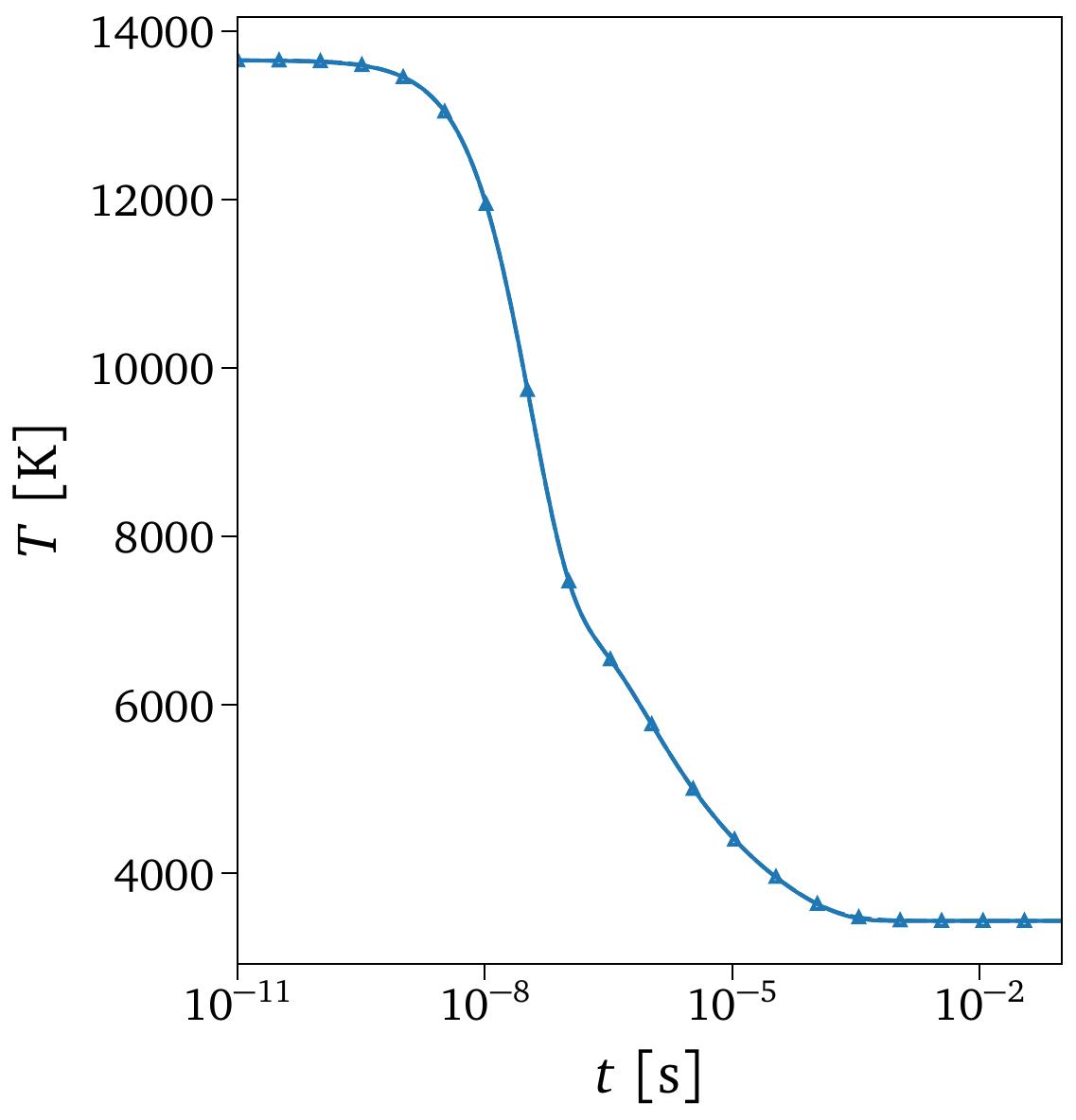}
    \end{subfigure}
    \begin{subfigure}[htb!]{0.32\textwidth}
        \centering
        \includegraphics[width=\textwidth]{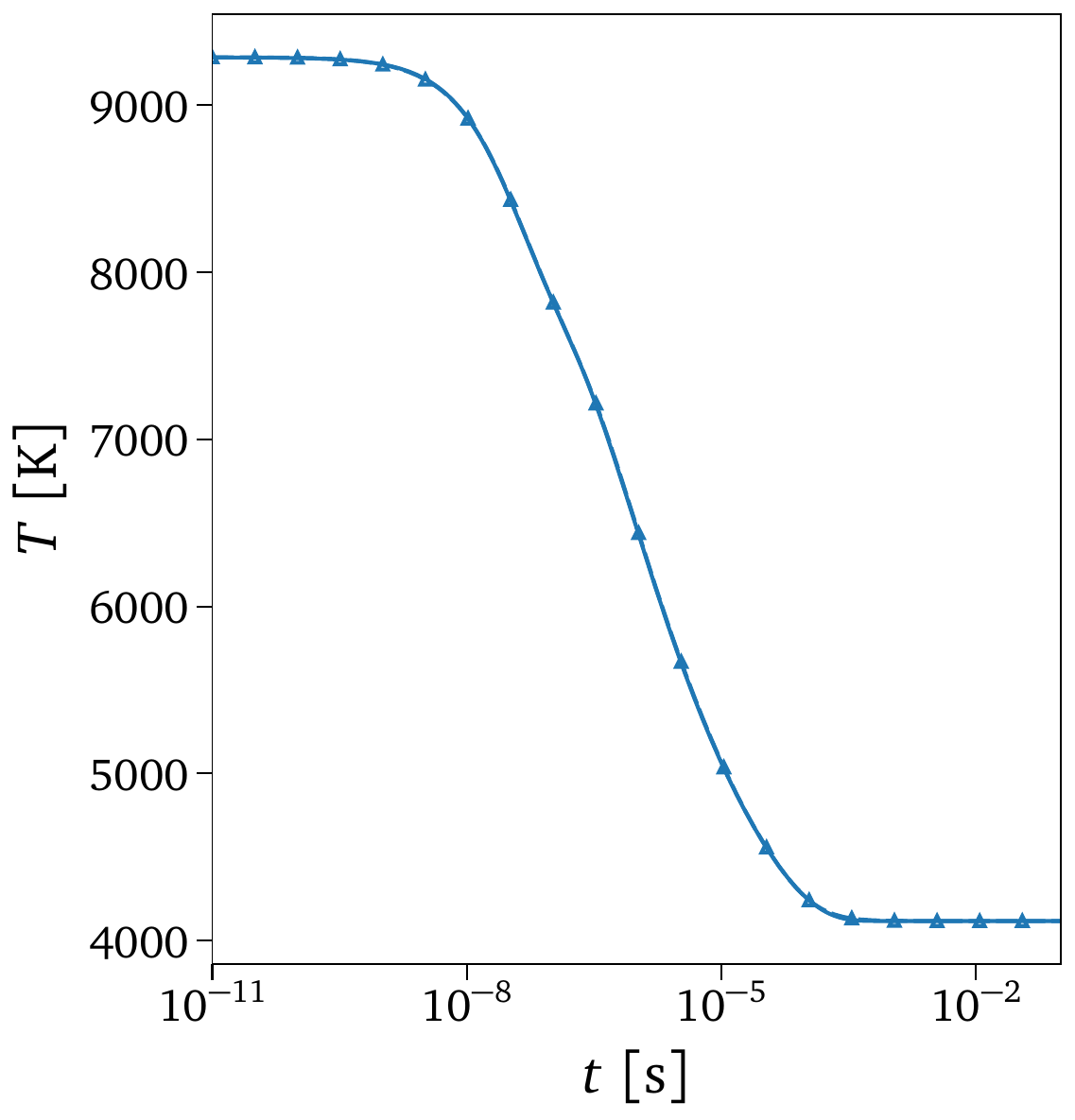}
    \end{subfigure}
    \\
    \begin{subfigure}[htb!]{0.32\textwidth}
        \centering
        \includegraphics[width=\textwidth]{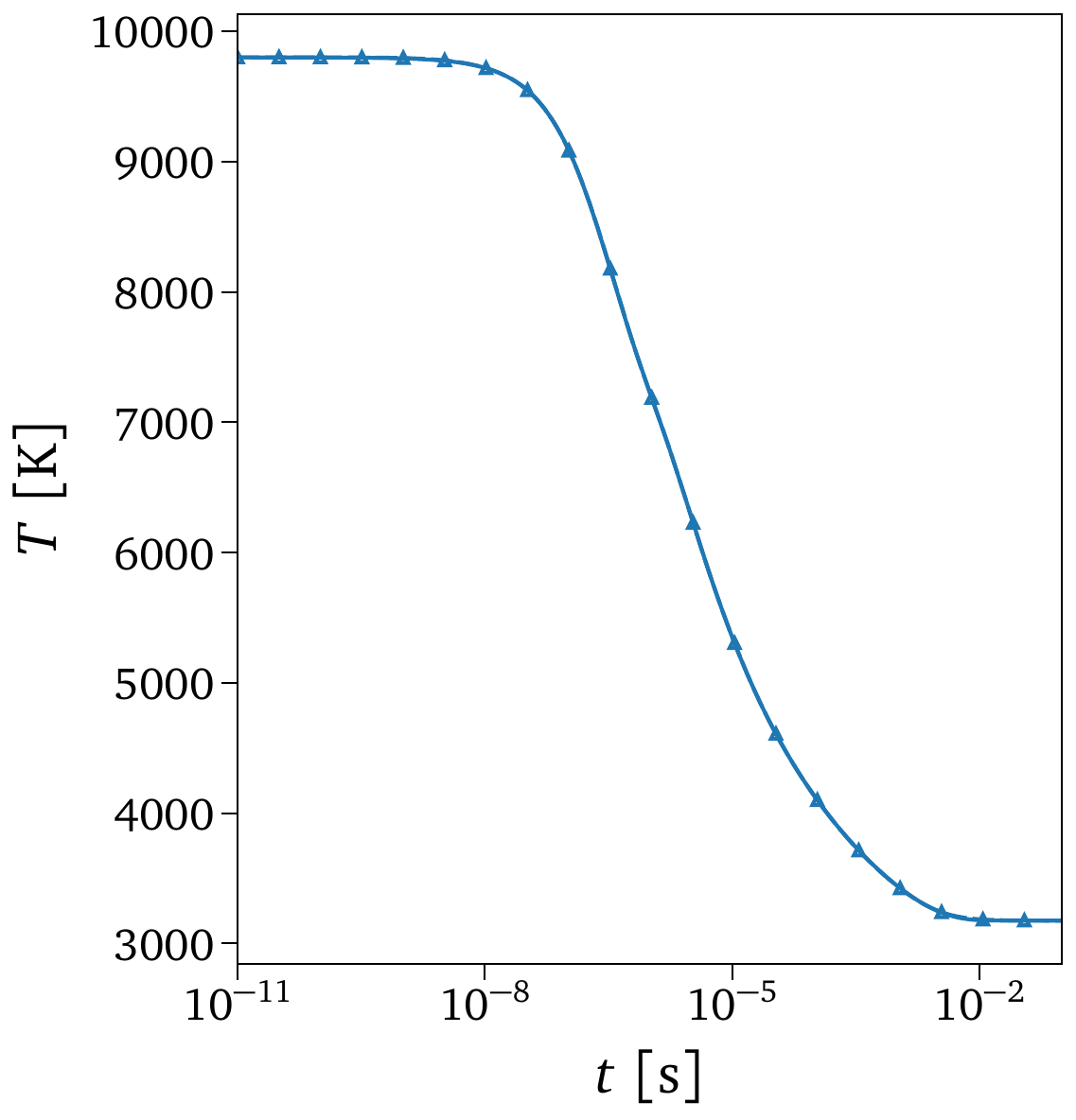}
    \end{subfigure}
    \begin{subfigure}[htb!]{0.32\textwidth}
        \centering
        \includegraphics[width=\textwidth]{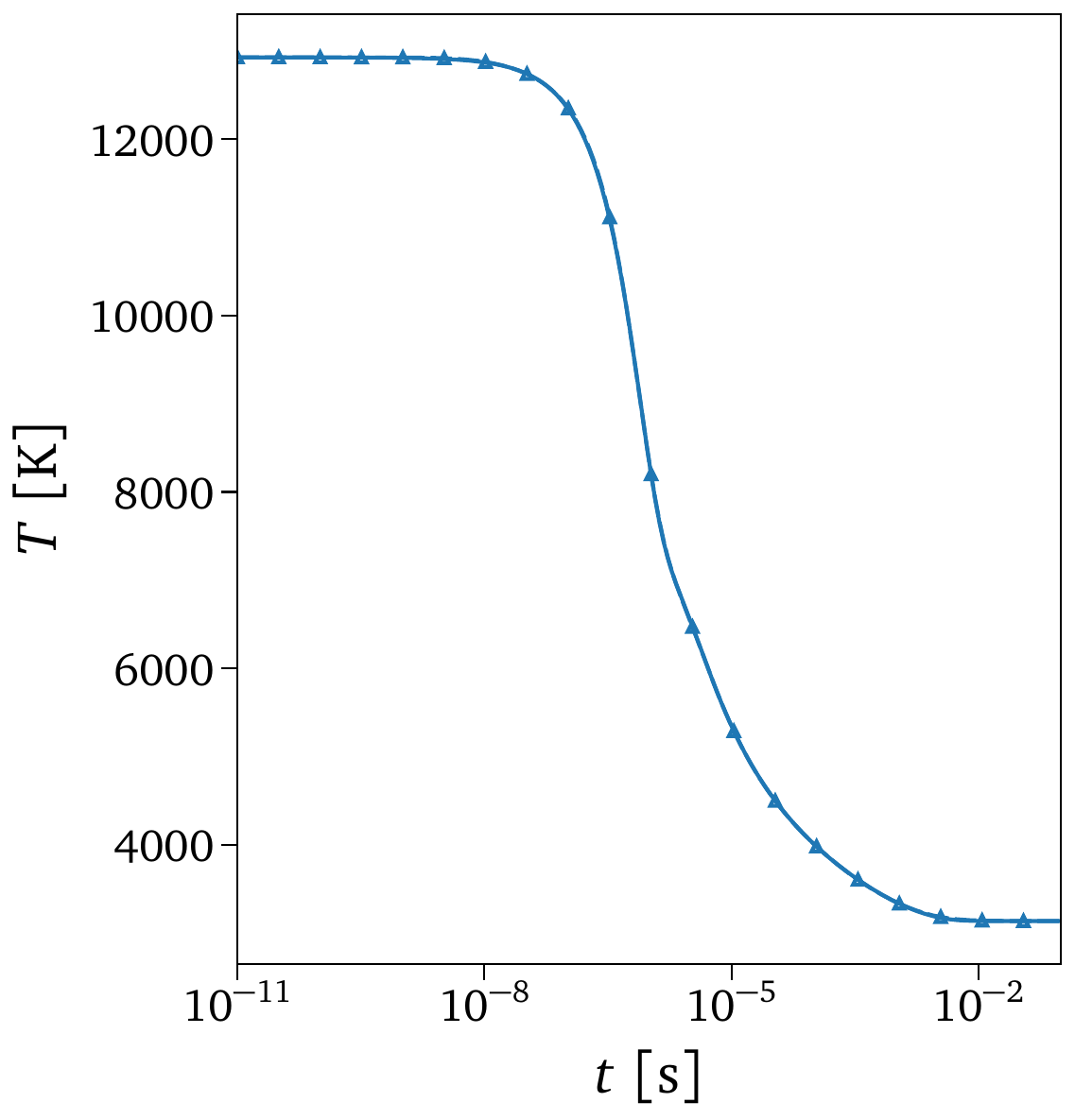}
    \end{subfigure}
    \caption{\textit{Comparison of inferred solutions for the nonequilibrium oxygen mixture}. The reference solutions (solid lines) are compared against predictions from the trained MENO (dashed lines with markers) for four different test cases. Results are shown for temperature $T$.}
    \label{fig:noneq.temp}
\end{figure}

\begin{figure}[!htb]
    \centering
    \begin{subfigure}[htb!]{0.32\textwidth}
        \centering
        \includegraphics[width=\textwidth]{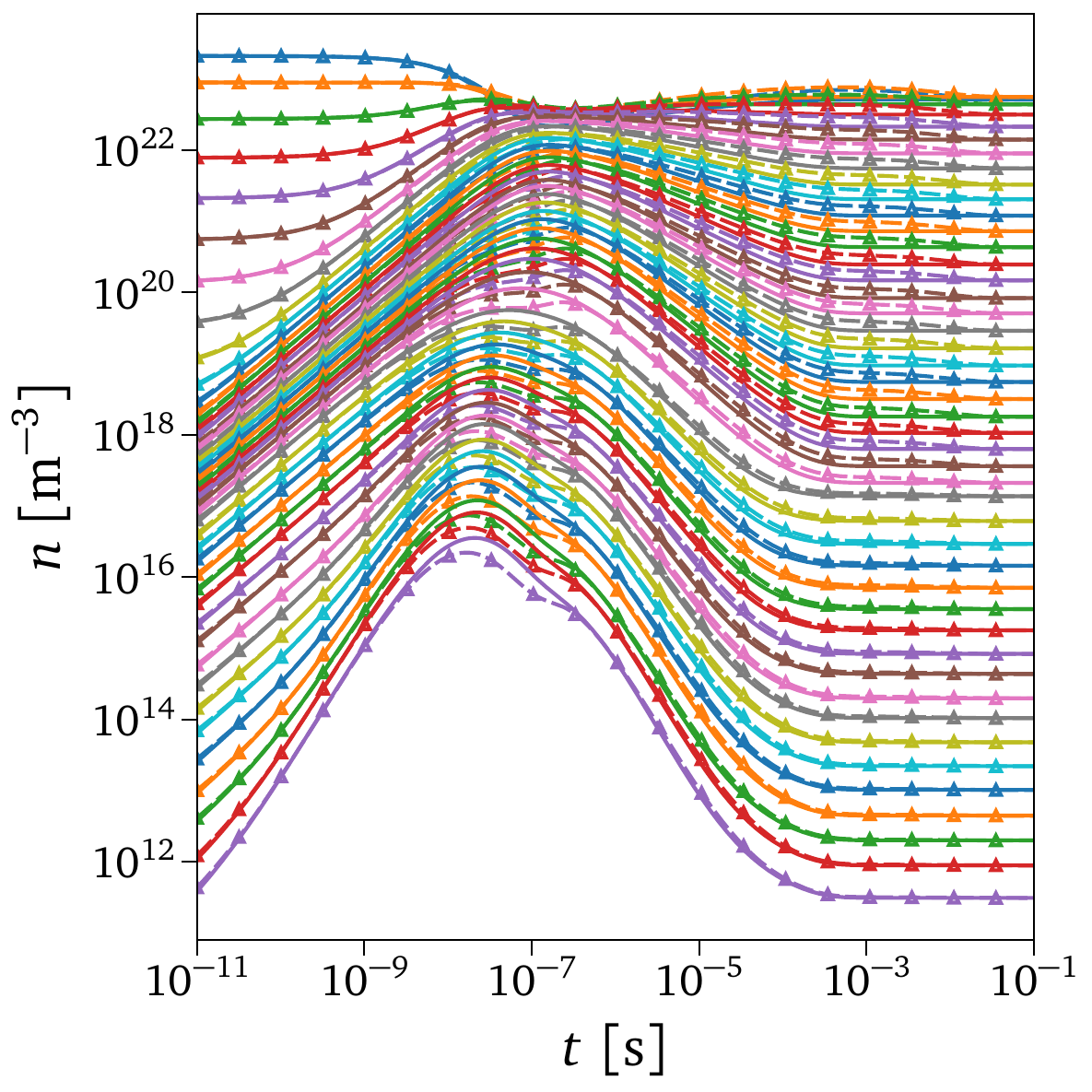}
    \end{subfigure}
    \begin{subfigure}[htb!]{0.32\textwidth}
        \centering
        \includegraphics[width=\textwidth]{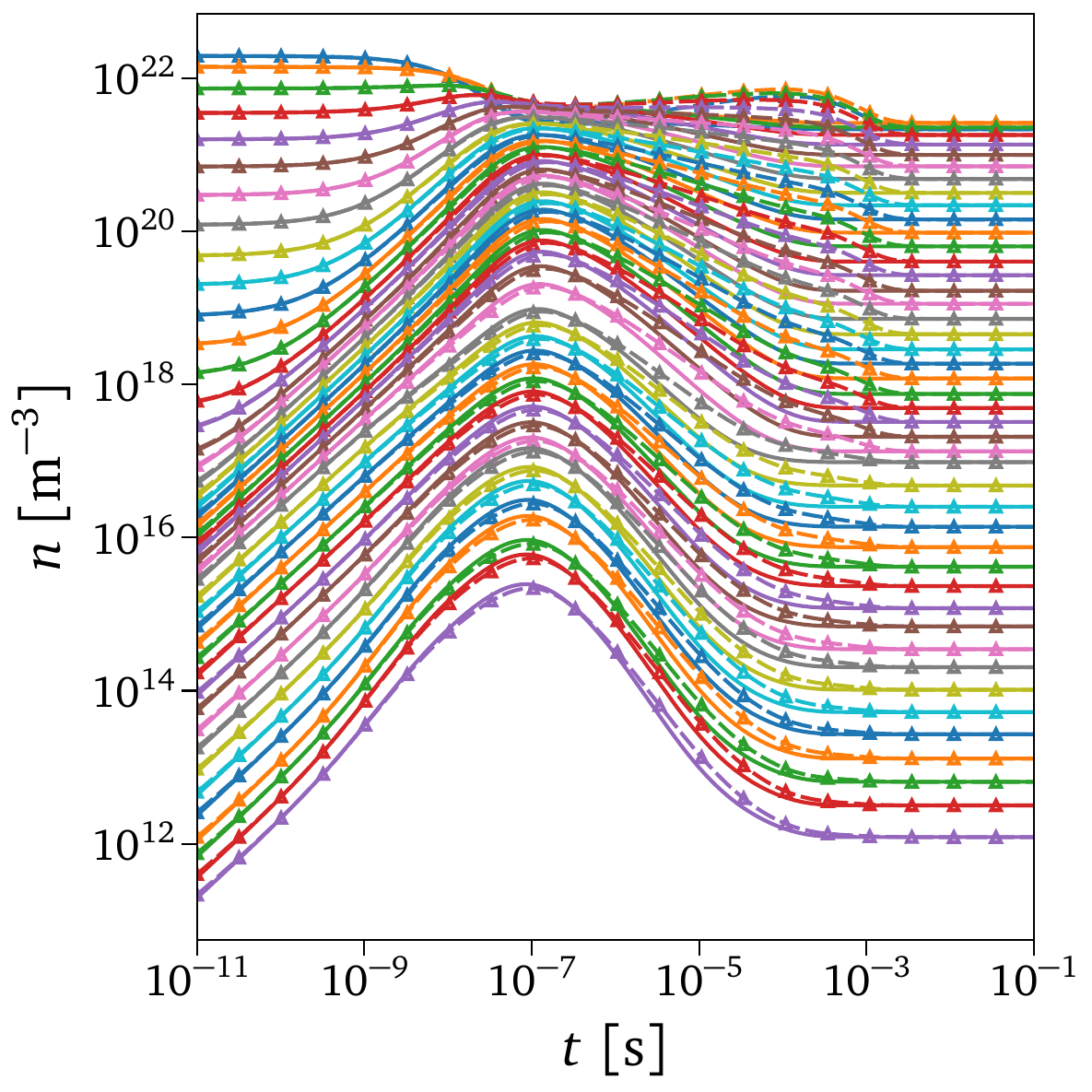}
    \end{subfigure}
    \\
    \begin{subfigure}[htb!]{0.32\textwidth}
        \centering
        \includegraphics[width=\textwidth]{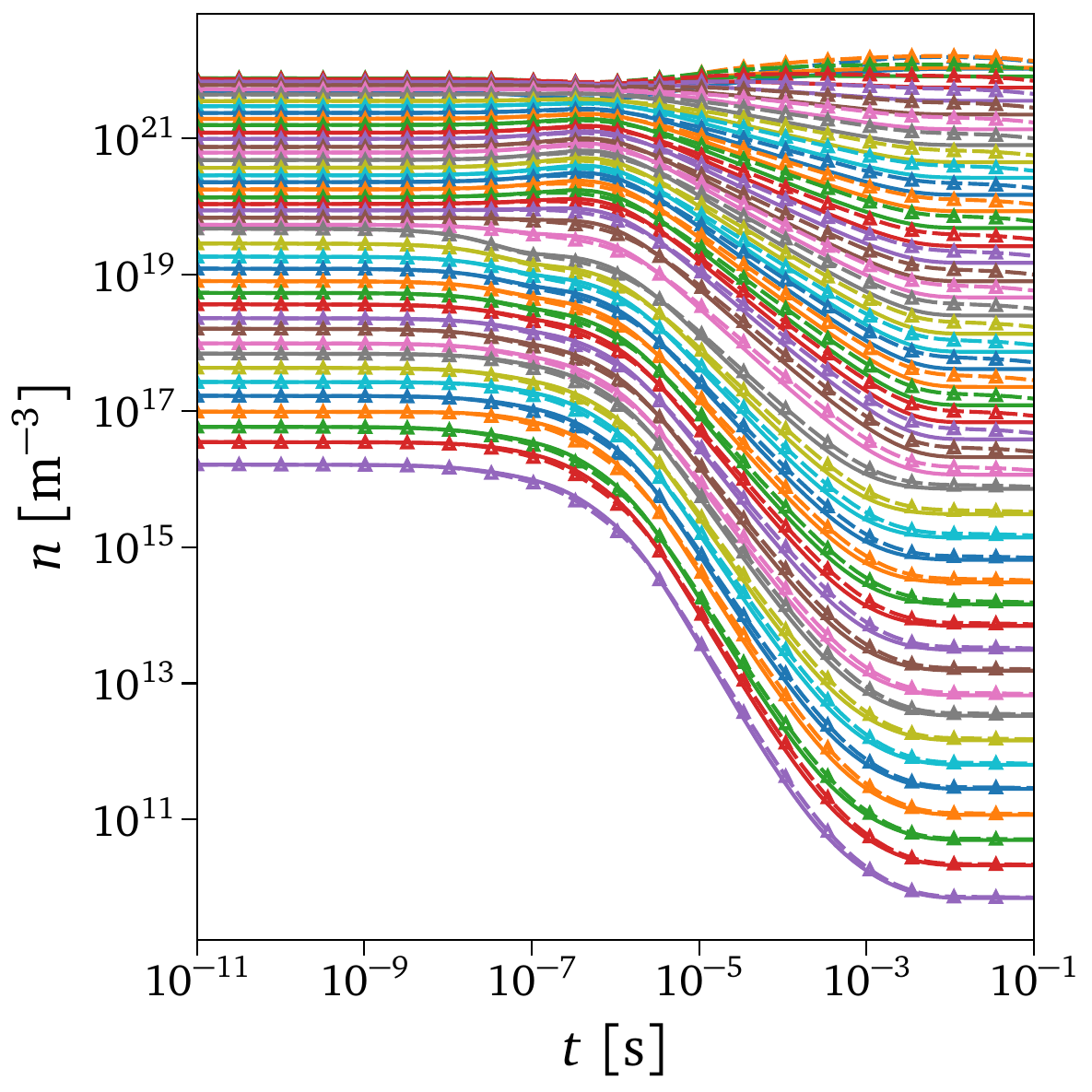}
    \end{subfigure}
    \begin{subfigure}[htb!]{0.32\textwidth}
        \centering
        \includegraphics[width=\textwidth]{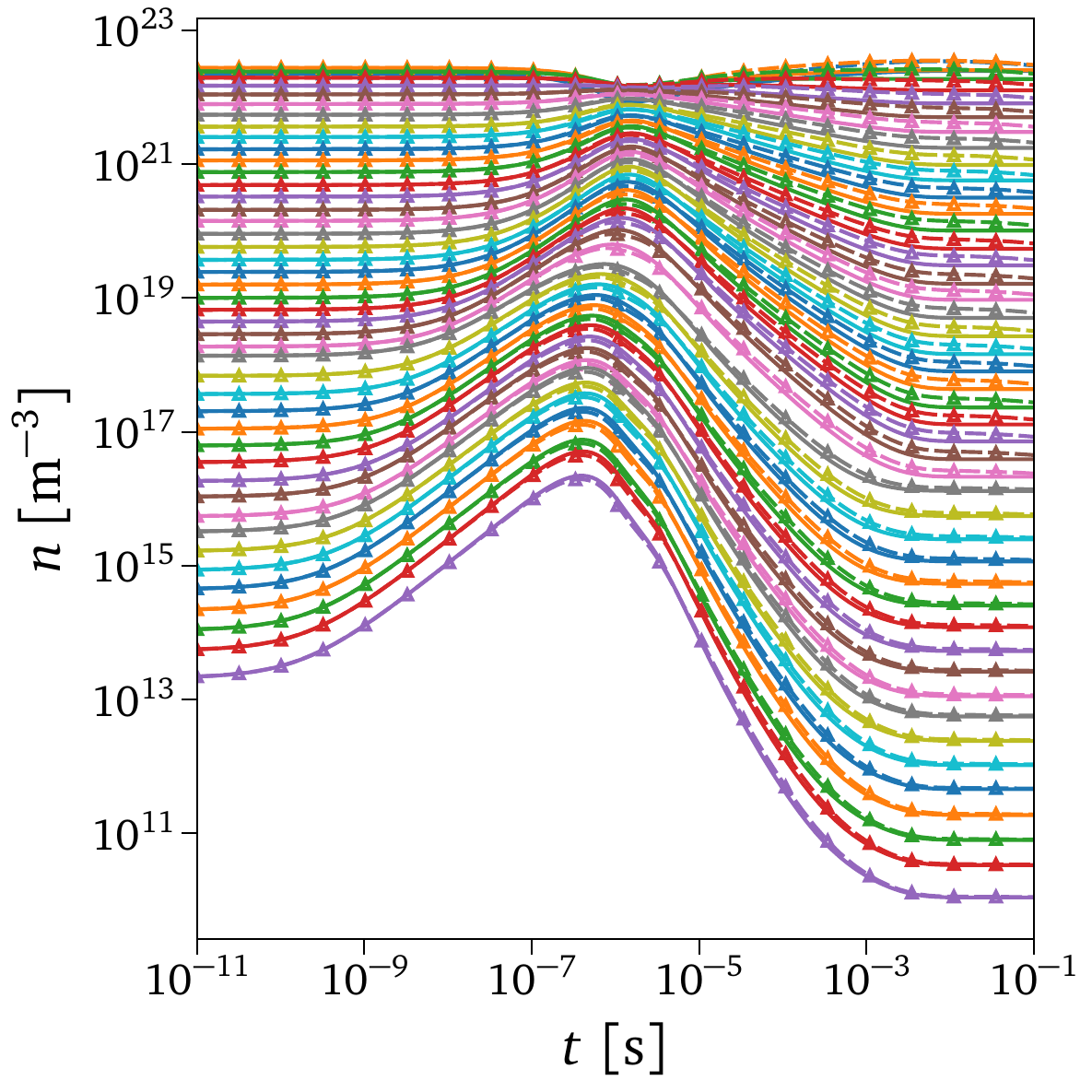}
    \end{subfigure}
    \caption{\textit{Comparison of MENO (no corrections) and true solutions for the nonequilibrium oxygen problem.} The reference solutions (solid lines) are compared against predictions from the uncorrected MENO model (dashed lines with markers) for four different test cases. Results are shown for O$_2$ group populations.}
    \label{fig:noneq.nd_nocorr}
\end{figure}

\paragraph{Accuracy}
Figure \ref{fig:noneq.temp} shows the time evolution of the temperature for the same four representative test cases presented in figure \ref{fig:noneq.nd}.
Similarly, figure \ref{fig:noneq.nd_nocorr} illustrates the temporal evolution of the coarse-grained O$_2$ group populations for those cases.
In this latter figure, predictions from the uncorrected MENO model are compared with the reference solutions to assess baseline accuracy prior to incorporating graph-based correction mechanisms.

\begin{figure}[!htb]
    \centering
    \includegraphics[width=0.35\textwidth]{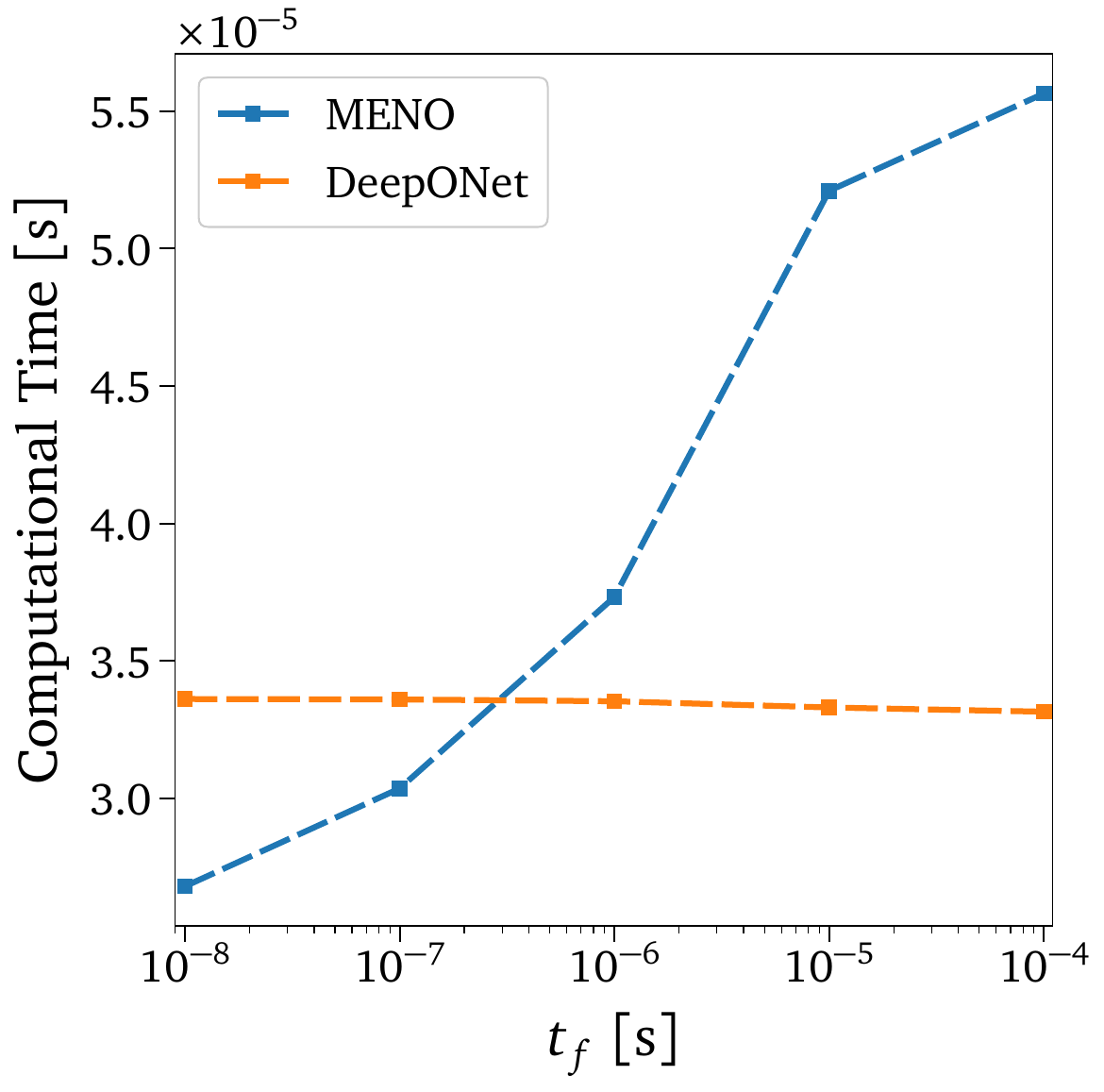}
    \caption{\textit{Surrogate inference time.} Comparison of the mean per-sample inference time between the trained MENO framework and the vanilla DeepONet, evaluated on the same GPU device.}
    \label{fig:noneq.don_vs_meno.time}
\end{figure}

\paragraph{Speedup}
Figure~\ref{fig:noneq.don_vs_meno.time} compares the average inference time per sample for the trained MENO framework and the vanilla DeepONet, both evaluated on the same GPU.
Tables~\ref{table:noneq.speedup.cpu} and~\ref{table:noneq.speedup.gpu} report the computational speedup of MENO relative to the BDF integrator, for both CPU and GPU implementations. Speedup values are presented for a range of BDF orders, relative tolerances, and final integration times.

\clearpage

\clearpage

\begin{table}[!htb]
    \centering
    \begin{tabular}{c|c||c|c|c|c|c}
        \arrayrulecolor{black}\midrule
        \multicolumn{7}{c}{Speedup - CPU} \\
        \arrayrulecolor{black}\midrule
        \multicolumn{2}{c||}{BDF Scheme} & \multicolumn{5}{c}{Final Time [s]} \\
        \arrayrulecolor{gray}\midrule
        Order & RTol & $10^{-8}$ & $10^{-7}$ & $10^{-6}$ & $10^{-5}$ & $10^{-4}$ \\
        \arrayrulecolor{black}\midrule \midrule
        \multirow{3}*{2} & $10^{-6}$ & $4.38\times 10^{1}$ & $7.01\times 10^{1}$ & $9.98\times 10^{1}$ & $1.23\times 10^{2}$ & $1.67\times 10^{2}$ \\
        & $10^{-7}$ & $6.55\times 10^{1}$ & $1.21\times 10^{2}$ & $1.85\times 10^{2}$ & $2.37\times 10^{2}$ & $3.04\times 10^{2}$ \\
        & $10^{-8}$ & $1.17\times 10^{2}$ & $2.35\times 10^{2}$ & $3.46\times 10^{2}$ & $3.32\times 10^{2}$ & $3.17\times 10^{2}$ \\
        \arrayrulecolor{gray}\midrule
        \multirow{3}*{3} & $10^{-6}$ & $3.58\times 10^{1}$ & $5.00\times 10^{1}$ & $6.35\times 10^{1}$ & $6.96\times 10^{1}$ & $9.19\times 10^{1}$ \\
        & $10^{-7}$ & $4.02\times 10^{1}$ & $6.01\times 10^{1}$ & $8.18\times 10^{1}$ & $9.60\times 10^{1}$ & $1.33\times 10^{2}$ \\
        & $10^{-8}$ & $4.87\times 10^{1}$ & $8.15\times 10^{1}$ & $1.20\times 10^{2}$ & $1.48\times 10^{2}$ & $2.12\times 10^{2}$ \\
        \arrayrulecolor{gray}\midrule
        \multirow{3}*{4} & $10^{-6}$ & $3.29\times 10^{1}$ & $4.37\times 10^{1}$ & $5.54\times 10^{1}$ & $7.34\times 10^{1}$ & $1.21\times 10^{2}$ \\
        & $10^{-7}$ & $3.73\times 10^{1}$ & $5.06\times 10^{1}$ & $6.53\times 10^{1}$ & $9.83\times 10^{1}$ & $1.84\times 10^{2}$ \\
        & $10^{-8}$ & $4.15\times 10^{1}$ & $5.94\times 10^{1}$ & $8.23\times 10^{1}$ & $1.40\times 10^{2}$ & $2.54\times 10^{2}$ \\
        \arrayrulecolor{black}\midrule
    \end{tabular}
    \caption{\textit{CPU speedup of MENO relative to the BDF integrator for the nonequilibrium oxygen mixture.} Speedup values are reported for varying BDF orders and relative tolerances, across multiple final integration times. All simulations were performed using a single CPU core.}
    \label{table:noneq.speedup.cpu}
\end{table}

\begin{table}[!htb]
    \centering
    \begin{tabular}{c|c||c|c|c|c|c}
        \arrayrulecolor{black}\midrule
        \multicolumn{7}{c}{Speedup - GPU} \\
        \arrayrulecolor{black}\midrule
        \multicolumn{2}{c||}{BDF Scheme} & \multicolumn{5}{c}{Final Time [s]} \\
        \arrayrulecolor{gray}\midrule
        Order & RTol & $10^{-8}$ & $10^{-7}$ & $10^{-6}$ & $10^{-5}$ & $10^{-4}$ \\
        \arrayrulecolor{black}\midrule \midrule
        \multirow{3}*{2} & $10^{-6}$ & $1.22\times 10^{3}$ & $1.90\times 10^{3}$ & $2.60\times 10^{3}$ & $3.05\times 10^{3}$ & $4.14\times 10^{3}$ \\
        & $10^{-7}$ & $1.83\times 10^{3}$ & $3.27\times 10^{3}$ & $4.84\times 10^{3}$ & $5.88\times 10^{3}$ & $7.55\times 10^{3}$ \\
        & $10^{-8}$ & $3.26\times 10^{3}$ & $6.36\times 10^{3}$ & $9.01\times 10^{3}$ & $8.25\times 10^{3}$ & $7.86\times 10^{3}$ \\
        \arrayrulecolor{gray}\midrule
        \multirow{3}*{3} & $10^{-6}$ & $1.00\times 10^{3}$ & $1.35\times 10^{3}$ & $1.65\times 10^{3}$ & $1.73\times 10^{3}$ & $2.28\times 10^{3}$ \\
        & $10^{-7}$ & $1.12\times 10^{3}$ & $1.63\times 10^{3}$ & $2.13\times 10^{3}$ & $2.39\times 10^{3}$ & $3.29\times 10^{3}$ \\
        & $10^{-8}$ & $1.36\times 10^{3}$ & $2.21\times 10^{3}$ & $3.12\times 10^{3}$ & $3.69\times 10^{3}$ & $5.26\times 10^{3}$ \\
        \arrayrulecolor{gray}\midrule
        \multirow{3}*{4} & $10^{-6}$ & $9.19\times 10^{2}$ & $1.18\times 10^{3}$ & $1.44\times 10^{3}$ & $1.83\times 10^{3}$ & $3.00\times 10^{3}$ \\
        & $10^{-7}$ & $1.04\times 10^{3}$ & $1.37\times 10^{3}$ & $1.70\times 10^{3}$ & $2.44\times 10^{3}$ & $4.55\times 10^{3}$ \\
        & $10^{-8}$ & $1.16\times 10^{3}$ & $1.61\times 10^{3}$ & $2.15\times 10^{3}$ & $3.48\times 10^{3}$ & $6.31\times 10^{3}$ \\
        \arrayrulecolor{black}\midrule
    \end{tabular}
    \caption{\textit{GPU speedup of MENO relative to the BDF integrator for the nonequilibrium oxygen mixture.} Speedup values are reported for varying BDF orders and relative tolerances, across multiple final integration times. All MENO simulations were executed on a single GPU, while BDF runs were performed on a single CPU core.}
    \label{table:noneq.speedup.gpu}
\end{table}

\clearpage

\subsection{Example 3: Collisional-radiative argon plasma}\label{suppl:num_exp:plasma_cr}

\paragraph{Physical model}
The collisional-radiative model accounts for five types of fundamental processes~\cite{Vlcek1989AData,Bultel2002InfluenceModel2,Kapper2011IonizingStructure,Kapper2011IonizingEffects}:
\begin{enumerate}
  \item Excitation and ionization by electron impact:
    \begin{align}\label{eq:kin.e}
        \ce{
            Ar$(i)$ + e$^-$ &<=> Ar$(j)$ + e$^-$
        }\eqspace, \\
        \ce{
            Ar$(i)$ + e$^-$ &<=> Ar^+ + e$^-$ + e$^-$
        }\eqspace,
    \end{align}
  \item Excitation and ionization by heavy-particle impact:
    \begin{align}\label{eq:kin.h}
        \ce{
            Ar$(i)$ + Ar$(1)$ &<=> Ar$(j)$ + Ar$(1)$
        }\eqspace, \\
        \ce{
            Ar$(i)$ + Ar$(1)$ &<=> Ar^+ + e$^-$ + Ar$(1)$
        }\eqspace,
    \end{align}
  \item Spontaneous emission/absorption (bound-bound):
    \begin{equation}\label{eq:rad.bb}
        \ce{
            Ar$(i)$ + h $\nu_{ij}$ <=> Ar$(j)$
        }\eqspace,
    \end{equation}
  \item Photo-ionization/radiative recombination (free-bound and bound-free):
    \begin{equation}\label{eq:rad.bf}
        \ce{
            Ar$(i)$ + h $\nu_{i}$ <=> Ar^+ + e$^-$
        }\eqspace,
    \end{equation}
  \item Bremsstrahlung radiation (free-free):
    \begin{equation}\label{eq:rad.ff}
        \ce{
            e$^-$ + Ar^+ <=> e$^-$ + Ar^+ + h $\nu$
        }\eqspace.
    \end{equation}  
\end{enumerate}
Here, indices $i \in \{1,\dots,31\}$ and $j > i$ refer to the electronic levels of Ar; the two levels of Ar$^+$ are omitted for brevity.
Superelastic collisions are neglected, as excited states remain sparsely populated under the conditions considered, rendering their contribution to heavy-particle collisions negligible.
Accordingly, only the ground state, Ar$(1)$, is included in these interactions.
Radiative absorption is modeled using escape factors, $\Lambda$, which range from 0 to 1, corresponding to optically thick and optically thin plasma, respectively.
A value of $\Lambda = 0$ indicates that all emitted photons are reabsorbed locally, while $\Lambda = 1$ implies complete photon escape without reabsorption.
In this work, the plasma is assumed to be optically thin for all radiative transitions, with the exception of high-frequency bound–bound emissions to the ground state Ar(1), which are treated as fully reabsorbed by the gas~\cite{Kapper2011IonizingStructure,Kapper2011IonizingEffects}.
Planck’s constant and the radiation frequency are denoted by $h$ and $\nu$, respectively.

The plasma is described using a single-fluid formulation, in which all species share a common velocity field, and thermal nonequilibrium is captured by distinguishing between electron and heavy-particle temperatures, $T_\mathrm{e}$ and $T_\mathrm{h}$, respectively.
Charge neutrality is maintained by the short Debye length and the absence of applied electric fields.
The assumption of strong collisional coupling justifies a single-fluid velocity. Viscous and diffusive transport effects are neglected.
As a result, the dynamics are governed by the two-temperature Euler equations~\cite{Kapper2011IonizingEffects}, expressed in conservation form as:
\begin{equation}\label{eq:euler}
    \frac{\partial \mU}{\partial t}+\frac{\partial \mF_\alpha}{\partial x_\alpha} = \mS
\end{equation}
for $\alpha \in \{1,2,3\}$, with $\mathbf{x}=[x_1,x_2,x_3]=[x,y,z]$ being the Cartesian coordinates and $t$ denoting time. Summation is implied over repeated indices. The conservative variables $\mU$, inviscid fluxes $\mF_\alpha$, and source terms $\mathbf{S}$ are defined as: 
\begin{equation}\label{eq:euler.terms}
    \mU = \begin{bmatrix}
    \rho_s^i \\
    \phantom{;} \rho u_\beta \phantom{;} \\
    \rho E \\
    \rho e_\mathrm{e}
    \end{bmatrix}\eqspace,\quad
    \mF_\alpha = \begin{bmatrix}
    \rho_s^i u_\alpha \\
    \phantom{;} \rho u_\alpha u_\beta + p\delta_{\alpha\beta} \phantom{;} \\
    \rho H u_\alpha \\
    \rho e_\mathrm{e}u_\alpha
    \end{bmatrix}\eqspace,\quad
    \mS = \begin{bmatrix}
    \omega_s^i \\
    0 \\
    \Omega_{\mathrm{bb}}^{\mathrm{R}}+\Omega_{\mathrm{ff}}^{\mathrm{R}} \\
    \phantom{;}\Omega_{\mathrm{el}}^{\mathrm{C}}+\Omega_{\mathrm{in}}^{\mathrm{C}}+\Omega_{\mathrm{ion}}^{\mathrm{C}}+\Omega_{\mathrm{bf}}^{\mathrm{R}}+\Omega_{\mathrm{ff}}^{\mathrm{R}}-p_{\mathrm{e}} \partial u_\alpha / \partial x_\alpha\phantom{;}
    \end{bmatrix}\eqspace,
\end{equation}
with $\beta \in \{1,2,3\}$, $s \in \mathcal{S}$, and $i \in \mathcal{I}_s$. Here, $\mathcal{S} = \{\mathrm{e}^-, \mathrm{Ar}, \mathrm{Ar}^+\}$ denotes the set of chemical species, and $\mathcal{I}_s$ the set of electronic states for species $s$.
The symbols $\rho_s^i$ and $p_s^i$ represent the partial density and pressure of state $i$ of species $s$, while $\rho$ and $p$ denote the total plasma density and pressure.
$u_\alpha$ corresponds to the $\alpha$-component of the mass-averaged velocity with $\delta_{\alpha \beta}$ being the Kronecker delta, whereas $\rho E$, $\rho H$, and $\rho e_\mathrm{e}$ denote the total energy, total enthalpy, and electron internal energy densities, respectively.
Mass source terms are denoted by $\omega_s^i$, while energy source terms include radiative contributions from bound-bound ($\Omega_{\mathrm{bb}}^{\mathrm{R}}$), bound-free ($\Omega_{\mathrm{bf}}^{\mathrm{R}}$), and free-free ($\Omega_{\mathrm{ff}}^{\mathrm{R}}$) transitions, as well as collisional energy transfer from elastic ($\Omega_{\mathrm{el}}^{\mathrm{C}}$), inelastic ($\Omega_{\mathrm{in}}^{\mathrm{C}}$), and ionizing ($\Omega_{\mathrm{ion}}^{\mathrm{C}}$) collisions.
The bound-free radiative term is omitted from the total energy conservation equation, as its contribution is negligible compared to the free-free and especially the bound-bound terms~\cite{Kapper2011IonizingStructure}.
The non-conservative term $-p_e \partial u_\alpha / \partial x_\alpha$ is related to the work of the self-consistent plasma electric field, which acts against charge separation, whose expression can be derived from the free-electron momentum equation when neglecting inertial and diffusive terms~\cite{Zeldovich1967PhysicsPhenomena}.

\paragraph{Computational method}
The cell-centered finite volume discretization~\cite{Hirsch_Book_2007} is employed, with inviscid fluxes computed using the Van Leer flux-splitting scheme~\cite{vanLeer1982Flux-vectorEquations2}.
To achieve second-order spatial accuracy, left and right states at cell interfaces are reconstructed via the MUSCL approach~\cite{VanLeer_JComP_1979}, applied to primitive variables (partial densities, velocity components, and temperatures).
To mitigate numerical instabilities associated with strong shock fronts, a reconstruction blending technique is implemented. In this approach, the reconstructed state at each cell interface, $\mathbf{U}_{f}$, is defined as a weighted combination of the high-order (MUSCL) reconstruction, $\mathbf{U}_{f}^{\mathrm{H}}$, and the low-order (first-order upwind) reconstruction, $\mathbf{U}_{f}^{\mathrm{L}}$:
\begin{equation}
\mathbf{U}_{f} = (1-\eta)\,\mathbf{U}_{f}^{\mathrm{H}}
    + \eta\,\mathbf{U}_{f}^{\mathrm{L}} \eqspace,
\end{equation}
where $\eta$ is a shock-sensing function that transitions smoothly between 1 (upstream and within the shock) and 0 (downstream of the shock) via a Gaussian smoothing profile. Shock locations are identified using the detection algorithm described in~\cite{Quirk1994ADebate}.

Time integration is performed using the operator-splitting technique proposed by Strang~\cite{Strang_SIAM_1968}.
This method integrates the transport operator, $\boldsymbol{\mathcal{T}}\left(\mathbf{U}\right) = \partial \mF_\alpha / \partial x_\alpha$, and the reaction operator, $\boldsymbol{\mathcal{R}}\left(\mathbf{U}\right) = \mathbf{S}$, sequentially in a symmetric fashion, as defined by equations~\eqref{eq:split.01}–\eqref{eq:split.04}.
The splitting formulation is second-order accurate, strongly stable, and symplectic for nonlinear equations. Its convergence and stability properties have been extensively studied for reacting flow simulations~\cite{Knio_JComP_1999,Singer_CTM_2006,Ren_JComP_2014,Wu_CPC_2019}.
The transport operator is advanced in time using a four-stage, third-order strong-stability-preserving Runge-Kutta (SSP-RK3) scheme~\cite{Durran2010NumericalDynamics}, which allows a Courant-Friedrichs-Lewy (CFL) number of up to 2.
For the reaction operator, integration is performed either through a second-order backward differentiation formula (BDF) solver provided by the LSODE library~\cite{Radhakrishnan_LSODE_1993}, or via the MENO surrogate model, which offers an efficient, data-driven alternative.

\begin{figure}[htb!]
    \centering
    \includegraphics[width=0.43\textwidth]{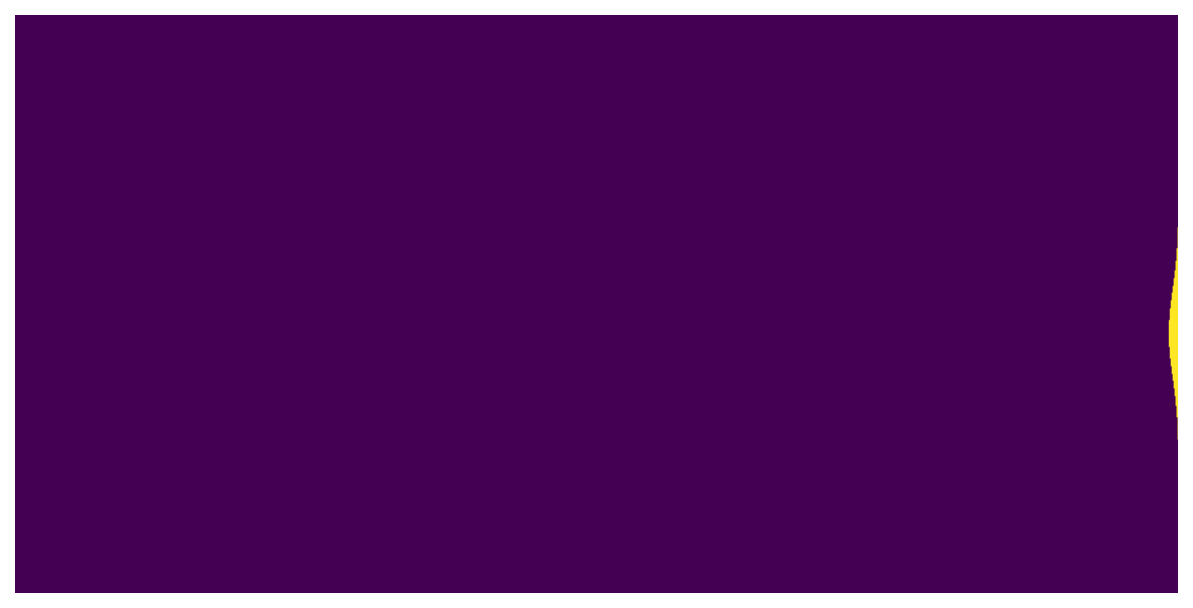}
    \caption{\textit{Initial pressure field for 2D simulations.} A localized Gaussian pressure perturbation is introduced near the right wall.}
    \label{fig:init_pert}
\end{figure}

For the 2D simulations, a weak Gaussian pressure perturbation is introduced near the right boundary to promote the development of transverse instabilities and facilitate the formation of multidimensional flow structures. The perturbation has an amplitude three times the freestream pressure and density, a spatial extent of $3 \times 10^{-3}$~m in the $x$-direction, and is centered at 40\% of the shock-tube height in the $y$-direction (see figure~\ref{fig:init_pert}). This perturbation accelerates the development of instabilities while remaining weak enough to avoid overwhelming the flow physics.

\begin{figure}[htb!]
    \centering
    \includegraphics[width=0.35\textwidth]{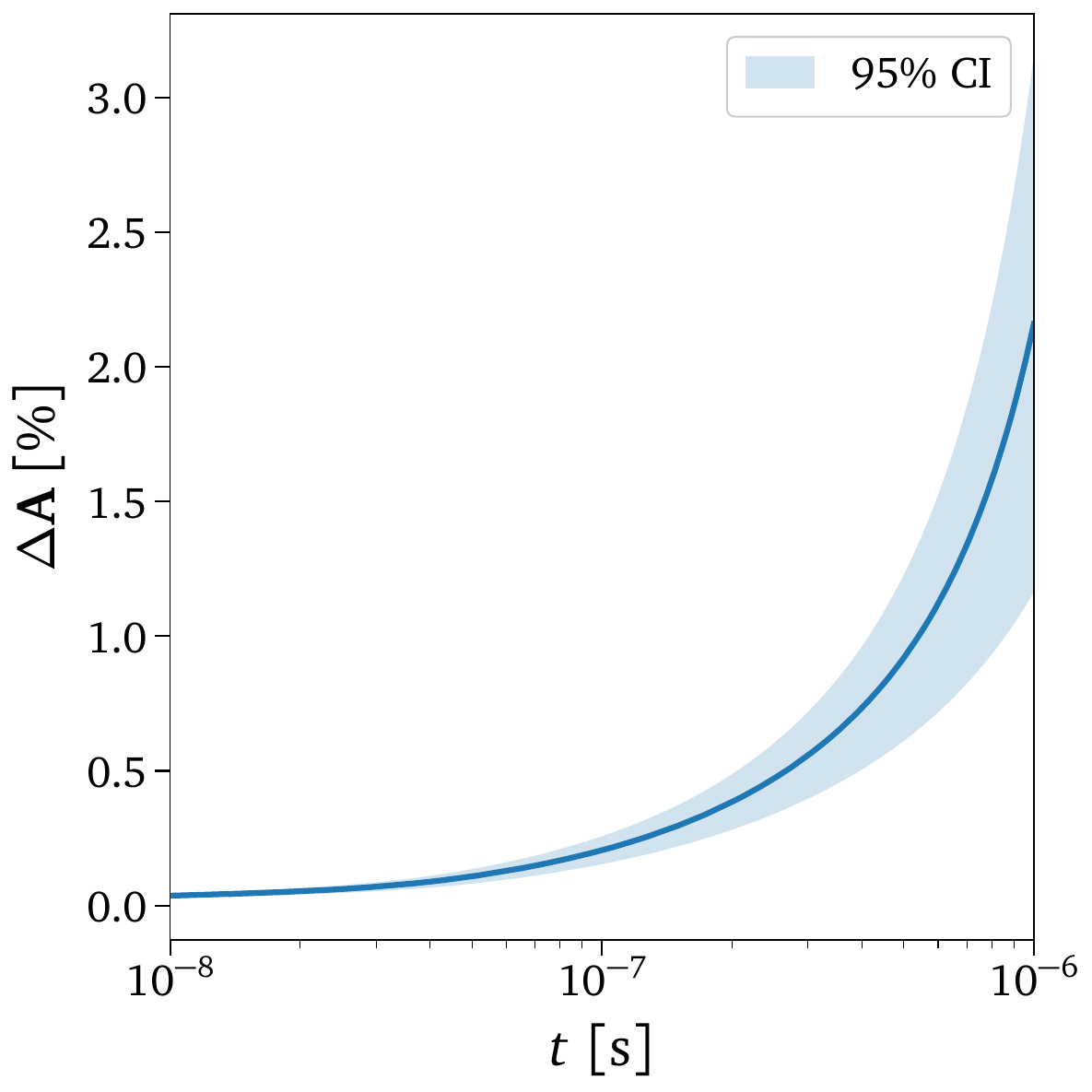}
    \caption{\textit{Variability of the operator \(\mA(t)\) in the argon plasma problem}. The mean variability of operator $\mA$ shown with a 95\% confidence interval, calculated using equation \eqref{eq:delta_a} across 100 test trajectories.}
    \label{fig:plasma_cr.0d.delta_a}
\end{figure}

\paragraph{Validity}
As with the previous test cases, figure~\ref{fig:plasma_cr.0d.delta_a} presents the variability metric \(\Delta \mA(t)\), defined in equation~\eqref{eq:delta_a}, evaluated across 100 test trajectories. The consistently low values of \(\Delta \mA(t)\), along with narrow 95\% confidence intervals, confirm that the operator \(\mA(t)\) remains nearly constant over the time span considered. This observation supports the validity of using the matrix exponential formulation for the argon plasma problem.

\paragraph{Data generation}
To generate a physically meaningful dataset for training and testing the 0D surrogate model, we adopt a systematic approach:
\begin{enumerate}
    \item \textit{Data collection.}
    We simulate a representative one-dimensional (1D) unsteady shock-tube problem (see section~\ref{sec:num_exp:plasma_cr:1d}) and collect the primitive thermochemical variables at each space-time point. The state vector is defined as $\vq = [\rho, w_s^i, T_\mathrm{h}, T_\mathrm{e}] \in \mathbb{R}^{37}$, where $w_s^i$ represents the mass fraction of state $i$ of species $s \in \mathcal{S}$.
    \item \textit{Data compression.}
    The spatiotemporal data is organized into a matrix $X \in \mathbb{R}^{37 \times N}$, where $N$ denotes the number of sampled space-time points and each column represents a full thermochemical state.
    Given the large volume of collected data, we begin by reducing the dimensionality associated with the species composition. To do so, we extract the mass fraction submatrix $X_w$, apply a logarithmic transformation and standard normalization, and then perform a truncated Singular Value Decomposition (SVD), $\hat{X}_w \approx U_r \Sigma_r V_r^T$, where $\hat{X}_w$ contains only the transformed species mass fractions.
    The final compressed state vector is defined as $\mathbf{z} = [\rho, \mathbf{z}_w, T_\mathrm{h}, T_\mathrm{e}] \in \mathbb{R}^8$, where $\mathbf{z}_w \in \mathbb{R}^5$ denotes the low-dimensional latent representation of the species mass fractions.
    The total density $\rho$, heavy-particle temperature $T_\mathrm{h}$, and electron temperature $T_\mathrm{e}$ are preserved in their original form and are not subject to compression.
    \item \textit{Initial conditions sampling.}
    We divide the dataset into two physically distinct regions:
    \begin{itemize}
        \item \textit{Post-shock region}, located downstream of the first electron avalanche;
        \item \textit{Induction region}, the area between the shock front and the onset of the first electron avalanche.
    \end{itemize}
    For each region, we fit an 8-dimensional multivariate kernel density estimator (KDE) to the corresponding latent representation of the data.
    From each fitted distribution, we sample 3\,000 initial thermochemical states for training and validation, and 50 initial states for testing.
    A blended sampling strategy is employed: 33\% of the samples are drawn from the KDE, while the remaining 67\% are sampled using LHS within the empirical bounds of the dataset.
    Each sampled latent vector $\vz$ is projected back to the full thermochemical state to define the initial condition for 0D training trajectories.
    \item \textit{Time sampling.}
    The temporal evolution of the system is simulated over the interval $[10^{-8}, 10^{-6}]$ seconds, encompassing the characteristic timescales of plasma dynamics relevant to multidimensional simulations. Each trajectory is sampled at 192 logarithmically spaced time points for training and 48 for validation, following the procedure described in the \textit{Data generation} paragraph of section~\ref{suppl:num_exp:pollu}.
\end{enumerate}

\paragraph{Architecture}
To enhance training stability and improve predictive accuracy, we apply several preprocessing transformations to both the inputs and outputs of the model.

The time variable $t$ is log-scaled and normalized using the initial heavy-particle pressure $p_{\mathrm{h}_0}$, which is computed from the initial thermochemical state vector $\vq(0)$.
Specifically, we define the transformed time as $\hat{t}_{\ln}=\ln(t\,p_{\mathrm{h}_0})$ a formulation that helps align the temporal trajectories by mitigating time shifts caused by varying initial species concentrations.
Since collision frequency scales with density, this transformation also reduces translational symmetries in the dataset, thereby promoting more consistent temporal alignment and improved learning performance across samples.
The parametric input vector $\vmu$ is defined as the latent representation of the initial state, $\vmu=\vz=\psi(\vq(0))$, where $\psi:\mathbb{R}^{37} \rightarrow \mathbb{R}^8$ is the encoding function described in the data compression procedure. Before being passed to the model, $\vmu$ is also standardized to have zero mean and unit variance.
The species attribute vector $\vchi_s=[\epsilon_s,j_s,Z_s]\in\mathbb{R}^{3}$ encodes physical characteristics of each argon electronic state, where $\epsilon_s$ is the energy level, $j_s$ is the core angular momentum~\cite{Kapper2011IonizingStructure}, and $Z_s$ is the charge number (used to distinguish Ar from Ar$^+$). This vector is standardized to have zero mean and unit variance prior to being fed into the model.

On the output side, the nonlinear state variables $\vq_{nl}$ are passed through an exponential function to enforce positivity, thereby ensuring physically meaningful predictions.
The linear state variables $\vq_{l}$, representing the number densities of the argon states and predicted via the neural matrix exponential formulation (equation~\eqref{eq:lin_sys.sol.ml}), are subsequently converted to mass fractions, which constitute the final output required by the CFD solver.

A complete overview of the MENO model architecture used for the argon plasma problem, including all components and hyperparameters, is provided in table~\ref{table:plasma_cr.arch}.
\begin{table}[!htb]
	\centering
	\begin{tabular}{c|c|ccccc}
        \arrayrulecolor{black}\cmidrule{3-7}
        \multicolumn{2}{c|}{} & Architecture & Parameters & Blocks & Layers & Activations \\
        \arrayrulecolor{black}\midrule
        \multirow{3}*{$\vq_{nl}\in\mathbb{R}^4$} & \multirow{3}*{$n_\mathrm{e}$, $n_{\mathrm{Ar}(1)}$, $T_\mathrm{h}$, $T_\mathrm{e}$} & \multirow{3}*{\textit{flexDeepONet}} 
            & \multirow{3}*{69\,127} & Branch & [128, 128, 64]
            & \multirow{3}*{tanh$\times$2 + linear} \\
        & & & & Trunk & [128, 128, 64] & \\
        & & & & Pre-Net & [128, 128, 1] & \\
        \arrayrulecolor{gray}\midrule
        \multirow{2}*{$\vq_l\in\mathbb{R}^{33}$} & $f_\chi$ & \multirow{2}*{\textit{Feedforward}}
            & \multirow{2}*{3\,157} & \multirow{2}*{-} & [32, 32, 10]
            & \multirow{2}*{tanh$\times$2 + linear} \\
        & $f_\mu$ & & & & [32, 32, 8] & \\
        \arrayrulecolor{black}\midrule
        \multicolumn{2}{c|}{} & MENO & 279\,653 & \multicolumn{3}{|c}{} \\
        \arrayrulecolor{black}\cmidrule{3-4}
	\end{tabular}
	\caption{\textit{Neural network architecture of MENO components for the argon plasma problem.}
    The total number of parameters is reported as computed by the TensorFlow library~\cite{TF_2016}.}
	\label{table:plasma_cr.arch}
\end{table}

\paragraph{Training}
The training procedure follows the same approach described in section~\ref{suppl:num_exp:noneq}, employing two distinct loss functions: MAE for temperature variables, and  MAPE for number densities (or mass fractions).

\begin{figure}[ht]
    \centering
    \begin{subfigure}[htb!]{0.32\textwidth}
        \includegraphics[width=0.95\textwidth]{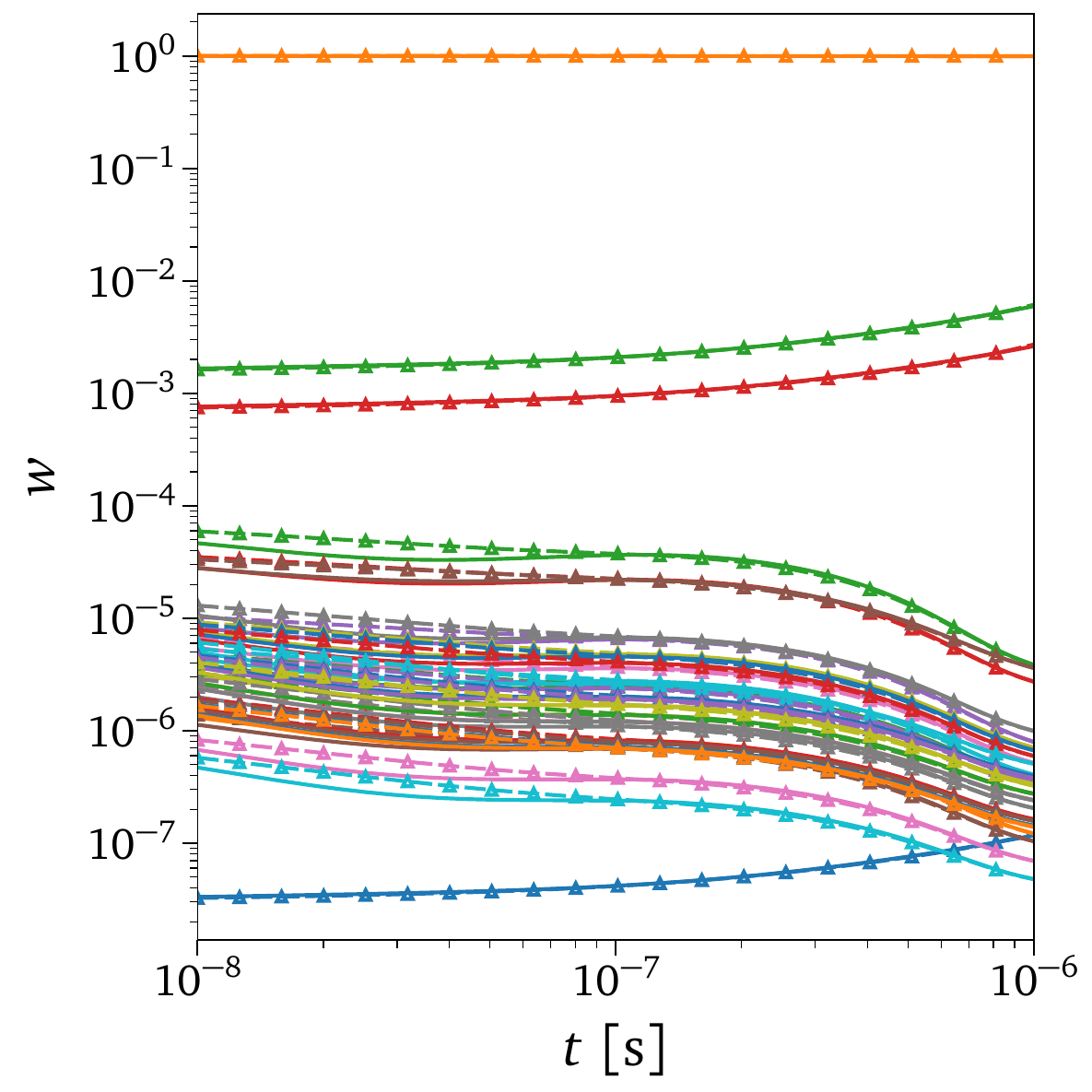}
    \end{subfigure}
    \begin{subfigure}[htb!]{0.32\textwidth}
        \includegraphics[width=0.95\textwidth]{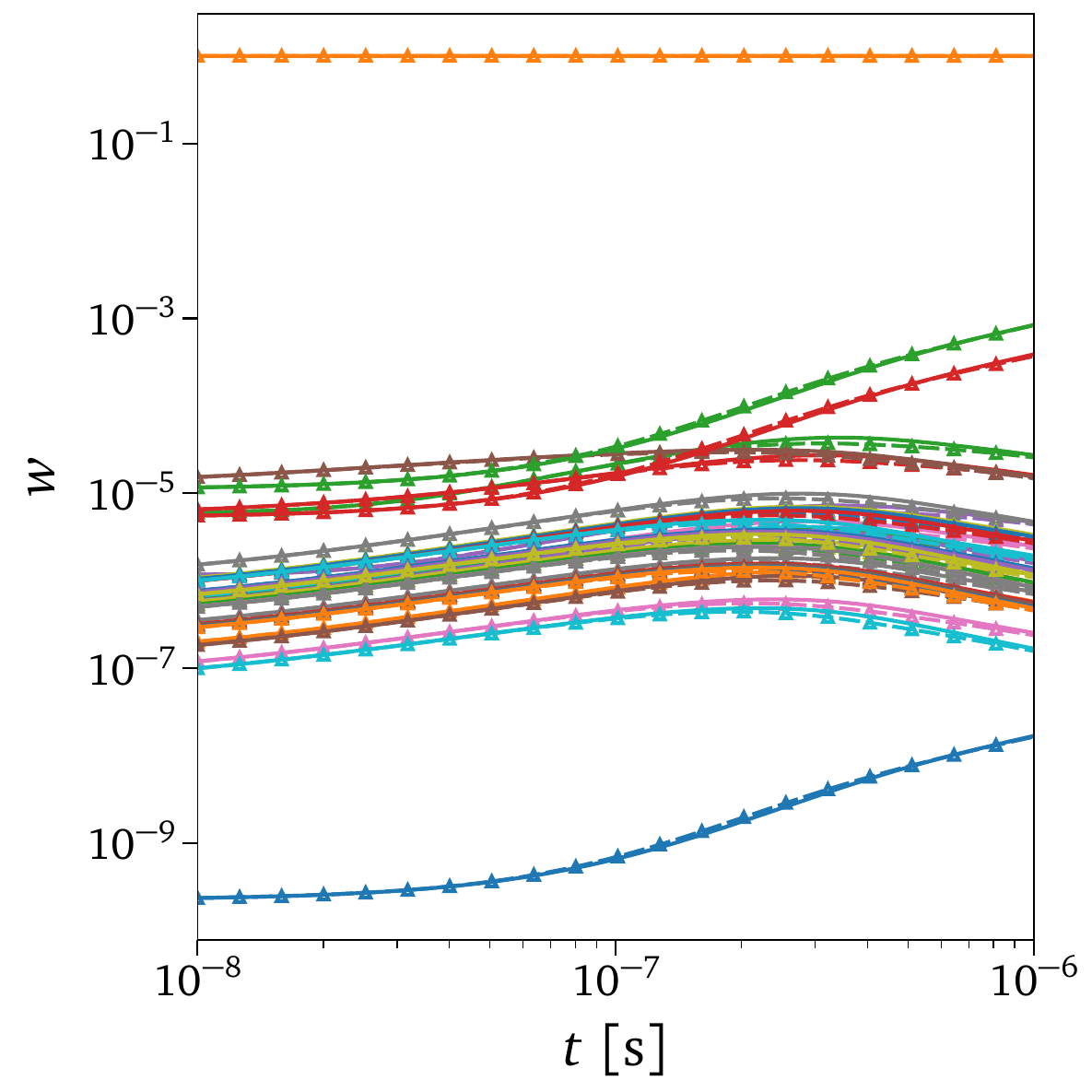}
    \end{subfigure}
    \begin{subfigure}[htb!]{0.32\textwidth}
        \includegraphics[width=0.95\textwidth]{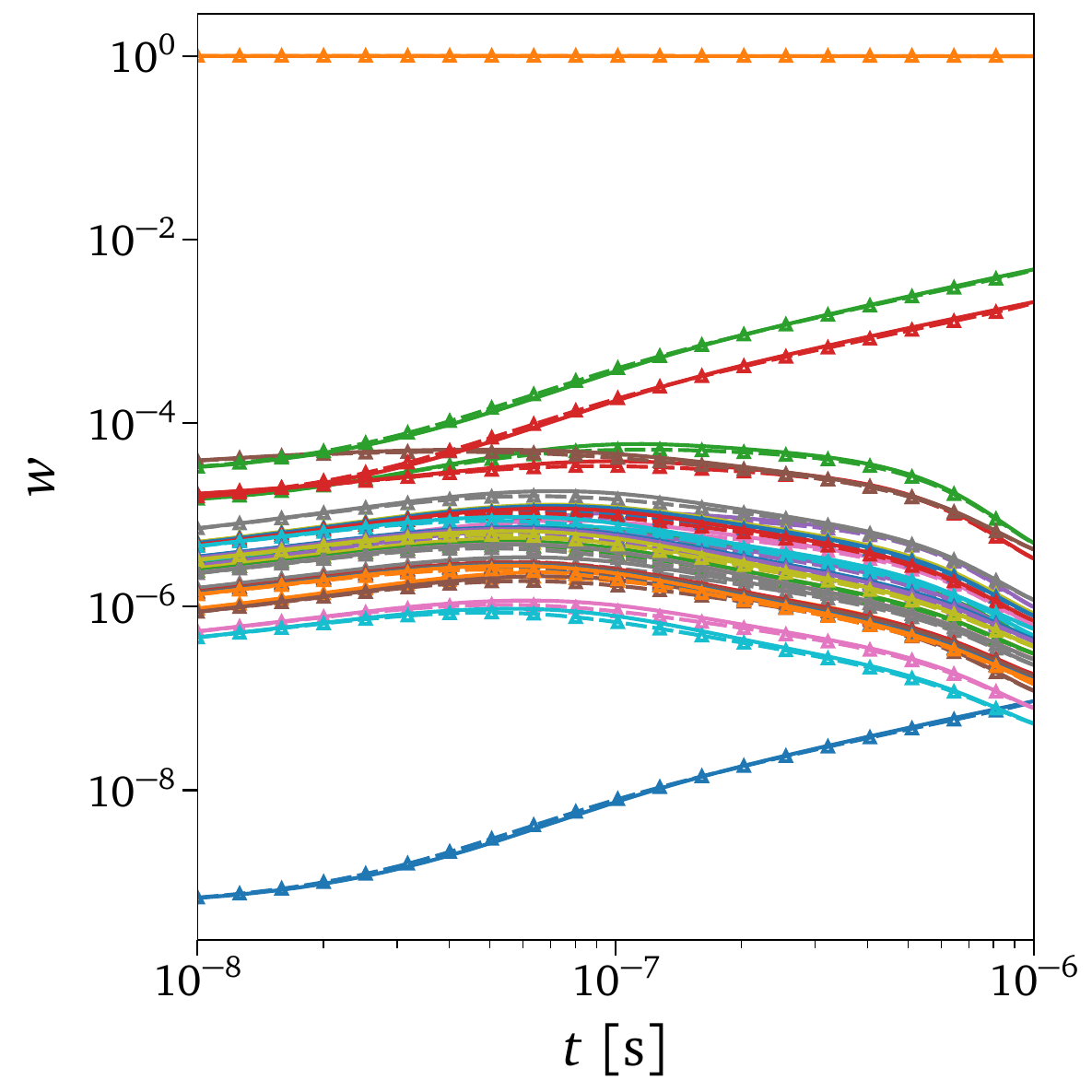}
    \end{subfigure}
    \\[2pt]
    \begin{subfigure}[htb!]{0.32\textwidth}
        \includegraphics[width=0.95\textwidth]{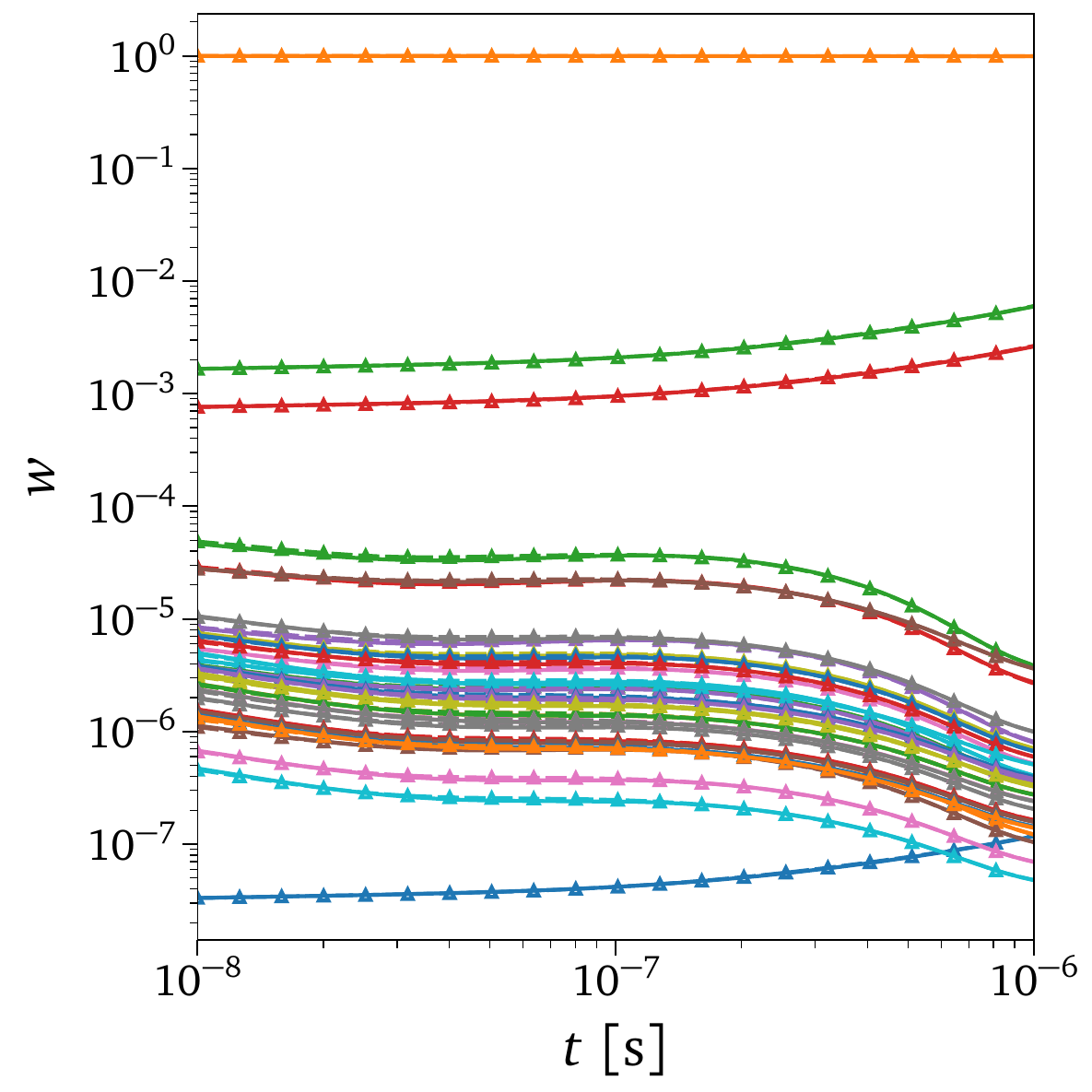}
    \end{subfigure}
    \begin{subfigure}[htb!]{0.32\textwidth}
        \includegraphics[width=0.95\textwidth]{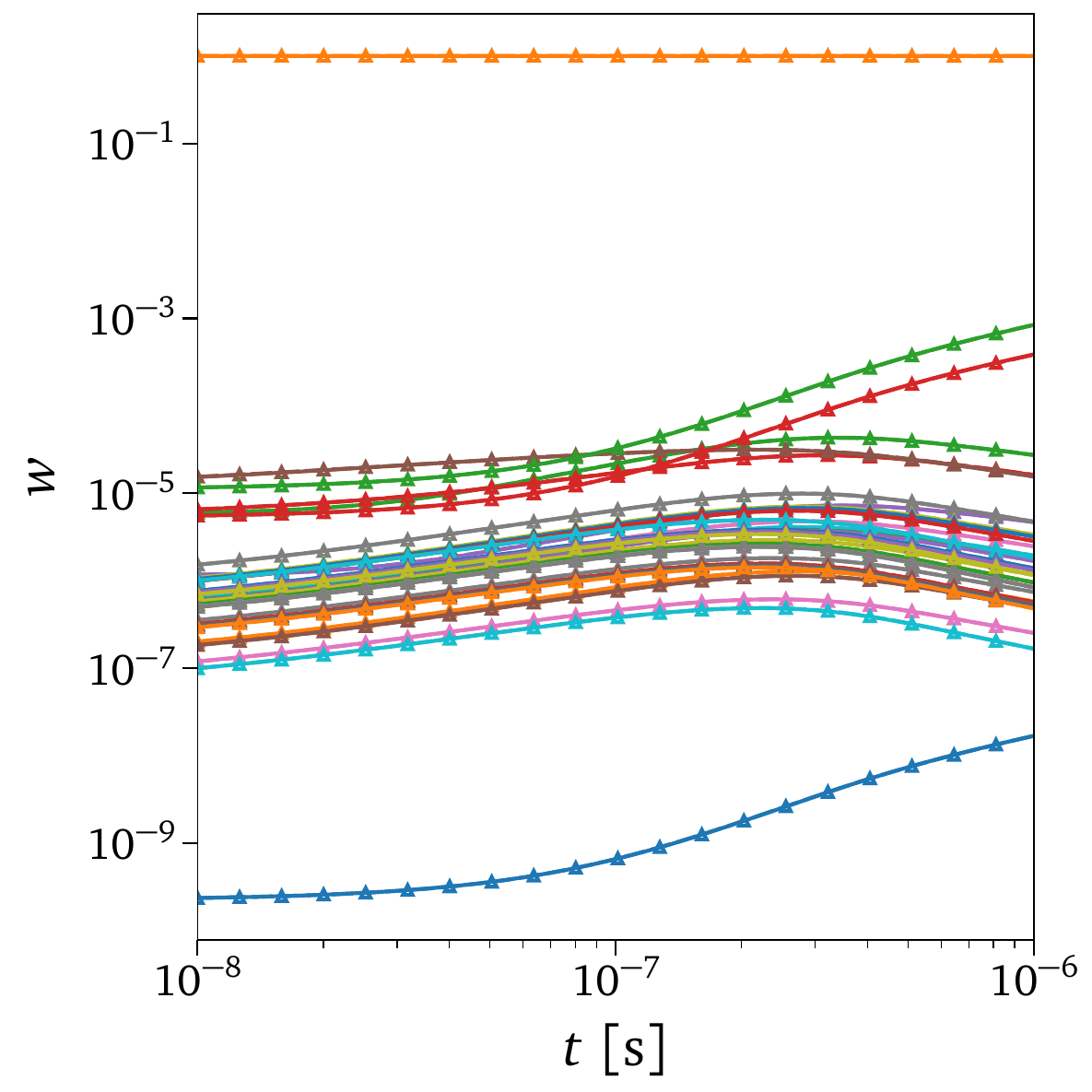}
    \end{subfigure}
    \begin{subfigure}[htb!]{0.32\textwidth}
        \includegraphics[width=0.95\textwidth]{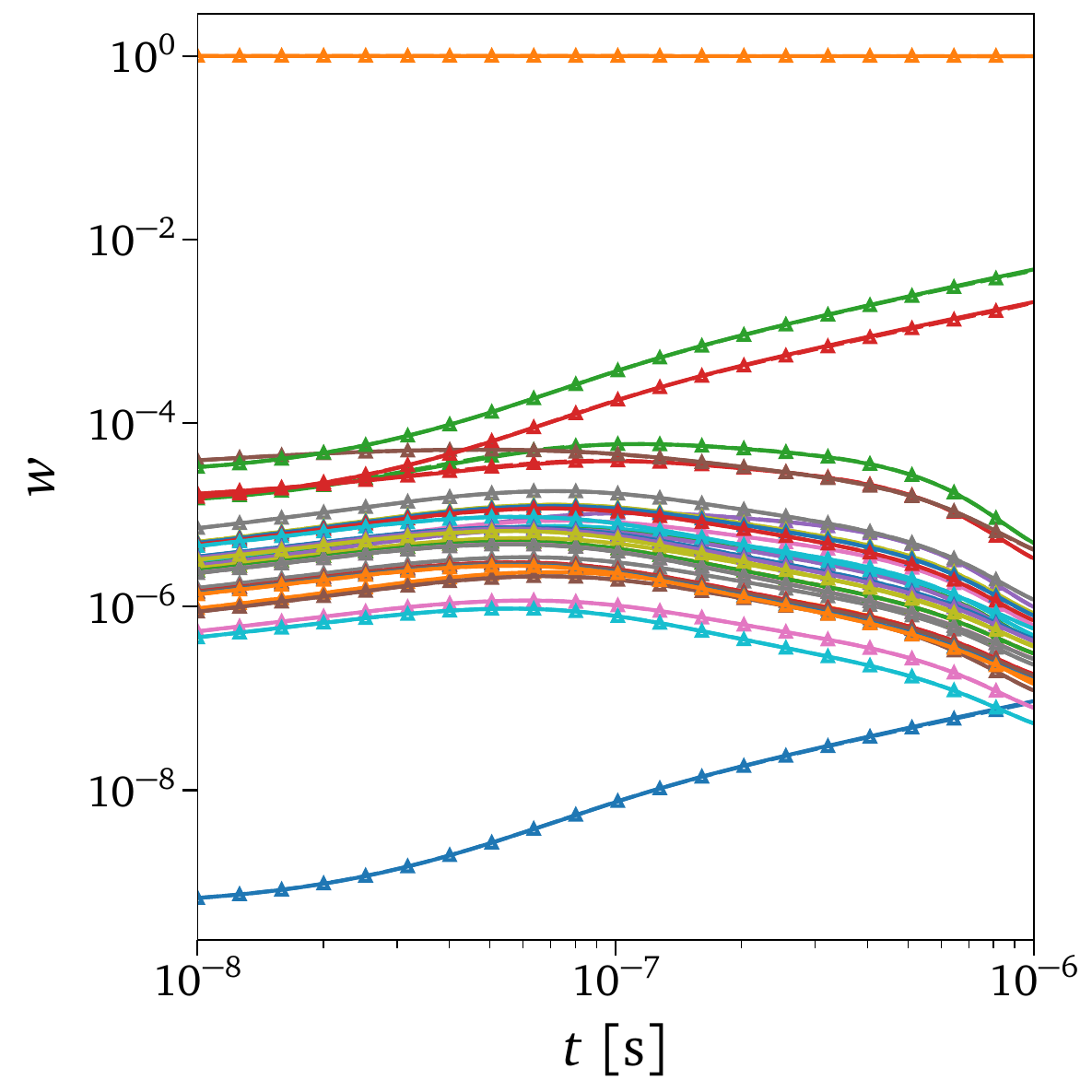}
    \end{subfigure}
    \caption{\textit{Comparison of MENO and reference solutions for the argon plasma problem.} Reference solutions (solid lines) are compared against predictions from the MENO model (dashed lines with markers) across three representative test cases (columns). Each subplot displays the time evolution of species mass fractions. Top row: MENO without corrections. Bottom row: fully trained MENO model.}
    \label{fig:plasma_cr.sol}
\end{figure}

\paragraph{Accuracy}
Figure~\ref{fig:plasma_cr.sol} presents three representative test cases, arranged by column, in which the surrogate model predictions for the pseudo-species mass fractions are compared against the reference solutions. The top row shows results from the MENO model without corrections, while the bottom row depicts predictions from the fully trained MENO model.

\begin{figure}[!htb]
    \centering
    \includegraphics[width=0.98\textwidth]{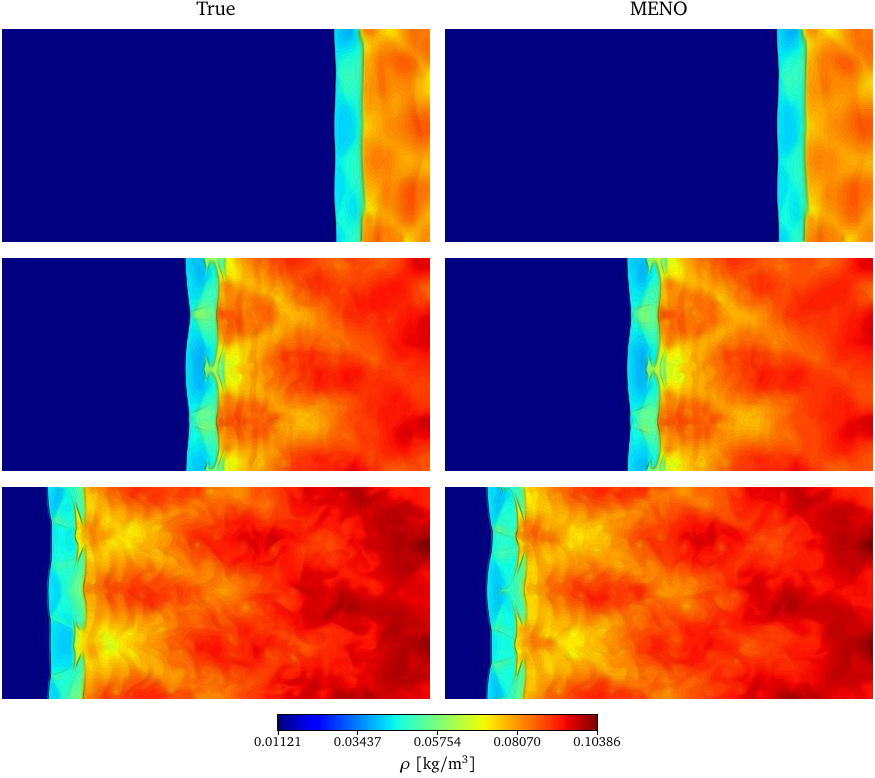}
    \vspace{-2mm}
    \caption{\textit{MENO vs. true solutions: Density snapshots from 2D simulation.} Comparison between MENO-based predictions, coupled with a CFD solver, and ground-truth results for the mass density $\rho$ at three time instants: \(t_i = [1,\,3,\,5] \times 10^{-4}\)~s.}
    \label{fig:2d.rho}
\end{figure}

\begin{figure}[!htb]
    \centering
    \includegraphics[width=0.98\textwidth]{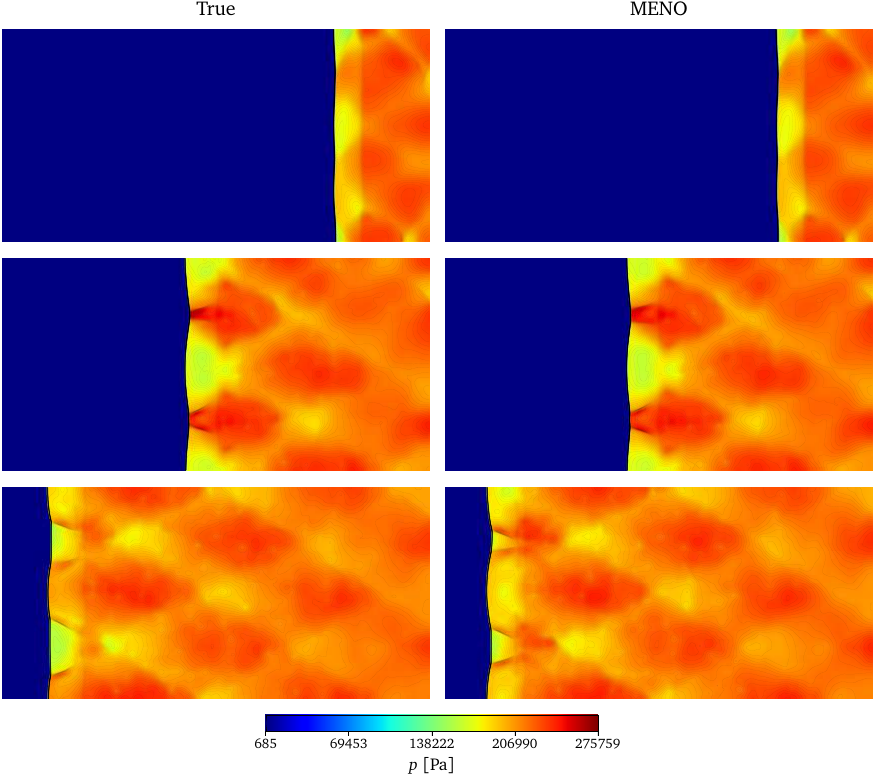}
    \vspace{-2mm}
    \caption{\textit{MENO vs. true solutions: Pressure snapshots from 2D simulation.} Comparison between MENO-based predictions, coupled with a CFD solver, and ground-truth results for the pressure $p$ at three time instants: \(t_i = [1,\,3,\,5] \times 10^{-4}\)~s.}
    \label{fig:2d.p}
\end{figure}

\begin{figure}[!htb]
    \centering
    \includegraphics[width=0.98\textwidth]{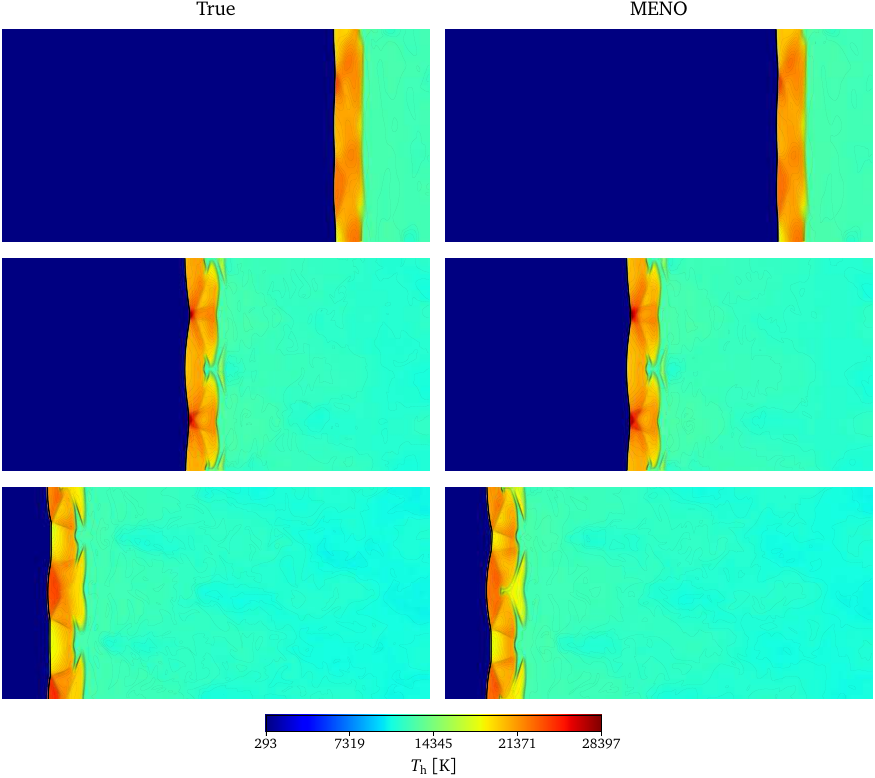}
    \vspace{-2mm}
    \caption{\textit{MENO vs. true solutions: Heavy-particle temperature snapshots from 2D simulation.} Comparison between MENO-based predictions, coupled with a CFD solver, and ground-truth results for the heavy-particle temperature $T_h$ at three time instants: \(t_i = [1,\,3,\,5] \times 10^{-4}\)~s.}
    \label{fig:2d.Th}
\end{figure}

\begin{figure}[!htb]
    \centering
    \includegraphics[width=0.98\textwidth]{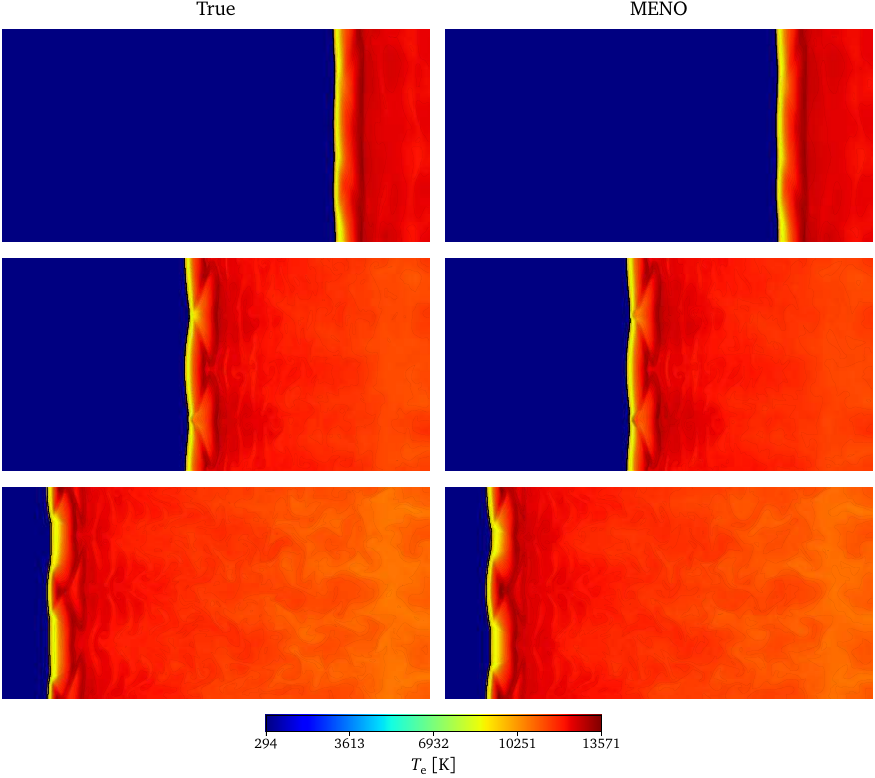}
    \vspace{-2mm}
    \caption{\textit{MENO vs. true solutions: Electron temperature snapshots from 2D simulation.} Comparison between MENO-based predictions, coupled with a CFD solver, and ground-truth results for the electron temperature $T_e$ at three time instants: \(t_i = [1,\,3,\,5] \times 10^{-4}\)~s.}
    \label{fig:2d.Te}
\end{figure}

Additional comparisons for the 2D test case (see figure \ref{fig:2d.xe}) are shown in figures~\ref{fig:2d.rho}–\ref{fig:2d.Te}, highlighting the agreement between the ground-truth solutions and the fully trained MENO model coupled with the CFD solver. These figures include spatial distributions at three time instants (\(t_i = [1,\,3,\,5] \times 10^{-4}\)~s) of key thermochemical quantities: mass density $\rho$, pressure $p$, heavy-particle temperature $T_\mathrm{h}$, and electron temperature $T_\mathrm{e}$.

\clearpage

\paragraph{Speedup}
Tables \ref{table:plasma_cr.0d.speedup.cpu} and \ref{table:plasma_cr.0d.speedup.gpu} report the computational speedup of MENO relative to the BDF integrator, for both CPU and GPU implementations. Speedup values are presented for a range of BDF orders, relative tolerances, and final integration times.

\begin{table}[!htb]
    \centering
    \begin{tabular}{c|c||c|c|c}
        \arrayrulecolor{black}\midrule
        \multicolumn{5}{c}{Speedup - CPU} \\
        \arrayrulecolor{black}\midrule
        \multicolumn{2}{c||}{BDF Scheme} & \multicolumn{3}{c}{Final Time [s]} \\
        \arrayrulecolor{gray}\midrule
        Order & RTol & $10^{-8}$ & $10^{-7}$ & $10^{-6}$ \\
        \arrayrulecolor{black}\midrule \midrule
        \multirow{3}*{2} & $10^{-6}$ & $4.42\times 10^{1}$ & $5.56\times 10^{1}$ & $6.67\times 10^{1}$ \\
        & $10^{-7}$ & $6.72\times 10^{1}$ & $8.86\times 10^{1}$ & $1.09\times 10^{2}$ \\
        & $10^{-8}$ & $1.19\times 10^{2}$ & $1.52\times 10^{2}$ & $1.64\times 10^{2}$ \\
        \arrayrulecolor{gray}\midrule
        \multirow{3}*{3} & $10^{-6}$ & $3.83\times 10^{1}$ & $4.66\times 10^{1}$ & $5.52\times 10^{1}$ \\
        & $10^{-7}$ & $4.60\times 10^{1}$ & $5.58\times 10^{1}$ & $6.68\times 10^{1}$ \\
        & $10^{-8}$ & $6.07\times 10^{1}$ & $7.61\times 10^{1}$ & $9.19\times 10^{1}$ \\
        \arrayrulecolor{gray}\midrule
        \multirow{3}*{4} & $10^{-6}$ & $3.58\times 10^{1}$ & $4.40\times 10^{1}$ & $5.53\times 10^{1}$ \\
        & $10^{-7}$ & $4.17\times 10^{1}$ & $5.23\times 10^{1}$ & $6.74\times 10^{1}$ \\
        & $10^{-8}$ & $5.12\times 10^{1}$ & $6.60\times 10^{1}$ & $8.65\times 10^{1}$ \\
        \arrayrulecolor{black}\midrule
    \end{tabular}
    \caption{\textit{CPU speedup of MENO relative to the BDF integrator for the argon plasma problem.} Speedup values are reported for varying BDF orders and relative tolerances, across multiple final integration times. All simulations were performed using a single CPU core.}
    \label{table:plasma_cr.0d.speedup.cpu}
\end{table}

\begin{table}[!htb]
    \centering
    \begin{tabular}{c|c||c|c|c}
        \arrayrulecolor{black}\midrule
        \multicolumn{5}{c}{Speedup - GPU} \\
        \arrayrulecolor{black}\midrule
        \multicolumn{2}{c||}{BDF Scheme} & \multicolumn{3}{c}{Final Time [s]} \\
        \arrayrulecolor{gray}\midrule
        Order & RTol & $10^{-8}$ & $10^{-7}$ & $10^{-6}$ \\
        \arrayrulecolor{black}\midrule \midrule
        \multirow{3}*{2} & $10^{-6}$ & $7.61\times 10^{2}$ & $9.35\times 10^{2}$ & $1.10\times 10^{3}$ \\
        & $10^{-7}$ & $1.16\times 10^{3}$ & $1.49\times 10^{3}$ & $1.80\times 10^{3}$ \\
        & $10^{-8}$ & $2.06\times 10^{3}$ & $2.56\times 10^{3}$ & $2.70\times 10^{3}$ \\
        \arrayrulecolor{gray}\midrule
        \multirow{3}*{3} & $10^{-6}$ & $6.59\times 10^{2}$ & $7.85\times 10^{2}$ & $9.07\times 10^{2}$ \\
        & $10^{-7}$ & $7.92\times 10^{2}$ & $9.39\times 10^{2}$ & $1.10\times 10^{3}$ \\
        & $10^{-8}$ & $1.04\times 10^{3}$ & $1.28\times 10^{3}$ & $1.51\times 10^{3}$ \\
        \arrayrulecolor{gray}\midrule
        \multirow{3}*{4} & $10^{-6}$ & $6.16\times 10^{2}$ & $7.41\times 10^{2}$ & $9.09\times 10^{2}$ \\
        & $10^{-7}$ & $7.17\times 10^{2}$ & $8.80\times 10^{2}$ & $1.11\times 10^{3}$ \\
        & $10^{-8}$ & $8.81\times 10^{2}$ & $1.11\times 10^{3}$ & $1.42\times 10^{3}$ \\
        \arrayrulecolor{black}\midrule
    \end{tabular}
    \caption{\textit{GPU speedup of MENO relative to the BDF integrator for the argon plasma problem.} Speedup values are reported for varying BDF orders and relative tolerances, across multiple final integration times. All MENO simulations were executed on a single GPU, while BDF runs were performed on a single CPU core.}
    \label{table:plasma_cr.0d.speedup.gpu}
\end{table}

\clearpage
\putbib  
\end{bibunit}

\end{document}